\documentclass[a4paper,12pt]{report}

\usepackage{fancyheadings}
\usepackage{graphicx}
\usepackage[utf8]{inputenc}
\usepackage{slashed}
\usepackage{braket}
\usepackage{float}

\usepackage[top=3cm,bottom=2cm,left=4cm,right=2cm]{geometry}
\usepackage{indentfirst}
\usepackage{setspace}	
\usepackage{amsmath,amssymb,amsfonts,textcomp}	

\RequirePackage[colorlinks,citecolor=blue,urlcolor=blue,linkcolor=blue]{hyperref}

\newcommand{\nin}{\noindent}
\newcommand{\be}{\begin{equation}}
\newcommand{\ee}{\end{equation}}
\newcommand{\bea}{\begin{eqnarray}}
\newcommand{\eea}{\end{eqnarray}}

\newcommand{\nn}{\nonumber}

\newcommand{\ol}{\overline}

\usepackage{color}

\usepackage{feynmf}

\begin{document}
\onehalfspace
\begin{titlepage}
\vfill
\begin{center}
~\\
\vspace{5cm}
{\huge\sc {\bf Aspects of fermion dynamics from \\
Lorentz symmetry violation}}\\
\vspace{6cm}
{\large\sc Author: Julio Leite}\\
{\large\sc Supervisor: Jean Alexandre}\\

\vspace{3cm}
{\large {\it A thesis submitted in fulfilment of the requirements\\
for the degree of Doctor of Philosophy\\
in the}\\
Theoretical Particle Physics and Cosmology Group\\
Department of Physics, King's College London}


\vfill

London - UK\\
2016

\end{center}
\end{titlepage}

\pagenumbering{arabic}
\newpage
\onehalfspace



\newpage




\newpage

\thispagestyle{empty}
\begin{center}
\textbf{Abstract}
\mbox{}
\end{center}
\vspace{1.8cm} 

In this thesis we are interested in understanding how Lorentz symmetry violation can affect some features of fermion dynamics and, perhaps, help to solve some well-known problems in particle physics, such as the origin of neutrino masses and oscillations.

Firstly, we consider two Lorentz-Invariance-Violating (LIV) models and investigate the possibility of generating masses and oscillations dynamically for both Dirac and Majorana neutrinos, using non-perturbative methods, such as the Schwinger-Dyson and the effective potential approaches. In our studies, Lorentz symmetric models are extended by the inclusion of higher-order LIV operators, which improve the convergence of loop integrals and introduce a natural mass scale to the theories. We then present how Lorentz invariance can be recovered, for both models, after quantisation, in such a way that the dynamical masses and mixing are the only quantum effects that remain finite.

Additionally, we study how matter fields, especially fermions, behave when coupled to two modified gravity models. Such modified gravity models break the 4-dimensional diffeomorphism invariance and, consequently, induce local Lorentz violation. In particular, we consider Horava-Lifshitz gravity, which presents an improved ultraviolet behaviour when compared to General Relativity (GR), and thus addresses a fundamental problem in physics: the non-(perturbative-)renormalisability of the theory of GR. We calculate the LIV one-loop corrections to the matter sector dispersion relations, after integration over graviton components, and show that, by imposing reasonable constraints on the energy scales of our gravity models, our results are consistent with the current bounds on Lorentz symmetry violation.

\newpage

\thispagestyle{empty}

\begin{center}
\textbf{Acknowledgments}
\end{center}

\vspace{1.8cm}
\onehalfspace

\noindent First of all I would like to thank my supervisor Jean Alexandre for giving me invaluable support through these past four years. His expertise together with his enormous patience have helped me to go further and achieve more. \\

\noindent I am also very grateful to Nick Mavromatos for his support and for the exciting discussions we had.\\

\noindent I would like to express my gratitude to everyone in the Department of Physics, especially to the members of the Theoretical Particle Physics \& Cosmology (TPPC) group.\\

\noindent For all the love and encouragement, I thank my family. In particular, I want to thank my mother Ana Lucia for being such a determined and strong woman, who keeps teaching me about overcoming difficulties in life.\\

\noindent I am also indebted to Sarah Seco who has always been extremely supportive. \\

\noindent I would like to thank my examiners Paul Saffin and Joao Magueijo for the discussion and valuable suggestions.\\

\noindent Finally, I want to thank the National Council for Scientific and Technological Development (CNPq - BRAZIL) for the financial support.

\onehalfspace
\chapter*{Declaration}
\mbox{}

I, Julio Rafael da Silva Leite, declare that the work presented in this thesis is my own. Where the work of others has been consulted, it is always cited accordingly. 	
This thesis content is based on research presented in the papers \cite{ALM1,ALM2,AL} below and done by myself in collaboration with others.
\\

\begin{enumerate}
	\item  J.~Alexandre, J.~Leite and N.~E.~Mavromatos,
  ``Lorentz-violating regulator gauge fields as the origin of dynamical flavor oscillations,''
  Phys.\ Rev.\ D {\bf 87} (2013) 12,  125029
  [arXiv:1304.7706 [hep-ph]];
	
  \item J.~Alexandre, J.~Leite and N.~E.~Mavromatos,
  ``Quasirelativistic fermions and dynamical flavor oscillations,''
  Phys.\ Rev.\ D {\bf 90} (2014) 4,  045026
  [arXiv:1404.7429 [hep-th]].

  \item J.~Alexandre and J.~Leite,
  ``Effective fermion kinematics from modified quantum gravity,''
  Class.\ Quant.\ Grav.\  {\bf 33}, no. 19, 195005 (2016)
  doi:10.1088/0264-9381/33/19/195005
  [arXiv:1506.03755 [hep-ph]].

\end{enumerate}
\vspace{5cm}

\noindent Julio Rafael da Silva Leite\\
King's College London\\
August 2016

\chapter*{Notation and conventions}\label{notation}
\mbox{}

Natural units, where $c=1=\hbar$, are used throughout this thesis, in such a way that the Planck mass is given by $M_P^{2} = (16\pi G_N)^{-1}$, 
where $c$ is the speed of light in vacuum, $\hbar$ is the Planck's constant and $G_N$ is the Newton's gravitational constant.\\

Unless stated otherwise, we work in $3+1$ space-time dimensions.\\

Greek letters $\mu,\nu,...$ denote space-time indices, while Latin letters $i,j,...$ denote space indices only. Time components are represented by the index $0$.\\

The usual summation convention for repeated indices is applied.\\

We adopt the mostly-minus convention $(+,-,-,-)$ for the metric signature when working in flat space-time, so that $p_\mu p^\mu=p_0^2-\vec{p}^2$, where	$\vec{p}^2 =p_1^2+p_2^2+p_3^2$. Whereas, when gravity is present, the mostly plus metric convention $(-,+,+,+)$ is adopted; this is the case in section \ref{subsec:HL} of chapter \ref{chap3}, as well as for the whole chapter \ref{chap6}, where Horava-Lifshitz gravity is considered. This more closely follows the conventions adopted in the respective fields.

\tableofcontents 

\chapter{Introduction}\label{Intro}
\mbox{}

The Standard Model of particle physics (SM) together with the theory of General Relativity (GR) form the best description of nature to present date. The accuracy of these models has been tested to ever higher limits in recent years with the discovery of the Higgs boson, the last missing part of the SM, at the Large Hadron Collider (LHC), and the first detection of gravitational waves, a prediction of GR, by the twin Laser Interferometer Gravitational-wave Observatory (LIGO) detectors. Despite all the experimental successes they have gathered, it is clear that these models do not give us a complete understanding of nature and, therefore, need to be modified in order to provide satisfactory answers to many open questions in particle physics.

The first problem, from a theoretical viewpoint, with these theories is related to the non-(perturbative-)renormalisability of GR, which forbids a unified description of both theories in certain energy regimes. The search for a fully consistent description of quantum gravity, being a very active research area, has led to many new ideas. One particularly interesting concept which often appears in these studies is the violation of Lorentz symmetry. In some cases Lorentz symmetry is a main ingredient of a given theory, and is only broken spontaneously by vacuum expectations values of tensor fields. In other cases, however, Lorentz symmetry is explicitly broken from the beginning, but emerges as an accidental symmetry at low energies. In addition to shedding some light on the problem of quantum gravity, Lorentz-Invariance-Violating (LIV) theories can also be used as powerful tools when tackling other important questions, as discussed below.

Another well-known problem left unanswered by the SM and GR is that of the generation of neutrino masses and oscillations. While the generation of quark, lepton and vector boson masses as a result of their couplings to the Higgs boson (after spontaneous symmetry breaking) is now widely accepted, the origin of tiny neutrino masses cannot be naturally explained in the same way. Within this context, the so-called seesaw mechanism \cite{seesaw1,seesaw2,seesaw3}, which requires the existence of new (sterile) neutrino states without any standard model interactions, is generally seen as the most elegant and simple alternative to explain the smallness of neutrino masses. Nevertheless, the seesaw mechanism does not provide the full picture because it demands very heavy sterile neutrinos in order to lower the SM neutrino masses, but it does not explain how such large masses are generated. Furthermore, the possibility of generating neutrino masses dynamically without the involvement of sterile neutrinos cannot be ruled out and deserves consideration.

In this thesis, we propose different ways, based on LIV models, of exploring the problem of neutrino masses and oscillations with or without the introduction of sterile neutrinos. In our studies, the usual, Lorentz symmetric, models are extended by the inclusion of LIV operators which are suppressed by a large mass scale $M$. The LIV operators considered here contain, in general, higher-order space derivatives, which improve the convergence of loop integrals without introducing the extra degrees of freedom (ghosts) generally associated with the introduction of higher-order time derivatives. We take advantage of the fact that a large mass scale is naturally introduced to suppress the LIV operators, preventing significant LIV effects at low energies, to investigate the dynamical generation of neutrino masses and oscillations, with the use of non-perturbative methods.

In our first model, studied in chapter \ref{chap4}, we consider the coupling of two flavoured fermion (neutrino) fields to LIV vector gauge bosons and study the dynamical generation of masses and oscillations by using the Schwinger-Dyson approach. No vector boson mass is generated and, therefore, the LIV vector bosons can be viewed as regulator fields. In solving the dynamical equations for Dirac neutrinos, we find that, although neutrino masses are generated dynamically in different situations, flavour oscillations are only allowed to take place in the case where one of the mass eigenstates is massless and the mixing angle is maximal, $\theta =\pm \pi/4$. We then show that the Lorentz symmetric limit, {\it i.e.} when the LIV mass scale $M\to \infty$ and the gauge couplings go to zero simultaneously, can be taken, after quantisation, in such a way that the dynamically generated neutrino masses remain finite, whereas the gauge boson completely decouples from neutrinos. Finally, we extend our analysis to include Majorana neutrinos in two different contexts, demonstrating that Majorana masses can be generated when considering only left-handed neutrinos, as well as when sterile neutrinos are taken into account.

The second model, proposed to study the dynamical generation of neutrino masses and oscillations, is discussed in chapter \ref{chap5} and consists of fermions (neutrinos) with LIV kinematics coupled among themselves through four-fermion interaction terms. The fermions, in this case, present an almost relativistic behaviour in both IR and UV regimes, but depart from it in an intermediate regime defined by the LIV mass scale $M$, hence we call them ``quasi-relativistic'' fermions. We demonstrate, using the effective potential approach, that the four-fermion interaction allows for neutrino mass generation (and, in the two flavour case, oscillations as well) for any coupling strength $g$. Because the model does not impose a critical coupling below which no mass is generated, the Lorentz symmetric limit $M\to \infty$ and $g\to 0$ can be taken, after quantisation, in such a way that the masses generated by quantum corrections remain finite, while neutrinos become free relativistic particles with oscillating flavours. The values of the masses and the mixing angle generated by this mechanism can then be chosen according to phenomenology. Finally, we show how our findings can be extended to a seesaw-type model with Majorana fields.

We also consider modified gravity models and investigate the appearance, via quantum gravity corrections, of LIV effects in the matter sector of the SM, especially for fermions. The alternative theories of gravity considered here are based on a preferred time direction, thus breaking local Lorentz symmetry. In particular, we couple matter fields to the so-called Horava-Lifshitz gravity (HL) which is built upon an anisotropic (Lifshitz) scaling between space and time. The Lifshitz scaling is a crucial element of this theory because it allows for the introduction of higher-order space derivatives, while keeping the number of time derivatives to a minimum, thus improving the UV behaviour of the theory when compared to GR. However, for this scaling to be implemented, one needs to break the invariance under 4-dimensional diffeomorphisms which, in addition to breaking Lorentz symmetry locally, implies the appearance of a new degree of freedom in the gravity sector, known as the scalar graviton. 

The last of our studies, presented in chapter \ref{chap6}, concerns a classical fermion and a classical complex scalar field propagating on two different 4-dimensional diffeomorphism breaking gravity backgrounds. Although both gravity models are invariant under foliation-preserving diffeomorphisms, only one of them, which is the non-projectable version of HL gravity, exhibits an improved UV behaviour as a consequence of the anisotropy between space and time. In both cases, we derive how the matter field dispersion relations are modified by one-loop LIV corrections after integrating out gravitons. The first model involves quadratic divergences at one-loop, as in Einstein gravity, while for the second one we find logarithmic divergences only. The fact that these two models behave differently in the UV does not give rise to significant differences in the IR phenomenology for matter fields. On the one hand, we find that when assuming generic values for the parameters, both models identify $10^{10}$ GeV as the  characteristic scale up to which their respective results can be made consistent with current upper bounds on Lorentz symmetry violation. For the first model, treated as an effective field theory, the $10^{10}$ GeV upper bound is seen as the cut off of the theory, whereas, for the second one, $10^{10}$ GeV represents the maximum allowed value for the Horava-Lifshitz scale $M_{HL}$, {\it i.e.} the mass scale suppressing the higher-order operators. On the other hand, if one wished to do so, it is always possible to fine-tune the parameters in both models in such a way that the signal for Lorentz symmetry violation disappears.

\vspace{0.5cm}

In addition to the chapters \ref{chap4}, \ref{chap5} and \ref{chap6} where our main work is presented, we dedicate three other chapters (chapters \ref{chap1}, \ref{chap2} and \ref{chap3}) to review some of the fundamental theory upon which our work is based.

Chapter \ref{chap1} contains a review of the theory of neutrino masses and oscillations. After summarising some historical events that led to our current level of understanding about neutrinos, we show how they fit within the SM and why we need to extend the present picture. We then present the different neutrino mass terms which are theoretically possible and, finally, discuss the difference between neutrino flavour and mass states, and how their non-trivial relationship gives rise to the phenomenon of flavour oscillations. In chapter \ref{chap2}, we focus on well-known examples of dynamical mass generation for fermions. We start by deriving the Schwinger-Dyson (SD) equation for the fermion propagator in Quantum Electro-Dynamics (QED), following then with some examples where the dynamical mass is calculated by means of the SD approach. After that, we turn our attention to four-fermion interaction models, such as the Nambu-Jona-Lasinio (NJL) and Gross-Neveu models, and explore how masses can be generated dynamically using the effective potential approach. Additionally, chapter \ref{chap3} introduces the concept of Lorentz violation. We first present the Standard Model Extension (SME) which summarises the possibilities of local Lorentz violation as a result of spontaneous symmetry breaking, and pay special attention to its neutrino sector. We then discuss explicit Lorentz violation, global and local, in the context of Lifshitz-type theories, where we focus on Horava-Lifshitz gravity and its different versions.

Finally, in chapter \ref{Conclusions}, we present our concluding remarks and discuss possible directions for future work.

\chapter{The theory of neutrino masses and oscillations}\label{chap1}
\mbox{}

The standard model of particle physics has once again been confirmed to provide the best description of the subatomic world to present date with the discovery of its last missing piece, the Higgs boson, at the Large Hadron Collider (LHC)~\cite{Higgs1,Higgs2}. However, as many important questions in particle physics remain unanswered, such as the origin of neutrino masses, the SM needs to be modified to give a more complete description of nature.

In this chapter we consider one of the most active research areas in physics beyond the SM: neutrino physics. The focus being on two of its fundamental aspects, which are closely related to each other, the generation of neutrino masses and the phenomenon of neutrino (flavour) oscillations.\footnote{For more detailed information on neutrinos, see the following books~\cite{BilenkyBook, GiuntiBook, ZuberBook} upon which most parts of this chapter are based.}

In the next section we present some important historical events that led to the discovery of neutrinos and the development of the respective theory. In the section~\ref{sec:N+SM} we show how these particles fit within the SM. Finally, in the last sections, we introduce the different possible neutrino mass terms and the basics of neutrino mixing and oscillations.

\section{A summary of neutrinos history}\label{sec:NHist}
\mbox{}

Neutrinos, as well as other particles in the SM, such as the Higgs boson, were first postulated theoretically and only many years later observed in nature. The theoretical prediction of the existence of neutrinos was made in 1930 by W. Pauli as a way of providing a consistent explanation for the observed experimental results regarding the study of $\beta$-decays. At that time, $\beta$-decays were believed to be processes in which a nucleus $^A_Z X$, where $A$ represents the mass number (number of nucleons) and $Z$ the atomic or proton number of the element $X$, decays into another nucleus $_{Z+1}^A X$ while emitting an electron with a given kinetic energy, {\it i.e.} $^A_Z X \to _{Z+1}^A X+e^- $. If this was indeed the case, the resulting electron should have a fixed kinetic energy which could be easily calculated in terms of the masses of the electron and the initial and final nuclei. However, instead of a discrete spectrum, experiments revealed a continuous spectrum. Given such an inconsistency, in 1930, W. Pauli proposed a ``desperate remedy'' to solve this problem by postulating that, in addition to the electron, another particle is emitted in a $\beta$-decay. According to his proposal, such a new particle, which was later named neutrino by E. Fermi, should be lightweight, spin-1/2 and neutral. Following Pauli's idea, the $\beta$-decay of $^A_Z X$ should be described by
	\be 
		^A_Z X \to _{Z+1}^A X+e^- +\overline{\nu}~,
	\ee
where $\nu$ represents the new particle postulated by him, the neutrino.

Not so long after Pauli's theoretical prediction, E. Fermi proposed the first theory of $\beta$-decay which is based on the following four-fermion interaction term~\cite{Fermibeta}: $\propto G_F (\overline{p} \gamma_\alpha n)(\overline{e}\gamma^\alpha \nu)$, where $G_F$ is the Fermi constant, and $p$, $n$, $e$ and $\nu$ represent the proton, neutron, electron and neutrino, respectively. Over the years, Fermi's theory of the $\beta$-decay was improved step by step, culminating in what we know today as the the electro-weak (EW) theory present in the SM.

Experimental progress, on the other hand, was also made, with electron neutrinos ($\nu_e$) being detected for the first time in 1956~\cite{1stneutdet1,1stneutdet2}. In the following decade, a second kind of neutrino, the muon neutrino ($\nu_\mu$), was detected~\cite{muonneut}. Finally, in 2000, the last of the three known neutrinos, the tau neutrino ($\nu_\tau$), was found~\cite{tauneut}.

Although neutrinos, as we will see in the next section, are considered to be massless particles in the SM, experimental results started to indicate that they actually have a very little mass instead of none at all. The first evidence for this can be understood in terms of the ``solar neutrino problem'', consisting in the fact that only about one third of the expected neutrino flux coming from the Sun has been observed on Earth~\cite{solarneutpro}. As a way out of this problem, Pontecorvo suggested that neutrinos oscillate ({\it i.e.} change flavours while propagating), and therefore only a fraction of the expected amount of a given neutrino flavour is observed on Earth, while the rest of them change flavours, thus making them invisible in an experiment built to observe one kind of flavour neutrino only. In the formalism developed by Pontecorvo, neutrino oscillations only take place if neutrinos are massive particles. 

It was only in 1998 that neutrino oscillations could be confirmed experimentally in the Super-Kamiokande experiment which considered atmospheric neutrinos. After this first concrete evidence for neutrino oscillations, many other experimental results have reinforced the idea of neutrino oscillations in different situations, e.g. results by the SNO collaboration with solar neutrinos~\cite{SNO} and the KamLAND experiment results with reactor neutrinos~\cite{KamLAND}. Therefore, with the confirmation of neutrino oscillations, neutrino masses should be added to the SM. Nonetheless, as we will see in the following sections, the usual mechanism behind mass generation in the SM, the Higgs mechanism, cannot be the main one behind neutrino mass generation. As a consequence, the SM needs to be extended to incorporate neutrino masses and oscillations.

\section{The neutrino sector of the standard model}\label{sec:N+SM}
\mbox{}

The standard model, which is built upon the local gauge $SU(3)_c\times SU(2)_L \times U(1)_Y$ invariance of massless fields, explains how quarks and leptons interact according to three of the four fundamental forces of nature: strong, weak and electromagnetic force. Nonetheless, since most of the particles observed in nature are, actually, massive, this symmetry should be broken in order to allow for mass terms to appear. This is achieved with the introduction of the Higgs field which breaks the $SU(2)_L\times U(1)_Y$ symmetry spontaneously by acquiring a non-trivial vacuum expectation value (vev), {\it i.e.} through the well-known Higgs mechanism. While the $SU(3)_c$ gauge symmetry is responsible for the strong interaction which affects the quarks directly and is mediated by gluons, the $SU(2)_L\times U(1)_Y$ symmetry is associated with the unified electro-weak force felt by quarks and leptons. 

The SM leptons, neutrinos and charged leptons, do not interact via the strong force, and as such they transform as singlets with respect to the $SU(3)_c$ gauge symmetry. Because our aim here is the study of neutrinos, we omit here the role played by strong interactions. We therefore focus on the EW sector of the SM, proposed by Glashow, Weinberg and Salam in the sixties, which had its beginnings with Fermi's $\beta$-decay theory, as discussed above.

Experimental results have revealed the existence of three generations of charged leptons ($l_\alpha$) with $\alpha=e, \mu, \tau$ (electron, muon, tau) and, associated with each of them, a different neutrino ($\nu_\alpha$). These six spin-$1/2$ particles (fermions) form the lepton content of the SM. These lepton fields, which are massless before the EW symmetry breaking, can be decomposed into left-handed (LH) and right-handed (RH) components. For a generic field $\psi$, we have 
	\be\label{LHRH}
		\psi = \psi_L + \psi_R~,~\mbox{where}~~ \psi_{L(R)} = P_{L(R)}\psi = \frac{1-(+)\gamma_5}{2}\psi~,
	\ee
where $\psi_L$ and $\psi_R$ are the LH and RH components of $\psi$, respectively.

Within the SM, as only LH fields interact via the weak force, LH neutrinos and LH charged leptons of the same generation are grouped together and transform as doublets under the $SU(2)_L$ symmetry 
	\be\label{doublet}
		{\bf L'}_{\alpha L} = \begin{pmatrix} \nu'_{\alpha L} \\ l'_{\alpha L} \end{pmatrix}, ~~~~\alpha = e, \mu, \tau~.
	\ee
RH charged leptons ($l'_{\alpha R}$), on the other hand, transform as singlets under the $SU(2)_L$ symmetry, whereas RH (or sterile) neutrinos, which have not been found in nature yet~\cite{white2}, are not present in the SM.

The Lagrangian for massless leptons in the SM can then be written as
	\be\label{masslesslep}
		\mathcal{L} = \sum_{\alpha=e,\mu,\tau}\left(\overline{{\bf L'_{\alpha L}}} i \slashed{D} {\bf L'}_{\alpha L} + \overline{l'_{\alpha R}} i \slashed{D} l'_{\alpha R}\right)=\sum_{\alpha=e,\mu,\tau}\left(\overline{l'_{\alpha L}} i \slashed{D} l'_{\alpha L} + \overline{l'_{\alpha R}} i \slashed{D} l'_{\alpha R}+\overline{\nu'_{\alpha L}} i \slashed{D} \nu'_{\alpha L}\right)~,
	\ee
where, instead of partial derivatives, covariant derivatives ($D_\alpha$) are introduced to preserve the gauge invariance of the theory
	\bea\label{covdev}
		D_\lambda {\bf L'}_{\alpha L} = \left(\partial_\lambda + i g_1\frac{1}{2} \vec{\sigma}\cdot \vec{A}_\lambda +ig_2 \frac{1}{2} Y_L B_\lambda\right){\bf L'}_{\alpha L}~\\
		D_\lambda l'_{\alpha R} = \left(\partial_\lambda +ig_2 \frac{1}{2} Y_R B_\lambda \right) l'_{\alpha R}~,
	\eea 
with $A^i_\lambda$ ($i=1,2,3$) and $g_1$ being the gauge fields and coupling constant related to the $SU(2)_L$ group, while $B_\lambda$ and $g_2$ are the Abelian gauge field and the coupling constant associated with the $U(1)_Y$ group. In addition, $\sigma^i$ represent the $2\times 2$ Pauli matrices and $Y$ is the hypercharge. 

The Lagrangian~(\ref{masslesslep}) clearly needs to be extended to include mass terms. However, it is not difficult to see that a mass term for fermions of the generic form: $m\overline{\psi} \psi = m(\overline{\psi_L}\psi_R + \overline{\psi_R} \psi_L)$ is not invariant under the SM gauge symmetries and, consequently, cannot be directly added to the SM Lagrangian. Therefore, although the SM gauge symmetries need to be broken in order to allow for massive fermions, the symmetry breaking cannot be explicit. This problem was finally overcome by means of spontaneous symmetry breaking with the Higgs mechanism, which we discuss below.

\subsection{The electro-weak spontaneous symmetry breaking}\label{sec:EWSB}

In order to give masses to the SM particles, a new elementary field $\Phi$ needs to be introduced in the SM. The Higgs field $\Phi$ should transform as a doublet under the $SU(2)_L$ symmetry:
	\bea\label{Higgs}
		\Phi = \begin{pmatrix} \phi^+ \\ \phi^0 \end{pmatrix}~,
	\eea 
where $\phi^+$ and $\phi^0$ describe an electrically charged and a neutral complex scalar field, respectively. Similarly to what happens in the lepton sector, in order to preserve the gauge invariance of the theory, we must use covariant derivatives with respect to the Higgs field, instead of partial ones. The need for covariant derivatives naturally introduces interactions between the $SU(2)_L\times U(1)_Y$ gauge fields and $\Phi$. Forgetting about the fermions in the SM for now, the Lagrangian for the Higgs field is then given by
	\be\label{HiggsL}
		\mathcal{L}_H = (D^\alpha \Phi^\dagger)( D_\alpha \Phi) + \mu^2 (\Phi^\dagger \Phi) - \lambda (\Phi^\dagger \Phi)^2~,  
	\ee
where $\mu^2$ and $\lambda$ are positive constants. 

The Lagrangian above is obviously invariant under the gauge symmetries of the SM, however, one can show, using the potential for $\Phi$, that the field has a non-trivial vev: $v=\mu/\sqrt{\lambda}$, which breaks the electro-weak, $SU(2)_L\times U(1)_Y$, symmetry spontaneously. Taking into account the Higgs vev, we can express the Higgs field, in the so-called ``unitary gauge'', as
	\be\label{HiggsUni}
		\Phi = \begin{pmatrix} 0 \\ \frac{v+H(x)}{\sqrt{2}}\end{pmatrix}~,
	\ee
where the neutral scalar field $H(x)$ is defined in such a way that its vev vanishes.

It can now be shown by replacing~(\ref{HiggsUni}) into~(\ref{HiggsL}) and re-arranging the terms in the standard way, that three of the initially four massless gauge fields become massive. The only one which remains massless is the field associated with the electromagnetic force, $A_\lambda$. The three massive fields mediate the weak interactions, and two are electrically charged: $W_\lambda =W^{+}_\lambda$ and $W_\lambda^\dagger =W^{-}_\lambda$, while the other, $Z^0$, is neutral. In terms of the initial fields, the new fields can be written as
	\bea
		W_\lambda &=& \frac{1}{\sqrt{2}}(A_\lambda^1- i A_\lambda^2)~,\\
		Z_\lambda &=& \cos \theta_W A_\lambda^3 - \sin \theta_W B_\lambda~,\nn	\\
		A_\lambda &=& \sin \theta_W A_\lambda^3 + \cos \theta_W B_\lambda~, \nn
	\eea
where $\theta_W$ is the weak angle, defined by
	\bea
		\tan (\theta_W) &=& g_2/g_1~.
	\eea
	
After the EW symmetry breaking, the interaction terms which appear in Lagrangian~(\ref{masslesslep}) become
	\be\label{EWI}
			\mathcal{L}_I = \left(-\frac{g_1}{2\sqrt{2}} j^{CC}_\lambda W^\lambda + h.c.\right)-\frac{g_1}{2\cos{\theta_W}} j^{NC}_\lambda Z^\lambda - e j^{EM}_\lambda A^\lambda~,
	\ee
where $h.c.$ stands for Hermitian conjugate. The charge of a proton ($e$) and the charged, the neutral and the electromagnetic currents are respectively given by
	\bea\label{cdef}
		e &=& \frac{g_1 g_2}{\sqrt{g_1^2+g_2^2}}~,\\		
		j^{CC}_\lambda &=& 2 \sum_{\alpha=e,\mu,\tau} \overline{\nu'_{\alpha L}} \gamma_\lambda l'_{\alpha L} ~,\nn\\
		j^{NC}_\lambda &=& \sum_{\alpha=e,\mu,\tau}\left(\overline{\nu'_{\alpha L}}\gamma_\lambda\nu'_{\alpha L}-\overline{l'_{\alpha L}}\gamma_\lambda l'_{\alpha L}\right)-2\sin^2(\theta_W)j^{EM}_\lambda~,\nn\\
		j^{EM}_\lambda &=& -\sum_{\alpha=e,\mu,\tau}\left( \overline{l'_{\alpha L}}\gamma_\lambda l'_{\alpha L}+\overline{l'_{\alpha R}}\gamma_\lambda l'_{\alpha R}\right)~.\nn
	\eea

In addition to generating masses for some of the gauge fields that mediate the weak interactions~(\ref{EWI}), charged lepton masses are also generated via the interaction between the Higgs field and such particles. According to the $SU(2)_L\times U(1)_L$, the following Yukawa interaction term between leptons and the Higgs field is allowed and must be included in the Lagrangian of the SM:
	\be\label{Yuk}
		\mathcal{L}_{Yuk} = -\sum_{\alpha,\beta = e, \mu, \tau} y_{\alpha \beta} \overline{{\bf L'_{\alpha L}}}\left(\Phi\right) l'_{\beta R} + h.c.~,
	\ee
where $y_{\alpha\beta}$ are the Yukawa couplings. When $\Phi$ acquires a non-vanishing vev,  we can substitute~(\ref{HiggsUni}) into~(\ref{Yuk}) to obtain
	\be\label{Yukmass}
		\mathcal{L}_{Yuk} = -\sum_{\alpha,\beta = e, \mu, \tau} \overline{l'_{\alpha L}}  M_{\alpha \beta}^{(cl)} l'_{\beta R} + h.c. +\cdots~,
	\ee
where $\cdots$ represent terms depending on the field $H$. The term shown in~(\ref{Yukmass}) is the mass terms for the charged leptons, where $M_{\alpha\beta}^{(cl)}=(v/\sqrt{2})y_{\alpha \beta}$ are the elements of the the $3\times 3$ complex mass matrix $M^{(cl)}$.

A complex matrix like $M^{(cl)}$ can be diagonalised by a bi-unitary transformation~\cite{BilenkyBook}. Let $U_L$ and $U_R$ be $3\times 3$ unitary matrices and $m^{(cl)}=m_\alpha \delta_{\alpha\beta}$, where the elements $m_\alpha$ are real. $M^{(cl)}$ can then be expressed in terms of $U_L$, $U_R$ and $m^{(cl)}$ as
	\be\label{cldiag}
		M^{(cl)} = U_L m^{(cl)} (U_R)^\dagger~.
	\ee
Replacing (\ref{cldiag}) into~(\ref{Yukmass}), the mass term becomes	
	\be\label{diagmm}
		\mathcal{L}_m^{(cl)} = - \sum_{\alpha} \overline{l_{\alpha L}}  m_{\alpha}^{(cl)} l_{\alpha R} + h.c.= - \overline{L_L} m^{(cl)} L_R + h.c.~,
	\ee
where we have defined the new (unprimed) fields as
	\be\label{leptransf}
	l_{\alpha L(R)} = \sum_{\beta} (U_{L(R)})_{\alpha \beta} l'_{\beta L(R)}~~\mbox{or}~~L_{L(R)} = U_{L(R)} L'_{L(R)}~,
	\ee
with
	\be
		L_{L(R)} = \begin{pmatrix} l_{eL(R)}\equiv e_{L(R)} \\ l_{\mu L(R)}\equiv\mu_{L(R)} \\ l_{\tau L(R)}\equiv\tau_{L(R)} \end{pmatrix}~.
	\ee
Thus, the unprimed fields ($l=l_L + l_R$) represent the charged lepton fields with definite mass, {\it i.e.} the electron ($e = e_L +e_R$ ), muon ($\mu=\mu_L+\mu_R$) and tau ($\tau=\tau_L+\tau_R$).

It is natural to re-express all contributions in the SM in terms of the ``unprimed'' charged lepton fields by using the relation~(\ref{leptransf}). For all terms where the only leptons are the charged ones, this transformation can be easily performed, and all contributions coming from the $U$ matrices disappear due to the unitarity of $U_L$ and $U_R$. On the other hand, when considering the charged current (CC) interaction term, which mixes charged leptons and neutrinos, upon the substitution~(\ref{leptransf}), we find:
	\be\label{ccint}
		\mathcal{L}_I^{(CC)} \to -\frac{g_1}{\sqrt{2}} \left(\overline{\nu'_{L}}\gamma_\lambda U_{L}^\dagger L_L\right) W^\lambda+h.c.~=-\frac{g_1}{\sqrt{2}}\left( \overline{\nu_{L}}\gamma_\lambda L_L \right) W^\lambda+h.c.~, 
	\ee
where we have defined the neutrinos which enter the CC interaction together with the charged leptons as the flavour (unprimed) neutrinos
	\be\label{neutransf}
		\nu_L = U_L \nu'_L~~\mbox{with}~~ \nu_L = \begin{pmatrix} \nu_{eL}\\ \nu_{\mu L} \\ \nu_{\tau L} \end{pmatrix}~.
	\ee
Thus, with the field redefinitions~(\ref{leptransf}) and~(\ref{neutransf}), we can write all terms in the SM as functions of the unprimed lepton fields, with the advantage that in this basis the mass matrix for charged lepton is real and diagonal. For massless neutrinos in the SM, the flavour neutrinos are also ``mass eigenstates''; however, as we will show below, when extending the SM in order to account for neutrino masses, flavour neutrinos are not, in general, mass eigenstates, but a mixture of massive neutrinos. 
	
Because no RH neutrino fields are present in the SM, no mass term can be generated for neutrinos by the mechanism described through the equations~(\ref{Yuk}) and (\ref{Yukmass}). Nonetheless, as already discussed, the observation of neutrino oscillations strongly suggests that neutrinos are massive particles. Thus, in the next section, we discuss some ways of going beyond the SM to provide neutrino fields with mass terms.

\section{Neutrino masses}\label{sec:NMass}
\mbox{}

The nature of neutrinos, {\it i.e.} whether they are Dirac or Majorana fermions, is still unknown. If, similar to the other fermions in the SM, neutrinos are Dirac fermions, it is necessary to add RH neutrino fields to the SM in order to generate Dirac mass terms, as in~(\ref{Yukmass}). If, however, neutrinos and anti-neutrinos are indistinguishable,{\it i.e.} they are Majorana fermions, we can construct Majorana mass terms which do not require the introduction of RH neutrinos in the SM, but bring other consequences, as we discuss later. Finally, a more general and perhaps efficient alternative is to include ``Dirac and Majorana'' mass terms as in the seesaw mechanism~\cite{seesaw1,seesaw2,seesaw3}. We consider below all of these alternatives in some detail. 

\subsection{Dirac mass term}\label{sec:DMT}

In analogy with the other fermions in the SM, let us assume that neutrinos are Dirac fermions to show how their masses can be generated by means of the Higgs mechanism. As seen previously, Dirac mass terms, as in~(\ref{Yuk}) for charged leptons, can only be constructed if RH neutrino fields ($N_{lR}$) are included in the SM. Because neutrinos are electrically neutral and colour blind, and only LH fields enter the weak interactions, RH neutrinos have no interactions in the SM, and they therefore are known as sterile neutrinos, while the LH neutrinos ($\nu_{lL}$) are known as active neutrinos.

With LH and RH neutrinos, the following Yukawa term, which is invariant under the SM gauge symmetries, can be constructed
	\be\label{YukNeut}
		\mathcal{L}_{Yuk}^{\nu} = -\sum_{\alpha,\beta=e, \mu, \tau}\overline{{\bf L}_{\alpha L}}\left(y^{\nu}_{\alpha \beta} \Phi^C\right) \nu_{\beta R} + h.c.~,
	\ee
where $	\Phi^C = i \sigma^2 \Phi^\star~$ is the charge conjugate of $\Phi$. As soon as the Higgs field acquires a non-trivial vev~(\ref{HiggsUni}), a Dirac mass term for neutrinos is generated
	\be\label{DMT}
		\mathcal{L}_{m}^{D} = -\sum_{\alpha,\beta=e, \mu, \tau}\overline{\nu_{\alpha L}} M^{D}_{\alpha \beta} \nu_{\beta R} + h.c. = -\overline{\nu_{L}} M^{D} \nu_{R} + h.c.,
	\ee
with $M^{D}_{\alpha\beta}= (v/\sqrt{2}) y^{\nu}_{\alpha\beta}$ being the elements of the $3\times 3$ complex mass matrix $M^D$, and $\nu_{L(R)}$ is defined according to~(\ref{neutransf}).

Although the generation of a Dirac mass term for neutrinos by means of the Higgs mechanism seems to be the most natural extension of the SM, it presents an enormous drawback when it comes to experimental results. Neutrino masses are bounded by measurements to be many orders of magnitude smaller than the masses of the charged leptons (and quarks). In this way, it seems very unnatural that the mechanism generating masses for the other SM fermions also generates neutrino masses. This would mean that the Yukawa couplings associated with neutrinos are extremely smaller than the Yukawa couplings associated with the other fermions. As a result, it is widely believed that a Dirac mass term coming from the Yukawa interaction term~(\ref{YukNeut}) cannot be the (only) reason behind the smallness of neutrino masses. Thus, generation of such tiny masses is expected to be a feature of physics beyond the SM.

\subsection{Majorana mass term}\label{sec:MMT}

The fact that neutrinos are, as the name suggests, neutral particles allows them to be Majorana fermions, which in physical terms means that they are their own anti-particles. If this is indeed the case, a different mass term can be constructed, the so-called Majorana mass term. 

As previously seen, a fermion mass term is expressed as products of LH and RH fields: $\propto \overline{\psi_R} \psi_L + \overline{\psi_L} \psi_R$. In the case of Dirac mass terms~(\ref{DMT}), the LH and RH fields are independent of each other, and the sum of them forms a Dirac fermion ($\psi=\psi_L+\psi_R$). Nonetheless, it is also possible to construct a mass term where the RH field depends on the LH component and vice-versa. In order to construct it, let us first consider a LH field $\psi_L$ and define its charge conjugate
	\be\label{cc}
			(\psi_L)^c \equiv C \overline{\psi_L}^T~,
	\ee
where $C$ is a unitary matrix\footnote{According to the Dirac representation of the $4\times 4$ $\gamma$ matrices, where $\gamma^0 \equiv \begin{pmatrix} 1 & 0\\0& -1\end{pmatrix}$, $\gamma^i \equiv \begin{pmatrix} 0 & \sigma^i\\-\sigma^i & 0\end{pmatrix}$ and $\sigma^i$ are the $2\times 2$ Pauli matrices, one can define the unitary matrix $C$ as $C \equiv i\gamma^2\gamma^0$.}. Using the relations~(\ref{LHRH}) and (\ref{cc}), we can show that 
	\bea
		\gamma_5 \psi_L = -\psi_L~,\\
		\gamma_5 (\psi_L)^c = (\psi_L)^c~, 
	\eea
which, therefore, means that the charge conjugate of a LH field is, actually, a RH field. Similarly, it can be shown that the charge conjugate of a RH field is a LH field. Thus, without introducing new RH fields as in the Dirac case, we can construct a Majorana mass term with LH fields only:
\be
		\mathcal{L}^M_{m} = -\frac{1}{2} m\overline{\psi_L} (\psi_L)^c + h.c.~,
	\ee
where the presence of the factor $1/2$	will be justified shortly.

If we now extend the mass term above to the physical case of three LH neutrinos ($\nu_{lL}$), as in the SM, we obtain
	\bea\label{mmt}
		\mathcal{L}_{m}^M &=& -\frac{1}{2} \sum_{\alpha, \beta=e,~\mu,~\tau} \overline{\nu_{\alpha L}}~M^L_{\alpha \beta}~(\nu_{\beta L})^c + h.c.~\\
		&=& -\frac{1}{2}\overline{\nu_{L}}~M^L~(\nu_{L})^c + h.c.~,\nn
	\eea
where the second line is expressed in the matrix form, with $M^L$ being a $3\times 3$ complex Majorana mass matrix and $\nu_{L}$ as given in~(\ref{neutransf}). Furthermore, since $C^T = -C$ and using the fact that a scalar is equal to its transpose, we find
	\be\label{symm}
		\overline{\nu_L} M^L (\nu_L)^c = -\overline{\nu_L} (M^L)^T C^T 	\overline{\nu_L}^T =  \overline{\nu_L} (M^L)^T (\nu_L)^c~,
	\ee
which implies that $M^L= (M^L)^T$, {\it i.e.} the Majorana mass matrix, differently from the Dirac mass matrix, is symmetric.

A symmetric complex matrix, such as $M^L$, can be written in terms of a diagonal mass matrix with elements $m^{L}_{ij}=m_i \delta_{ij}$ and a unitary matrix $U_L^M$ \cite{BilenkyBook}:
	\be\label{diagmmm}
		M^L = U_L^M m^{L} (U_L^M)^T~.
	\ee
Replacing~(\ref{diagmmm}) into~(\ref{mmt}) and rearranging the terms, we obtain
	\bea\label{mmtdiag}
		\mathcal{L}_{m}^M &=& -\frac{1}{2} \left[\overline{\left((U_L^M)^\dagger \nu_L\right)} m^{L} \left( (U_L^M)^\dagger \nu_L\right)^c  + \overline{\left((U_L^M)^\dagger \nu_L\right)^c} m^{L} \left((U_L^M)^\dagger \nu_L\right)\right]~\\
		&=& -\frac{1}{2} \overline{\nu^{M}} m^{L} \nu^{M} =-\frac{1}{2} \sum_{i=1,2,3}\overline{\nu_i^{M}} m_i \nu_i^{M} ~,
	\eea
where
	\be\label{massf} 
		\nu^{M} = \nu_L^M +\nu_R^M = \left( (U_L^M)^\dagger \nu_L\right) +\left( (U_L^M)^\dagger \nu_L\right)^c = \begin{pmatrix}\nu_1^{M} \\ \nu_2^{M} \\ \nu_3^{M} \end{pmatrix}~,
	\ee 
with $\nu_i^{M}$ being the massive neutrinos, {\it i.e.} neutrino fields with a definite mass $m_i$.

In general, a fermion field with LH and RH components related to each other by charge conjugation such as $\nu^M_i$, satisfies the so-called ``Majorana condition''
	\be
		(\nu_i^M)^c = \nu_i^M~, 
	\ee
and is, therefore, known as a Majorana field.\footnote{Starting with a Dirac field: $\psi = \psi_L + \psi_R$ where $\psi_L$ and $\psi_R$ are independent, one can always define two Majorana fields $\psi^M_1 = \psi_L + (\psi_L)^c$ and $\psi^M_2 = \psi_R + (\psi_R)^c$.}

Finally, let us consider the kinetic term for the LH flavour neutrinos $\nu_{L}$~(\ref{neutransf})
	\be\label{Mk1}
		\mathcal{L}_k = \overline{\nu_L} i \slashed{\partial} \nu_L~.
	\ee 
From~(\ref{massf}) $\nu_L = U^M_L \nu_L^M$ where $U^M_L$ is unitary, we can write the kinetic term above in terms of $\nu^M_{jL}$
	\be
		\mathcal{L}_k = \overline{\nu_L^M} i \slashed{\partial} \nu_L^M = \sum_{j=1,2,3}\overline{\nu_{jL}^M} i \slashed{\partial} \nu_{jL}^M~.
	\ee
Moreover, up to a total derivative, one can show that
	\be
		\overline{\nu_{jL}^M} i \slashed{\partial} \nu_{jL}^M = \overline{(\nu_{jL}^M)^c} i \slashed{\partial} (\nu_{jL}^M)^c = \overline{\nu_{jR}^M} i \slashed{\partial} \nu_{jR}^M~,
	\ee
thus, the kinetic term can be expressed as
	\be\label{Mk2}
		\mathcal{L}_k = \frac{1}{2}\overline{\nu}^M i \slashed{\partial} \nu^M = \frac{1}{2}\sum_{j=1,2,3} \overline{\nu_j}^M i \slashed{\partial} \nu_j^M~,
	\ee
with $\nu^M$ and $\nu_j^M$ the Majorana fields defined in~(\ref{massf}). Therefore, since the factor $1/2$ appears naturally when writing the kinetic term for the Majorana fields~$\nu_j^M$, we need to add the same factor in the mass term~(\ref{mmt}) in order to get the correct equation for the fermions. Writing together the kinetic and mass term for the free Majorana fields $\nu_{i}^M$, we have
	\be
		\mathcal{L}^M_{free} = \frac{1}{2} \sum_{j=1,2,3}\overline{\nu_j}^M\left( i\slashed{\partial} - m_j\right)\nu_j^M~.
	\ee
\\

\noindent{\bf Lepton number violation}
\\

In order to understand one important consequence related to the presence of Majorana mass terms in a theory, let us consider a generic fermion field $\psi$ under a global phase transformation $e^{i\alpha}$:
	\be\label{gtrans}
		\psi \to e^{i\alpha} \psi~,~\mbox{consequently},~\overline{\psi}\to e^{-i\alpha}\overline{\psi}~,	\psi^c \to e^{-i\alpha} \psi^c~,\overline{\psi^c} \to e^{i\alpha} \overline{\psi^c}~.
	\ee
Whereas terms of the form $\overline{\psi}\psi$ and $\overline{\psi^c}\psi^c$ are invariant under the global transformation above, terms like $\overline{\psi^c}\psi$ and $\overline{\psi}\psi^c$ are not.

The SM is (accidentally) symmetric under such global transformations~(\ref{gtrans}), which, according to Noether's theorem, implies in the conservation of a quantum number, the so-called lepton number in this case. On the other hand, if the SM is extended by the inclusion of Majorana mass terms~(\ref{mmt}), this symmetry will be broken, and lepton number-violating processes, such as neutrinoless double beta decays ($\beta\beta_{0\nu}$), will be allowed to take place. Therefore, a definitive way of proving that neutrinos are Majorana fermions would be the observation of processes such as $\beta\beta_{0\nu}$.\\

\noindent{\bf The Weinberg dimension $5$ operator}\\

Let us now suppose that $\beta\beta_{0\nu}$ have been observed in nature, showing that neutrinos are Majorana fermions. How could a Majorana mass term such as~(\ref{mmt}) be generated without breaking explicitly the SM gauge structure and without the introduction of new fundamental fields? 

It is obvious that a term of the form~(\ref{mmt}) cannot be added to the SM without breaking explicitly the SM gauge structure; therefore, analogously to the mass generation mechanism for the other fermions in the SM, let us suppose that such term is generated by coupling the LH neutrinos with the Higgs field. In this case, we will find that no renormalisable operator ($d\leq 4$) invariant under the SM symmetries can be constructed. For this task, the lowest dimension operator which can be constructed is the Weinberg dimension $5$ operator~\cite{Weinberg}:
	\be\label{Weinberg}
		\mathcal{L}_{W5} =- \frac{1}{2}\sum_{\alpha,\beta=e,\mu,\tau} c_{\alpha \beta}\left(\overline{{\bf L_{\alpha L}}^c} \Phi \right)\left( (\Phi^c)^\dagger {\bf L}_{\beta L} \right) +h.c.~,
	\ee 
where $c_{\alpha\beta}=\tilde{c}_{\alpha\beta}/\Lambda$ with $[\tilde{c}_{\alpha\beta}]=0$, and $\Lambda$ is a mass scale suppressing the operator. After the EW symmetry breaking, the term above will generate the Majorana mass term~(\ref{mmt}) with $M^M_{\alpha \beta} = (v^2/\Lambda)(\tilde{c}_{\alpha\beta}/2)$. 

If the operator~(\ref{Weinberg}) was completely described by the SM physics, we would expect the mass scale $\Lambda$, suppressing the Weinberg operator, to be the EW symmetry breaking scale: $(v^2/\Lambda)\approx v$. Thus, in order to obtain small neutrino masses, the coefficients $\tilde{c}_{\alpha\beta}$ would need to be unnaturally tiny. This problem is analogous to the one found in the previous section with the unnaturally small Yukawa couplings. 

Therefore, it is believed that $\Lambda$ comes from beyond the SM physics instead, so that $\Lambda \gg v$, which implies that the masses generated by~(\ref{Weinberg}) can be naturally small as required by experiments. For instance, to obtain sub-eV neutrino masses with $\tilde{c}_{\alpha\beta}=\mathcal{O}(1)$ and $v\approx 246 GeV$, the value of $\Lambda$ should be of the order of the Grand Unified Theory (GUT) scale. In conclusion, we have once again found that the SM physics does not seem to provide all that is necessary in the quest for a satisfactory neutrino mass generation mechanism. In this context, in the end of the next section, we show how the seesaw mechanism uses physics beyond the SM to provide a mass scale large enough to generate sub-$eV$ neutrino masses.

\subsection{The Dirac and Majorana mass term and the seesaw mechanism}\label{sec:Seesaw}

To finish our discussion on possible mass terms for neutrinos, we put together the ideas described in sections~\ref{sec:DMT} and~\ref{sec:MMT} to construct the ``Dirac and Majorana'' mass term. We then use this term to present possibly the simplest mechanism for neutrino mass generation, the so-called seesaw (type I) mechanism.

In addition to the three generations of LH active (flavour) neutrinos $\nu_{lL}$, let us introduce three generations of RH (flavour) neutrino fields $N_{lR}$. A Dirac mass term of the form~(\ref{DMT}), with mass matrix $M^D$, can be constructed by mixing both LH and RH fields. Moreover, two Majorana mass terms of the form~(\ref{mmt}) can be constructed, one of which contains only LH fields and a symmetric mass matrix $M^{L}$, and the other with only RH fields and a symmetric mass matrix $M^R$. These three mass terms, involving three active and three sterile neutrinos, can be written as
	\bea\label{dmmt1}
		\mathcal{L}_m^{D+M} &=& -\frac{1}{2} \sum_{\alpha, \beta = e, \mu, \tau} \left[ \overline{\nu_{\alpha L}}~M^{L}_{\alpha \beta}~(\nu_{\beta L})^c + \overline{N_{\alpha R}}~M^{R}_{\alpha \beta}~(N_{\beta R})^c + 2\overline{\nu_{\alpha L}} M_{\alpha \beta}^D N_{\beta R} + h.c.\right] \nn \\
		&=& -\frac{1}{2}\overline{\nu_{L}}~M^{L}~(\nu_{L})^c -\frac{1}{2}\overline{N_{R}}~M^{R}~(N_{R})^c -\overline{\nu_{L}} M^D N_{R} + h.c.~.
	\eea
The Dirac and Majorana mass term above can be further simplified by defining
	\bea\label{66}
		n_L = \begin{pmatrix}  \nu_L \\ (N_R)^c \end{pmatrix}~~~\mbox{and}~~~ M^{D+M} =\begin{pmatrix} M^{L} & M^D \\ (M^D)^T & M^{R} \end{pmatrix}~,
	\eea
so that~(\ref{dmmt1}) becomes
	\be\label{dmmt}
		\mathcal{L}^{D+M}_m = -\frac{1}{2} \overline{n_L} M^{D+M} (n_L)^c +h.c.~,
	\ee
where $M^{D+M}$ is now a $6\times 6$ complex mass matrix written in terms of $3\times3$ block matrices. From the definition~(\ref{66}), it is clear that, since the Majorana mass matrices are symmetric, $M^{D+M}$ is also a symmetric matrix and can be diagonalised by a bi-unitary transformation similar to~(\ref{diagmmm}).

Having constructed the most general Dirac and Majorana mass term in~(\ref{dmmt1}) or~(\ref{dmmt}), let us focus on the specific case of the seesaw mechanism~\cite{seesaw1,seesaw2,seesaw3}. The seesaw mechanism uses heavy sterile neutrinos to explain the smallness of the active neutrino masses. In this mechanism, the introduction of RH sterile neutrinos ($N_{lR}$) allows for the generation of Dirac mass terms via the standard Yukawa interaction terms~(\ref{YukNeut}) and (\ref{DMT}). As previously discussed, such Yukawa terms will naturally generate neutrino masses many orders of magnitude above the values found in experiments. In order to get tiny masses for neutrinos, instead of unnaturally fine-tuning the Yukawa couplings, the seesaw mechanism requires the existence of sterile neutrinos with heavy masses which end up suppressing the Dirac mass term. Because sterile neutrinos do not interact with any SM gauge field (they transform as singlets under the SM symmetries), one may assume that they are heavy particles with their masses, given by~$M^{R}$, coming from some (unknown) beyond the SM mechanism. Considering only renormalisable operators ($d\leq 4$), the mass matrix $M^{L}$ should vanish because, as we have seen, the lowest order operator, invariant under the SM symmetries, that can generate a Majorana mass term for the active neutrinos is of dimension $5$. In this way, in the seesaw mechanism, the Dirac and Majorana mass term is given by~(\ref{dmmt}) with the following mass matrix
	\be\label{ssmt}
		M^{D+M} =\begin{pmatrix} 0 & M^D \\ (M^D)^T & M^{R} \end{pmatrix}~,
	\ee
with $M^D = (v/2) y^\nu_{\alpha\beta}$, where $v$ is vev of the Higgs field and $y^\nu_{\alpha\beta}$ are the Yukawa couplings. 

By block-diagonalizing the mass matrix above, using the fact that $M^{R}\gg M^D$, we find that, while the mass matrix of the sterile neutrinos is effectively $\approx M^{R}$, the active neutrino mass matrix is approximately given by
	\bea
		 -M^D (M^{R})^{-1} (M^D)^T~.
	\eea 
Therefore, as the name suggests, in the seesaw mechanism, the heavier the sterile neutrino masses $M^{R}$, the lighter the active neutrino masses $-M^D (M^{R})^{-1} (M^D)^T\ll M^D \ll M^{R}$.

The mechanism described above is not the only known version of the seesaw mechanism. In fact, more involved versions have been proposed in which, instead of heavy sterile neutrinos, other field configurations, such as a ``Higgs'' triplet or a heavy Majorana fermion triplet, are added to the SM. 

\section{Neutrino mixing}\label{sec:Mix}

Neutrino mass matrices as defined in terms of flavour neutrinos are complex and can have a very general form. It is natural, however, to express the neutrino fields in a basis where their mass matrices are diagonal and real, and therefore their diagonal elements are indeed the physical masses of the respective particles. For this, we use unitary mixing matrices which diagonalise the mass matrix, telling us how to go from flavour neutrinos to massive neutrinos and vice-versa. We present here the standard parametrisation of the mixing matrices in terms of mixing angles and (CP-violating) phases for both, Dirac and Majorana, neutrino cases.

\subsection{Dirac neutrinos}

Let us start with Dirac neutrinos. In this case, there will be a Dirac mass term for charged leptons~(\ref{Yukmass}) and another for neutrinos~(\ref{DMT}). The Dirac mass matrices in~(\ref{Yukmass}) and~(\ref{DMT}), $M^{(cl)}$ and $M^D$, are generic $3\times 3$ complex matrices which can be diagonalised by making use of two unitary matrices
	\bea\label{diagD}
		M^D &=& U_L^D m^D (U_R^D)^\dagger~,
	\eea
where $m^D$ is a real diagonal matrix related to the neutrinos. By substituting~(\ref{diagD}) into~(\ref{DMT}), we find
\bea\label{diagDmt}
		\mathcal{L}^{D}_{m} =- \overline{\nu_L^{(m)}}  m^{D} \nu_R^{(m)} + h.c. ~,\nn
	\eea
where 
	\bea\label{mneut}
		\nu_{L(R)}^{(m)}= \left( U_{L(R)}^{D}\right)^\dagger \nu_{L(R)}=\begin{pmatrix} \nu^{(m)}_{1 L(R)} \\ \nu^{(m)}_{2 L(R)} \\ \nu^{(m)}_{3 L(R)}~. \end{pmatrix}
	\eea 
The fields $\nu^{(m)}_{iL(R)}$ with $i=1,2,3$ represent neutrino fields with definite mass, or massive neutrinos, which can be written in terms of the flavour neutrinos $\nu_{L(R)}$ according to the expressions above. Therefore, a massive (flavour) neutrino can then be seen as a mixture of flavour (massive) neutrinos, described in terms of the unitary ``mixing'' matrices ($U$'s).

If we now replace the flavour neutrinos by the massive neutrinos according to~(\ref{mneut}) into all terms of the SM, it is easy to see that, except for the charged current interaction terms, the dependence on the $U$'s  matrices disappear. Nonetheless, for the charged current interaction~(\ref{ccint}), we find that the contributions coming from the $U$'s matrices cannot be completely eliminated
	\be
		\mathcal{L}_I^{CC} = -\frac{g}{\sqrt{2}} \overline{\nu_L^{(m)}} \gamma_\alpha [U_{PMNS}^\dagger] L_L W^\alpha + h.c.~,
	\ee
where we define $U_{PMNS} = (U_L^D)$, the neutrino mixing matrix or PMNS (Pontecorvo-Maki-Nakagawa-Sakata) matrix. The PMNS matrix is the lepton analogue of the CKM (Cabibbo-Kobayashi-Maskawa) matrix for quarks.

In general, a $n\times n$ unitary matrix contains $n^2$ real parameters, which can be separated into $n(n-1)/2$ mixing angles and $n(n+1)/2$ phases, but not all of the phases are physical. In fact, for Dirac fields, only $(n-1)(n-2)/2$ of them are physical, whereas for Majorana fields, $n(n-1)/2$ phases are physical. Thus, in the present case, considering $n=3$ Dirac neutrinos, the neutrino mixing matrix can be parametrised in terms of three mixing angles: $\theta_{12}, \theta_{23}$ and $\theta_{13}$, and a CP-violating (Dirac) phase $\alpha$:
	\bea\label{PMNS}
		U_{PMNS} = \begin{pmatrix} c_{13} c_{12}                                &    c_{13} s_{12}                            & s_{13}e^{-i\delta} \\
															 -c_{23}s_{12}-s_{23}c_{12}s_{13} e^{i\delta} & c_{23}c_{12}-s_{23}s_{12}s_{13} e^{i\delta} & c_{13}s_{23} \\
															 s_{23}s_{12}-c_{23}c_{12}s_{13} e^{i\delta}   & -s_{23}c_{12}-c_{23}s_{12}s_{13}e^{i\delta} & c_{13}c_{23}
							\end{pmatrix}~,
	\eea
with $c_{ij} = \cos(\theta_{ij})$ and $s_{ij} = \sin(\theta_{ij})$.

Sometimes, however, as in chapters \ref{chap4} and \ref{chap5}, it is useful to consider two generations of neutrinos only. Then, when assuming $n=2$ Dirac neutrinos, we can parametrise the $2\times 2$ unitary mixing matrix in terms of one mixing angle, $\theta$, only
	\be\label{mm2}
		U(n=2) = \begin{pmatrix} \cos{\theta} & \sin{\theta} \\ -\sin{\theta} & \cos{\theta} \end{pmatrix}~.
	\ee

\subsection{Majorana neutrinos}

In section~\ref{sec:MMT}, we have shown that a neutrino mass term can be written without the need to introduce RH fields. The Majorana mass matrix $M^L$, found in~(\ref{mmt}), is symmetric and can be diagonalised by a unitary matrix $U^M_L$ according to~(\ref{diagmmm}). Thus, as we have seen, it is possible to define massive Majorana neutrinos $\nu^M = \nu^M_L + \nu^M_R$ in terms of the flavour neutrinos according to (\ref{massf}). When using (\ref{massf}) to replace flavour neutrinos by massive neutrinos in all terms of the SM, we note that the CC interaction term will not be diagonalised and can be written as
	\be
		\mathcal{L}_I^{CC} = -\frac{g}{\sqrt{2}} \overline{\nu_L^{M}} \gamma_\alpha [(U_L^M)^\dagger] L_L W^\alpha + h.c.~.	
	\ee

In this case, however, different from what happened with Dirac neutrinos, from the six phases present in the mixing matrix, only three can be eliminate (or made redundant) by field re-definitions. In this way, if neutrinos are Majorana fermions, the mixing matrix $U_L^M$ contains three mixing angles and a Dirac phase as in~(\ref{PMNS}), plus two extra CP-violating phases: $\alpha_2$ and $\alpha_3$, and it can be written as the product $U_{PMNS} S^M$ with
	\be\label{Mphases}
		S^M = diag(1, e^{i \alpha_2}, e^{i \alpha_3} )~.
	\ee
Nonetheless, as we will show in the next section, the Majorana phases do not play a role in neutrino oscillations.

\section{Neutrino oscillations}\label{sec:NOsc}
\mbox{}

Neutrinos, as seen in section~\ref{sec:N+SM}, interact via weak force and are consequently produced and detected through charged and neutral current interaction processes. It is an observed fact, however, that neutrinos with a given flavour, produced at an initial time $t_i=0$, can be later observed, at a time $t_f=t$, to have a different flavour, even if propagating through vacuum. This phenomenon is known as neutrino or flavour oscillations. Neutrino oscillations happen because flavour neutrinos are not pure states, but, actually, a mixture of massive states. In this section, we are interested in studying the time evolution of flavour neutrinos to calculate the probability of a given flavour neutrino, propagating in vacuum, to become another flavour neutrino. 

We review here the standard derivation of the neutrino oscillation probability. In this derivation, among other simplifications, we assume that all massive neutrinos have the same momentum (equal momentum assumption). This assumption, although not physically accurate, is not relevant in the calculation of oscillation probabilities as it leads to the correct expression.\footnote{For different (and more involved) derivations of the oscillation probability see, for example,~\cite{GiuntiBook}.}

A neutrino state with flavour $\alpha$ and momentum $\vec{p}$, $\ket{\nu_\alpha}$, is represented by the following mixture (superposition) of neutrino states with definite masses, $\ket{\nu_i}$,
	\be\label{nsmix}
		\ket{\nu_{\alpha}} = \sum_{i=1}^3 \tilde{U}_{\alpha i}^\star \ket{\nu_i}~,
	\ee
where the flavour neutrinos as well as the massive neutrinos are orthonormal among themselves: $\braket{ \nu_{i} | \nu_{j}} = \delta_{ij}~$ and $\braket{ \nu_{\alpha} | \nu_{\beta} } = \delta_{\alpha \beta}~$. The flavour and massive neutrino states in~(\ref{nsmix}) are mixed according to the mixing matrix $\tilde{U}$. In the Dirac case $\tilde{U} = U_{PMNS}$, defined in~(\ref{PMNS}), whereas in the case of Majorana neutrinos $\tilde{U} = U_{PMNS}S^M$ where $S^M$ contains the Majorana phases according to~(\ref{Mphases}). Although, as we show below, Majorana phases do not play a role in oscillations, we start by considering a generic mixing matrix $\tilde{U}$.

Neutrino states evolve in time according to the time evolution operator $\exp(-i \mathcal{H} t)$, where $\mathcal{H}$ is the Hamiltonian of the system. Since massive neutrinos are eigenstates of the Hamiltonian, {\it i.e.} $\mathcal{H}\ket{\nu_i} = E_i \ket{\nu_i}$ where $E_i$ is the neutrino energy, their time evolution is simply given by
	\be\label{te}
		\ket{\nu_i(t) } = \exp(-i \mathcal{H} t )\ket{\nu_i}~=\exp(-i E_i t )\ket{\nu_i} ~,
	\ee
where $\ket{\nu_i}\equiv \ket{\nu_i(t=0)}$. Flavour neutrinos, on the other hand, which are present in weak processes are not Hamiltonian eigenstates, so that to find out how they evolve with time we make use of the relation~(\ref{nsmix}):
	\be
		\ket{\nu_{\alpha} (t)} = \sum_{i} \tilde{U}_{\alpha i}^\star \exp(-iE_i t) \ket{\nu_i}~.
	\ee
From~(\ref{nsmix}) and the unitarity of the mixing matrix, it is possible to write $\ket{\nu_i}$ as a mixture of flavour neutrinos, in such a way that the expression above becomes
	\be
		\ket{\nu_{\alpha} (t)} = \sum_{\beta}\left(\sum_{i} \tilde{U}_{\alpha i}^\star \exp(-iE_i t) \tilde{U}_{\beta i}\right) \ket{\nu_\beta}~=\sum_\beta \left(\mathcal{A}_{\nu_\alpha \to \nu_\beta}\right)\ket{\nu_\beta}~.
	\ee
	
Therefore, $\mathcal{A}_{\nu_\alpha \to \nu_\beta}=\braket{\nu_\beta | \nu_{\alpha} (t)}$ is the transition amplitude for a neutrino, propagating in vacuum, to change its flavour after a time $t$. As a result, the oscillation probability is
	\be\label{oscp1}
		\mathcal{P}_{\nu_\alpha \to \nu_\beta} = |\mathcal{A}_{\nu_\alpha \to \nu_\beta}|^2 = \sum_{i,j} \tilde{U}^\star_{\alpha i} \tilde{U}_{\beta i}\tilde{U}_{\alpha j}\tilde{U}^\star_{\beta j} \exp\left[-i(E_i-E_j) t\right]~.
	\ee

At this point, it is important to note that in the Majorana case, where $\tilde{U} = U_{PMNS} S^M$ with $U_{PMNS}$ and $S^M$ given by~(\ref{PMNS}) and~(\ref{Mphases}), respectively, the Majorana phases will be cancelled in the the combination $\tilde{U}^\star_{\alpha i} \tilde{U}_{\beta i}\tilde{U}_{\alpha j}\tilde{U}^\star_{\beta j}$. Consequently, as already mentioned, it is necessary to consider processes such as $\beta\beta_{0\nu}$ to investigate the nature of neutrinos, because neutrino oscillations do not help us to distinguish between Majorana and Dirac neutrinos. Therefore, without loss of generality, we can assume that the mixing matrix $\tilde{U}$ relevant to neutrino oscillations is the PMNS matrix given in~(\ref{PMNS}). 
	
Because it is only possible to detect neutrinos with energy $E_i$ many order of magnitude greater than their masses $m_i$ (sub-$eV$), we can approximate the neutrino energy to: $E_i =\sqrt{\vec{p}^2+m_i^2}\simeq E+m_i^2/(2E)$ with $E=|\vec{p}|$ (equal momentum assumption). In this way, the difference between massive neutrino energies is
	\be
		E_i-E_j \simeq \frac{m_i^2 - m_j^2}{2 E}=\frac{\Delta m_{ij}^2}{2 E}~.
	\ee
Moreover, we can assume that neutrinos, being very light particles, travel with the speed of light, and, as a consequence, the time $t$ elapsed between emission and detection can be approximated by the distance $L$ between the source and the detector: $t\simeq L$ ($c=1$). Thus, the oscillation probability~(\ref{oscp1}) can be rewritten (for ultra-relativistic neutrinos) as
	\be\label{oscp2}
		\mathcal{P}_{\nu_\alpha \to \nu_\beta} (E,L)= \sum_{i,j} \tilde{U}^\star_{\alpha i} \tilde{U}_{\beta i}\tilde{U}_{\alpha j}\tilde{U}^\star_{\beta j} \exp\left(-i\frac{\Delta m_{ij}^2 L}{2E}\right)~.
	\ee
	
A more useful way of writing the oscillation probability~(\ref{oscp2}) is given by separating its real and imaginary parts:
	\bea\label{oscp3}
		\mathcal{P}_{\nu_\alpha \to \nu_\beta} (E,L)&=& \delta_{\alpha \beta}  - 4\sum_{i>j} \mbox{Re} \left( \tilde{U}^\star_{\alpha i} \tilde{U}_{\beta i}\tilde{U}_{\alpha j}\tilde{U}^\star_{\beta j}\right) \sin^2\left(\frac{\Delta m_{ij}^2 L}{4 E}\right)\\
		&&+2\sum_{i>j} \mbox{Im} \left( \tilde{U}^\star_{\alpha i} \tilde{U}_{\beta i}\tilde{U}_{\alpha j}\tilde{U}^\star_{\beta j}\right) \sin\left(\frac{\Delta m_{ij}^2 L}{2 E}\right)~.\nn
	\eea
From the expression above, it is not difficult to see that neutrino oscillations only take place if neutrinos are massive particles and their masses are different, which justifies the need for neutrino masses in the SM.

Comparing experimental results with the theory of neutrino oscillations, it is possible to determine the phenomenological values of not only the mixing angles and the Dirac phase of the PMNS matrix, but also of the neutrino mass squared differences (not the absolute value of each neutrino mass), as seen in~(\ref{oscp3}). The best estimates for these physical parameters can be found in~\cite{PDG}.

In the case of anti-neutrinos, on the other hand, instead of~(\ref{nsmix}), the relation which tells us about how flavour anti-neutrinos mix with massive anti-neutrinos is
	\be
		\ket{\overline{\nu_\alpha}} = \sum_i U_{\alpha i} \ket{\overline{\nu_i}}~.
	\ee
The oscillation probability for anti-neutrinos, which can be found by following the same steps of the neutrino case, is 
		\be\label{aoscp2}
		\mathcal{P}_{\overline{\nu_\alpha} \to \overline{\nu_\beta}} (E,L)= \sum_{i,j} \tilde{U}_{\alpha i} \tilde{U}^\star_{\beta i}\tilde{U}^\star_{\alpha j}\tilde{U}_{\beta j} \exp\left(-i\frac{\Delta m_{ij}^2 L}{2E}\right)~,
	\ee
or, separating the real and imaginary parts,
	\bea\label{aoscp3}
		\mathcal{P}_{\overline{\nu_\alpha} \to \overline{\nu_\beta} } (E,L)&=& \delta_{\alpha \beta}  - 4\sum_{i>j} \mbox{Re} \left( \tilde{U}^\star_{\alpha i} \tilde{U}_{\beta i}\tilde{U}_{\alpha j}\tilde{U}^\star_{\beta j}\right) \sin^2\left(\frac{\Delta m_{ij}^2 L}{4 E}\right)\\
		&&-2\sum_{i>j} \mbox{Im} \left( \tilde{U}^\star_{\alpha i} \tilde{U}_{\beta i}\tilde{U}_{\alpha j}\tilde{U}^\star_{\beta j}\right) \sin\left(\frac{\Delta m_{ij}^2 L}{2 E}\right)~.\nn
	\eea
	
The oscillation probabilities for neutrinos~(\ref{oscp3}) and antineutrinos~(\ref{aoscp3}) are related to each other via CP transformations. If the theory is CP symmetric, both oscillations probabilities should be the same: $\mathcal{P}_{\nu_\alpha \to \nu_\beta } (E,L)=\mathcal{P}_{\overline{\nu_\alpha} \to \overline{\nu_\beta} } (E,L)$. On the other hand, in the case of CP violation, we can define the quantity $A_{\alpha\beta}^{CP}$ as a ``measure'' of CP violation
	\bea\label{CP}
		A_{\alpha\beta}^{CP} &=& 	\mathcal{P}_{\nu_\alpha \to \nu_\beta}-\mathcal{P}_{\overline{\nu_\alpha} \to \overline{\nu_\beta}}\\
		&=& 4\sum_{i>j} \mbox{Im} \left( \tilde{U}^\star_{\alpha i} \tilde{U}_{\beta i}\tilde{U}_{\alpha j}\tilde{U}^\star_{\beta j}\right) \sin\left(\frac{\Delta m_{ij}^2 L}{2 E}\right)~.\nn
	\eea\\

\noindent {\bf Two-flavour case}\\

Let us obtain now the oscillation probability for the simplest case of two flavours only, which will be useful in the next chapters.
In this case the mixing matrix is given by (\ref{mm2}), such that it contains one mixing angle only and no Dirac phase. Replacing then (\ref{mm2}) into the expression (\ref{oscp3}) for $\alpha\neq \beta$, we find that the oscillation probability becomes
	\be\label{nosc2}
		\mathcal{P}(\nu_{\alpha} \to \nu_{\beta} ) = \sin^2(2 \theta) \sin^2\left[\frac{(m_1^2 - m_2^2) L}{4 E} \right].
	\ee
In such a case, because the mixing matrix does not contain Dirac phases, no difference between the oscillation probability for neutrinos and anti-neutrinos is expected: $\mathcal{P}(\overline{\nu_\alpha} \to \overline{\nu_\beta}) =\mathcal{P}(\nu_{ \alpha} \to \nu_{ \beta} )$.

\chapter{Dynamical mass generation}\label{chap2}
\mbox{}

Dynamical mass generation takes place when a mass term is generated as a consequence of the interactions present in a given model, even though the original (classical) action describing the physical system may not contain a (bare) mass term. In general, the absence of a mass term in the classical description of a theory is related to the invariance of its action under a symmetry, such as the chiral symmetry. When this is the case, such a symmetry needs to be broken dynamically for a mass term to be generated. By dynamical symmetry breaking, we mean that the symmetry is spontaneously broken ({\it i.e.} while the action of the theory is invariant under a given symmetry, its vacuum solution is not) by a condensate which is formed as a result of the interactions in the theory, instead of by an elementary scalar field, such as the Higgs field. Moreover, because perturbative corrections to mass terms are, in general, proportional to the bare mass, when the bare mass term is absent, such corrections are not able to introduce a mass in the theory. Thus, dynamical mass generation is usually a non-perturbative feature. 

In this chapter we present two non-perturbative methods which are useful within the context of dynamical mass generation.
First, we derive the Schwinger-Dyson equation for the fermion propagator and present examples in which fermion masses are generated dynamically by solving the SD equation. Second, we consider four-fermion interaction models and, by making use of the effective potential approach, we show how masses are generated dynamically for the fermions in the theory.

\section{Schwinger-Dyson equation for the fermion propagator}\label{sec:SD}
\mbox{}

In this section, considering a theory involving an interaction between a fermion field $\psi$ and an Abelian gauge field $A_\mu$, as in quantum electrodynamics (QED), we derive a non-perturbative equation for the fermion propagator, {\it i.e.} the Schwinger-Dyson equation for the fermion propagator.

\subsection{Definitions and useful relations} 

We present here the necessary tools for the derivation of the SD equation for the fermion propagator. We start with a generic bare QED action in $3+1$ dimensions
  \bea\label{qedaction} 
    S_{QED}[A_\mu, \psi, \bar{\psi}] &=&\int d^4 x~\mathcal{L}_{QED}\\
    &=&\int d^4 x\left[ -\frac{1}{4} F_{\mu\nu}F^{\mu\nu}-\frac{\xi}{2}\partial_\mu A^\mu(x)\partial_\nu A^\nu(x) + \bar{\psi}(x)(i \slashed{D}-m_0)\psi(x) \right]~,\nn
  \eea 
where $m_0$ is the bare fermion mass, $\xi$ is the gauge-fixing parameter, $F_{\mu\nu}$ is the field strength tensor, and $D_\mu$ is the covariant derivative, defined by
  \be
    F_{\mu\nu} = \partial_\mu A_\nu(x) - \partial_\nu A_\mu(x)~~~\mbox{and}~~~ D_\mu = \partial_\mu + i e A_\mu(x).
  \ee 

Using the path integral approach to quantum field theory, we define the generating functional of correlation functions $Z$ and the energy functional (or generating functional for connected graphs) $E$
  \be\label{z} 
    Z[{\bf J}] = e^{-i E[{\bf J}]} =\int \mathcal{D}[{\bf A}] ~e^{i\int d^4 x\left(\mathcal{L}_{QED} + J^\mu(x)A_\mu(x) + \bar{\psi}(x)\eta(x) +\bar{\eta}(x)\psi(x)\right)}~,
  \ee 
with $[{\bf J}] \equiv [A_\mu, \psi, \bar{\psi}]$ and $\mathcal{D}[{\bf A}] \equiv \mathcal{D}A_\mu\mathcal{D}\bar{\psi}\mathcal{D}\psi$.
Furthermore, performing a Legendre transform on $ E[{\bf J}]$, one can define the effective action $\Gamma$:
  \be\label{G}
		\Gamma[{\bf A^c}] = - E[{\bf J}] -\int d^4 x \left[ J^\mu(x)A_\mu^c(x) + \bar{\psi^c}(x)\eta(x) +\bar{\eta}(x)\psi^c(x) \right]~,
  \ee 
which is a functional of the ``classical fields'' $[{\bf A^c}] \equiv [A_\mu^c,\psi^c,\bar{\psi^c}]$, defined as
	\be\label{cl} 
		X^c =\frac{1}{Z} \int \mathcal{D}[{\bf A}]~X~e^{i\int d^4 x\left(\mathcal{L}_{QED} + J_\mu(x)A^\mu(x) + \bar{\psi}(x)\eta(x) +\bar{\eta}(x)\psi(x)\right)}~,
	\ee 
for $X=A_\mu, \psi$ or $\bar{\psi}$.

Taking functional derivatives of $E$~(\ref{z}) and $\Gamma$~(\ref{G}) with respect to their variables, we obtain
	\bea\label{relations} 
		\frac{\delta E[{\bf J}]}{\delta J^\mu(x)} &=& - A_\mu^c(x)~;~~~ \frac{\delta E[{\bf J}]}{\delta \bar{\eta}(x)} =  -\psi^c(x)~; ~~~ \frac{\delta E[{\bf J}]}{\delta \eta(x)} =  	 \bar{\psi^c}(x)~;  \\
		\frac{\delta \Gamma[{\bf A^c}]}{\delta A^c_\mu(x)} &=&  J^\mu(x)~; ~~~ \frac{\delta \Gamma[{\bf A^c}]}{\delta \psi^c(x)} =  \bar{\eta}(x)~; ~~~ \frac{\delta \Gamma[{\bf A^c}]}      {\delta \bar{\psi^c}(x)} =  -\eta(x)~. \nn
		\eea 
Particularly, using the expressions above, one can show that
	\be\label{EGrel} 
		\frac{\delta^2 E[{\bf J}]}{\delta \bar{\eta} (x_1) \delta \eta(x_2) } = \left(
		\frac{\delta^2 \Gamma[{\bf A^c}]}{\delta \bar{\psi^c} (x_1) \delta \psi^c(x_2) }\right)^{-1}~.
	\ee 

\subsection{Deriving the Schwinger-Dyson equation}\label{subsec:SD}

Having defined our tools, we now derive the SD equation for the fermion propagator~\cite{IZ}.\\
First, noting that the integral of a derivative vanishes, we can write
	\be\label{sd1d}
		\int \mathcal{D}[{\bf A}] \left(\frac{\delta S_{QED}[{\bf A}] }{\delta \bar{\psi}(x_1)} + \eta(x_1)\right) ~e^{i\int d^4 x\left(\mathcal{L}_{QED} + J_\mu(x)A^\mu(x) + \bar{		\psi}(x)\eta(x) +\bar{\eta}(x)\psi(x)\right)}=0~.
	\ee
According to the relations in~(\ref{relations}), the equation above can also be written as
	\be
		\left\{\frac{\delta S_{QED} }{\delta \bar{\psi}(x_1)} \left[\frac{\delta}{i \delta J_\mu},\frac{\delta}{i \delta \bar{\eta}},-\frac{\delta}{i\delta \eta} \right] + \eta(x_1) 		\right\} Z[{\bf J}] =0~,
	\ee
so that
	\be 
		\left\{\eta(x_1)+\left[i \slashed{\partial}-m_0-e\gamma^\mu\left(\frac{\delta}{i\delta J^\mu(x_1)}\right)\right]\left(\frac{\delta}{i\delta \bar{\eta}(x_1)}\right)\right\}Z[{\bf 		J}] =0~.
	\ee
From now on, we work with the generating functional for connected graphs $E$ ($=i \ln Z$) instead of $Z$. Taking a functional derivative of the last expression with respect to $\eta(x_2')$, and then setting all sources to zero: ${\bf J } \to 0$, we obtain
	\be\label{sdd} 
		\delta(x_1-x_2')+(i \slashed{\partial}-m_0)\left(\frac{\delta^2 E[{\bf{J}}]}{\delta \bar{\eta}(x_1)\delta \eta(x_2')}\right)_0=-ie\gamma^\mu\left(\frac{\delta^3 E[{\bf J}]}{\delta J^		 \mu(x_1)\delta \bar{\eta}(x_1)\delta \eta(x_2')}\right)_0~.
	\ee 

The term on the right-hand side of the equation above is more involved, so let us consider it separately
{\scriptsize
	\bea 
		&&\frac{\delta }{\delta J^\mu(x_1)}\left(\frac{\delta^2 E[{\bf{J}}]}{\delta \bar{\eta}(x_1)\delta \eta(x_2')}\right)_0 = \int d^4 y_1\left( \frac{\delta A_\nu^c (y_1)}{\delta J^\mu(x_1) }\right)_0\frac{\delta }{\delta A^c_\nu (y_1)}\left(\frac{\delta^2 \Gamma[{\bf{A^c}}]}{\delta \bar{\psi^c}(x_2')\delta \psi^c(x_1)}\right)_0^{-1}\\
		&&=- \int d^4 y_1 d^4 y_2 d^4 y_3\left(\frac{\delta A^c_\nu (y_1)}{\delta J^\mu(x_1) }\right)_0 \left(\frac{\delta^2 E[{\bf{J}}]}{\delta \bar{\eta}(x_1)\delta \eta(y_2)}\right)_0\left(\frac{\delta^3 \Gamma[{\bf{A^c}}]}{\delta A_\nu^c (y_1)\delta \bar{\psi^c}(y_2)\delta \psi^c(y_3)}\right)_0 \left(\frac{\delta^2 E[{\bf{J}}]}{\delta \bar{\eta}(y_3)\delta \eta(x_2')}\right)_0~,\nn
	\eea}
where we used the eq.~(\ref{EGrel}) to obtain the right-hand side of the first line, whereas to find the second line, we made use of the following relation
	\be 
		\frac{\delta}{\delta A_\nu^c} M^{-1} = - M^{-1} \left( \frac{\delta M}{\delta A_\nu^c} \right) M^{-1}~,~~~\mbox{with} ~~~M = \left(\frac{\delta^2 \Gamma[{\bf{A^c}}]}{\delta 		\bar{\psi^c}\delta \psi^c}\right)~.
	\ee
Finally, taking into account~(\ref{relations}) and~(\ref{EGrel}), we define
	\bea 
		-\left(\frac{\delta A^c_\nu (y_1)}{\delta J^\mu(x_1) }\right)_0 &=& \left(\frac{\delta^2 E [{\bf J}]}{\delta J^\mu(x_1) \delta J^\nu(y_1)}\right)_0 = -i D_{\mu\nu} (x_1,y_1)~    ;\\
		\left(\frac{\delta^2 E[{\bf{J}}]}{\delta \bar{\eta}(x_1)\delta \eta(x_2')}\right)_0 &=& iG(x_1,x_2')~;\\
		\left( \frac{\delta^3 \Gamma[{\bf{A^c}}]}{\delta A_\nu^c (y_1)\delta \bar{\psi^c}(y_2)\delta \psi^c(y_3)}\right)_0 &=& e \Gamma^\nu(y_1;y_2,y_3)~,
	\eea 
where $D_{\mu\nu}$, $G$ are the full propagators for the gauge and fermion fields, respectively, and $\Gamma^\nu$ is the irreducible vertex function. 

Therefore, eq.~(\ref{sdd}) can now be expressed as
	\bea 
		&&\delta(x_1-x_2') +i (i\slashed{\partial}-m_0)G(x_1,x_2')=\\
		&&e^2 \int d^4 y_1 d^4 y_2 d^4 y_3\gamma^\mu D_{\mu\nu} (x_1,y_1)G(x_1,y_2)\Gamma^\nu(y_1;y_2,y_3)G(y_3,x_2')~,\nn
	\eea 
which after multiplication by $G^{-1}(x_2',x_2)$ and integration over $x_2'$ becomes
	\be\label{SDEcoord}
		G^{-1}(x_1,x_2) +i(i\slashed{\partial}-m_0) \delta(x_1-x_2) =e^2\int d^4 y_1 d^4 y_2\gamma^\mu D_{\mu\nu} (x_1,y_1)G(x_1,y_2)\Gamma^\nu(y_1;y_2,x_2)~.
	\ee 
Finally, we perform a Fourier transform to obtain the SD equation for the fermion propagator in momentum space, {\it i.e.} 
	\be\label{SDE} 
		G^{-1}(p)-S^{-1}(p)=e^2\int \frac{d^4 k}{(2\pi)^4}  \gamma^\mu D_{\mu\nu}(k-p)G(p)\Gamma^\nu(k-p;k,p)~,
	\ee 
where $S(p)=i(\slashed{p}-m_0)^{-1}$ is the bare fermion propagator. 

In a similar way, one can derive the SD equation for the photon propagator or, in fact, for any n-point Green's function of the theory. The SD equation for the fermion propagator is only one of an infinite hierarchy of equations, in which for every n-point function one can find a corresponding Schwinger-Dyson equation involving an $n'$-point function, with $n'>n$. Therefore, as these are coupled equations, which always depend on higher-order n-point functions, it is usually necessary to truncate them in order to make them more manageable. 

Finally, it is worth emphasising that although from (\ref{SDE}) one may naively think that the SD equation for the fermion propagator gives a one-loop contribution, this is clearly not the case. The SD equation is a non-perturbative equation, containing dressed quantities which incorporate information about all orders in the loop expansion.

\subsection{Examples}\label{sec:examples}

\subsubsection{QED in 2+1 dimensions}

We briefly present now an example, based on~\cite{SDE2+1QED1,SDE2+1QED2}, where dynamical generation of fermion masses is observed when solving the SD equation~(\ref{SDE}) for QED in 2+1 dimensions ($QED_{2+1}$).

We consider an action similar to~(\ref{qedaction}), however, in 2+1 dimensions, with $m_0=0$ and containing N flavours of massless fermions
	\be\label{qed2+1action}
		S_{QED_{2+1}} =\int d t d^2 x \left[\sum_{i=1}^N \bar{\psi_i} (i\slashed{\partial}-e \slashed{A})\psi_i-\frac{1}{4}F_{\mu\nu}F^{\mu\nu}-\frac{\xi}{2}\left(\partial_\mu A^\mu(x)\right)^2\right]~,
	\ee 
In contrast with the 3+1-dimensional case where the coupling constant is dimensionless, in $QED_{2+1}$ we have that $[e^2]=[m]$, and the theory is super-renormalisable. Then, if a fermion mass is generated dynamically, we expect it to be proportional to the mass scale in the model, {\it i.e.} $e^2$.
Here, we study~(\ref{qed2+1action}) in the large-N limit (or $1/N$ expansion), with $\tilde{\alpha} =e^2 N$ fixed.

In $2+1$ dimensions, the fermion field $\psi$ can be entirely described by only two-components. In such a case, however, chiral symmetry cannot be defined in the usual way, since there is no matrix that anticommutes with all the $2\times2$ $\gamma$ matrices. In order to consistently define chiral symmetry in 2+1 dimensions, it is therefore necessary to work with four-component fermions instead. In this case we can define the three $4 \times 4$ $\gamma$ matrices, in  $2\times 2$ block form, as
\be\label{3gamma}
	\gamma^0 = \begin{pmatrix}  \sigma^3 & 0 \\ 0 & -\sigma^3 \end{pmatrix}~,~~~\gamma^1 = \begin{pmatrix}  i\sigma^1 & 0 \\ 0 & -i\sigma^1\end{pmatrix}~, ~~~\gamma^2 = \begin{pmatrix} i \sigma^2 & 0 \\ 0 & -i\sigma^2 \end{pmatrix}~,
\ee
in such way that the two $4 \times 4 $ matrices below anticommute with all matrices in (\ref{3gamma}), {\it i.e.}
\be
	\gamma^3 = \begin{pmatrix}  0 & i \\ i & 0 \end{pmatrix}~,~~~\gamma^5 = \begin{pmatrix}  0 & i \\ -i & 0 \end{pmatrix}~.
\ee
Consequently, a massless theory will be invariant under the ``chiral'' transformations associated with the matrices $\gamma^3$ and $\gamma^5$ above: $\psi \to e^{i \alpha \gamma^3}\psi$ and $\psi \to e^{i \beta \gamma^5}\psi$.

For this specific model, in the four-component formalism, two mass terms are possible. One of them, however, is chiral-symmetric and parity-violating, and therefore will not be considered here. The second possible mass term, on which we will be focusing from now on, breaks chiral symmetry and is parity-conserving: $m\bar{\psi}\psi$.

Going back to~(\ref{qed2+1action}), the photon propagator, in the Landau gauge ($\xi\to \infty$), can be written as~\cite{SDE2+1QED1,SDE2+1QED2}
	\be 
		D_{\mu\nu}(p) =(-i)\frac{\left(\eta_{\mu\nu} -\frac{p_\mu p_\nu}{p^2}\right)}{p^2[1+\Pi(p)]	}~,
	\ee 
with, when considering only the leading order term in the $1/N$ expansion,
	\be
		\Pi(p) =\frac{\tilde{\alpha}}{4\pi p^2}\left[2m+\frac{p^2-4m^2}{p}\sin^{-1}\left(\frac{p}{(p^2+4m^2)^{1/2}}\right)\right]~.
	\ee
Thus, in the massless limit, $\Pi(p) = \tilde{\alpha}/(8p)$. For the fermion fields, we consider the following fermion propagator 
\be 
G(p) = i/(\slashed{p}-m_d)~,
\ee
where $m_d$ is the dynamical mass, and we have neglected corrections to the wave-function renormalisation. In addition, we also approximate the vertex, by taking into account only its lowest order contribution: $\Gamma^\nu \approx \gamma^\nu$. 

Using the SD equation for the fermion propagator~(\ref{SDE}), at zero external momentum, to understand whether a mass term can be generated dynamically, we then obtain (in Euclidean space) 
	\be\label{sdepi}
	m_d = 2\frac{\tilde{\alpha}}{N} \int \frac{d^3 p}{(2\pi)^3} \frac{m_d}{p^2[1+\Pi(p)](p^2+m_d^2)}~.
	\ee
While $m_d=0$ is clearly a trivial solution, implying no dynamical mass generation, we need to solve the integral above to verify whether the SD equation admits other solutions with non-vanishing $m_d$. 

In the region $\tilde{\alpha}\gg p\gg m_d$ the integrand in (\ref{sdepi}) behaves as $1/p^3$, leading to a logarithmic divergence, which can be naturally cut off by $\tilde{\alpha}$ and $m_d$ in the UV and IR, respectively. Thus, we find
\be 
m_d \approx \tilde{\alpha} \exp\left[\frac{-\pi^2 N}{8}\right]~.
\ee 
An important point to be noted is that when $N\to \infty$, the solution $m_d$ goes to zero faster than any perturbative contribution governed by polynomials in $1/N$, making evident the non-perturbative nature of $m_d$.

\subsubsection{QED in an external magnetic field}

In this section we consider dynamical mass generation in a model where massless fermions interact with a constant magnetic field~\cite{ExtMagLad1,ExtMagLad2} in $3+1$ dimensions.
The action in consideration is the one given in~(\ref{qedaction}) with $m_0=0$, but, since we have an external magnetic field, the total potential should be replaced by a dynamical part $A_\mu$ plus a constant external contribution $A^{ext}_\mu$, {\it i.e.}
	\be\label{potential}
		A_\mu \to A_\mu + A^{ext}_\mu~~~~\mbox{with}~~~~ A^{ext}_\mu = (0, -\frac{B}{2} x_2, -\frac{B}{2}x_1 ,0)~,
	\ee
where $A^{ext}_\mu$ is expressed in the symmetric gauge and was chosen so that the constant magnetic field $B$ is in the $+x_3$ direction.

In such a configuration, it has been shown that the bare fermion propagator can be written as~\cite{ExtMagLad1,ExtMagLad2}
	\be 
			S(x,y) = \exp\left[\frac{i e}{2}(x-y)^\mu A^{ext}_\mu(x+y)\right] \tilde{S}(x-y)~.
	\ee 
Fermions in the presence of a magnetic field are known to present a discrete energy spectrum, with energy levels known as Landau levels. In this way, $\tilde{S}(x-y)$ can be expanded over the Landau levels which are separated from each other by $\sim \sqrt{|eB|}$. When in a strong magnetic field ($|eB|\gg m^2, k^2$), the contributions coming from levels other than the fundamental one, known as the Lowest Landau Level (LLL), can be consistently neglected. Therefore, the Fourier transform of $\tilde{S}(x-y)$ in the LLL, can be written as 
	\be\label{bfpLLL} 
		\tilde{S}(k) \approx i \exp\left(-\frac{k_\perp^2}{|eB|}\right) \frac{\slashed{k}_\parallel+m_0}{k_\parallel^2-m_0^2}\left[1-i\gamma^1\gamma^2 \text{sign}(eB)\right]~,
	\ee 
where $m_0$ is the bare fermion mass which will eventually be taken to zero, and $k_\perp = (k^1,k^2)$ and $k_\parallel = (k^0, k^3)$ are the fermion perpendicular and parallel (to the external magnetic field) momentum components, respectively. It is also worth mentioning the presence of the projection operator $\propto [1-i\gamma^1\gamma^2 \text{sign}(eB)]$ in~(\ref{bfpLLL}) which shows that the spin of the fermions in the LLL is polarised along the magnetic field. Moreover, from eq.~(\ref{bfpLLL}), we see that in the LLL ($|eB|\gg m^2, \vec{k}_\parallel^2, \vec{k}_\perp^2$), the contribution coming from the perpendicular momentum components is highly suppressed, thus the dynamics of the fermion is restricted to directions parallel to $B$. This feature is known as dimensional reduction: $3+1 \to 1+1$, since the fermions described in the LLL propagate essentially in $1+1$ dimensions.

In the present case, because the fermion propagator~(\ref{bfpLLL}) is not invertible due to the presence of the projection operator $\propto [1-i\gamma^1\gamma^2 \text{sign}(eB)]$, it is necessary to write the SD equation in an alternative way by eliminating the inverse propagators present in the expression~(\ref{SDEcoord}). This can be done by multiplying~(\ref{SDEcoord}) by $S(x,x_1)$ from the left, and by $G(x_2,y)$ from the right-hand side, and then integrating the resulting expression over $x_1$ and $x_2$, which gives
	\bea\label{SDEinv} 
		G(x,y) &=& S(x,y) - e^2 \int d^4 x_1 d^4 x_2 d^4 y_1 d^4 y_2 S(x,x_1) \gamma^\mu \\
		&\times& D_{\mu\nu}(x_1,y_1) G(x_1,y_2) \Gamma^\nu(y_1;y_2,x_2)G(x_2,y)~.\nn
	\eea 

In order to solve the equation above, we consider the simplest truncation as in the original papers by Gusynin {\it et al}~\cite{ExtMagLad1,ExtMagLad2}\footnote{More accurate results, with the use of the``improved ladder approximation'' (where the one-loop correction to the vacuum polarisation tensor is taken into account for the full photon propagator), were later published by the same authors~\cite{ExtMagImpLad1,ExtMagImpLad2}.}, {\it i.e.} the ``ladder approximation''. The full fermion propagator in the LLL, assuming $m_d^2\ll |eB|$, where corrections to the wave function renormalisation are neglected, can then be written as
	\be\label{ffpLLL} 
		\tilde{G}(k) \approx i \exp\left(-\frac{k_\perp^2}{|eB|}\right) \frac{\slashed{k}_\parallel+m_d}{k_\parallel^2-m_d^2}\left[1-i\gamma^1\gamma^2 sign(eB)\right]~,
	\ee
where $m_d$ is the dynamical mass. In addition, in the ladder approximation, the full photon propagator and vertex can be replaced by their bare versions
	\bea 
			D_{\mu\nu} (k) &=& -i\left(\frac{g_{\mu\nu}}{k^2}-\xi\frac{k_\mu k_\nu}{(k^2)^2}\right)~,  \\
			\Gamma^\mu &=& \gamma^\mu~.
	\eea
	
Substituting the expressions above in~(\ref{SDEinv}), we find
	\be 
		m_d = \frac{\alpha}{4\pi^3}m_d\int \frac{d^2 k_\parallel}{k_\parallel^2+m_d^2} \int d^2k_\perp \frac{ (1-\xi\frac{k_\perp^2}{4 |eB|}) \exp (-\frac{k_\perp^2}{2|eB|})}{k_\parallel^2+k_\perp^2}~.
	\ee 
The divergent integral above needs to be regularised, which can be done by using a physical UV cut off. In this case, the UV cut off is naturally chosen to be the magnetic scale or, more specifically, $\sqrt{|eB|}$. Finally, we find that a non-trivial solution corresponding to dynamical mass generation is	
	\be 
		m_d \simeq  \sqrt{|eB|} \exp\left[-\frac{\pi}{2} \left(\frac{\pi}{2 \alpha}\right)^{1/2}\right]~,
	\ee 
where $\alpha = e^2/4\pi$ is the renormalised coupling constant. 

The non-analyticity of the solution above with respect to the coupling constant $\alpha$, as in the previous example, leads us to the conclusion that this solution can only be obtained when using a non-perturbative method, such as the SD equation.

\section{The effective potential approach: four-fermion interactions}\label{sec:4f}
\mbox{}

Dynamical mass generation in models presenting four-fermion interactions was first investigated in the 1960's~\cite{NJL}. In addition to generating fermion masses without having to couple fermions to elementary scalar fields, models with four-fermion interactions which are invariant under continuous chiral transformations, such as the Nambu-Jona-Lasinio (NJL) model~\cite{NJL}, give rise to massless pseudo-scalar bound (Goldstone) states once the continuous symmetry is spontaneously broken. In this section we present the basic features of such models in $3+1$ dimensions, and then, using the effective potential approach, we show how fermion masses are dynamically generated.

\subsection{Four-fermion interaction models}
\mbox{}

As it has become common in particle physics, the NJL model was also inspired by a condensed matter theory: the BCS theory of superconductivity~\cite{BCS}. It was proposed as an effective theory able to generate masses for quarks as well as to provide an explanation for the appearance of bound states of quarks and anti-quarks, such as the pions, as a consequence of the Goldstone theorem. However, being non-renormalisable in $3+1$ dimensions in addition to not presenting colour confinement, the NJL model was somewhat abandoned in favour of quantum chromodynamics (QCD) which was later developed. Nevertheless, the NJL model shares interesting features with the low energy limit of QCD.

In contrast to NJL models, it is also possible to build four-fermion interaction models which do not produce Goldstone bosons once fermion masses are dynamically generated. In order to do so, the interaction term should break the continuous chiral symmetry explicitly. If the model is built in such a way not to allow for bare mass terms, another (non-continuous) symmetry should be present to prevent fermion masses to appear. This is what happens in the Gross-Neveu (GN) model~\cite{GN} where, contrary to the NJL model, the interaction term is not invariant under continuous chiral transformations
\be\label{ccs}
		\psi \to \exp(i \alpha \gamma_5)~\psi~,
	\ee 
but it is invariant under the discrete chiral transformation
	\be\label{dcs}
		\psi\to\gamma_5 \psi~,
	\ee
which, therefore, also prevents the appearance of a mass term in the bare model.

In general, the four-fermion interaction terms can be constructed from the usual bilinears, for which transformations under the continuous~(\ref{ccs}) and discrete~(\ref{dcs}) chiral transformations are shown in Table~\ref{bitransf}. 
	\begin{table}[h!]
		\centering
			\begin{tabular}{||c || c | c ||} 
				\hline
				  Bilinear                                           &  Discrete chiral transf.           &  Continuous chiral transf.  \\ 
				\hline\hline
				Scalar: $\bar{\psi}\psi$                & $-\bar{\psi}\psi$                    &  $\bar{\psi}\psi \cos{2\alpha} + i\bar{\psi}\gamma_5\psi\sin{2\alpha}$ \\ \hline
				Pseudoscalar: $i\bar{\psi}\gamma_5\psi$          &  $-i\bar{\psi}\gamma_5\psi$          &  $i\bar{\psi}\gamma_5\psi\cos{2\alpha}-\bar{\psi}\psi\sin{2\alpha}$ \\ \hline
				Vector: $i\bar{\psi}\gamma_\mu\psi$      &  $i\bar{\psi}\gamma_\mu\psi$         &  $i\bar{\psi}\gamma_\mu\psi$ \\ \hline
				Pseudovector: $i\bar{\psi}\gamma_\mu\gamma_5\psi$ &  $i\bar{\psi}\gamma_\mu\gamma_5\psi$ &  $i\bar{\psi}\gamma_\mu\gamma_5\psi$\\ \hline
				Tensor: $\bar{\psi}\sigma_{\mu\nu}\psi$  &  $-\bar{\psi}\sigma_{\mu\nu}\psi$    &  $\bar{\psi}\sigma_{\mu\nu}\psi\cos{2\alpha}+i\bar{\psi}\gamma_5\sigma_{\mu\nu}\psi\sin{2\alpha}$ \\ 
				\hline
			\end{tabular}
		\caption{Bilinears under discrete and continuous chiral transformations}
		\label{bitransf}
	\end{table}

From the Table~\ref{bitransf}, one can construct a four-fermion interaction term invariant under continuous chiral transformations (as well as discrete ones) by squaring either the vector or pseudovector contributions, for example. On the other hand, it is not difficult to see that all four-fermion interaction terms constructed by squaring any of the bilinears will be symmetric under discrete chiral transformations.

The four-fermion interaction terms, however, are not completely independent of each other, as they can be related by means of the Fierz identities
	\be\label{FierzId}
		(\bar{\psi}_a\mathcal{O}_i\psi_b)(\bar{\psi}_c\mathcal{O}^i\psi_d) = \sum_{k=S,Ps,V,Pv,T} C_{ik}(\bar{\psi}_a\mathcal{O}_k\psi_d)(\bar{\psi}_c\mathcal{O}^k\psi_b)~,
	\ee
where $\mathcal{O}_S=I$ ($I$ is the identity matrix), $\mathcal{O}_{Ps}=\gamma_5$, $\mathcal{O}_V=\gamma_\mu$, $\mathcal{O}_{Pv}=\gamma_\mu\gamma_5$ and $\mathcal{O}_T=\sigma^{\mu\nu}/(i\sqrt{2})$~,
and $C_{ik}$ are the elements of the $5\times 5$ ``matrix'' shown in Table~\ref{FierzT}~\cite{Miranskybook}.
	\begin{table}[b]
		\centering
			\begin{tabular}{||c || c  c  c  c  c||} 
				\hline
				&   S  &  V  &  T  &  Pv &  Ps \\
				\hline\hline
				S  & -1/4 & -1/4 & 1/4 & 1/4 & -1/4 \\ 
				V &  -1  &  1/2 &  0  & 1/2 &  1\\ 
				T  &  3/2 &   0  & 1/2 &  0  &  3/2\\ 
				Pv &   1  &  1/2 &  0  & 1/2 & -1\\ 
				Ps  & -1/4 & 1/4 & 1/4 & -1/4 & -1/4 \\
				\hline
			\end{tabular}
		\caption{Coefficients for the Fierz identities}
		\label{FierzT}
	\end{table}
	
As an example, considering the NJL model, where the interaction term is given by
	\be 
		[(\bar{\psi}\gamma_\mu\psi)^2-(\bar{\psi}\gamma_\mu\gamma_5\psi)^2 ]~,
	\ee 
one can, equivalently, use the Fierz identities~(\ref{FierzId}) together with the coefficients in Table~\ref{FierzT} to express such an interaction in terms of the scalar and pseudoscalar bilinears as
	\be 
		-2[(\bar{\psi}\psi)^2-(\bar{\psi}\gamma_5\psi)^2]~,
	\ee 
which is obviously invariant under the continuous chiral transformations~(\ref{ccs}).

We shall from now on focus on GN-type models, {\it i.e.} four-fermion interaction models which explicitly break the continuous chiral symmetry. The reason for this being that we are mainly interested in the dynamical mass generation aspect of the mechanism rather than on the appearance of Goldstone modes.

The Gross-Neveu model~\cite{GN} was originally considered in $1+1$ dimensions because in such low space-time dimensions the theory is renormalisable in the usual sense. Nevertheless, considering here the GN model as an effective field theory, we study it in $3+1$ dimensions and introduce a cut off to regularise any divergent integral that may appear in our calculations. It is worth mentioning though, that even in $3+1$ space-time dimensions Lorentz-violating versions of the four fermion interaction models, GN or NJL type, can be made ``renormalisable", once one adopts the Lifshitz scaling (anisotropic scaling between space and time)~\cite{ABH} or, equivalently, the ``weighted power counting" approach shown in~\cite{Anselmi4ferm}.

\subsection{Gross-Neveu-type model}

In the original work~\cite{GN}, the authors have studied, among other things, the dynamical mass generation in the large-N limit ($1/N$ expansion) in $1+1$ dimensions. They have shown that fermion masses are generated irrespective of the strength of the coupling constant $g$. We assume here, however, that $N=1$ and, as a consequence, there is a minimum value for $g$ below which no fermion mass can be generated dynamically.

The action for the Gross-Neveu model in 3+1 dimensions, where the interaction term is obtained by squaring the scalar bilinear in~(\ref{dcs}), reads
	\be\label{GNaction} 
		S_{NG} = \int d^4 x \left[\bar{\psi}(i\slashed{\partial})\psi + \frac{g^2}{2}(\bar{\psi}\psi)^2 \right]~,
	\ee 
where the coupling constant $g^2$ has dimension $[mass]^{-2}$.

Interestingly enough, the action below, which describes the interaction between a fermion field $\psi$ with an auxiliary (with no kinetic term) scalar field $\phi$ via a Yukawa term, shares its main features with~(\ref{GNaction}),
	\be\label{GNamod} 
		S'_{GN} = \int d^4 x \left[\bar{\psi}(i\slashed{\partial})\psi -\frac{1}{2}\phi^2- g\phi\bar{\psi}\psi \right]~.
	\ee
By requiring that the scalar field transforms as $\phi\to-\phi$, while the fermion field transforms as given by~(\ref{dcs}), we make the action above invariant under the discrete chiral symmetry which prevents the appearance of a mass term for $\psi$.	

To understand the similarity between the models in~(\ref{GNaction}) and~(\ref{GNamod}), let us write down the generating functional $Z$ for~(\ref{GNamod}) 
	\be\label{Zsc}
		Z[\eta,\bar{\eta}] = \int \mathcal{D}[\phi,\psi,\bar{\psi}] e^{i\int d^4 x \left[\bar{\psi}(i\slashed{\partial})\psi -\frac{1}{2}\phi^2- g\phi\bar{\psi}\psi + \bar{\eta}\psi+    \bar{\psi}\eta \right]}~,
	\ee
which, by integrating out the scalar field, reduces to
	\be
		Z[\eta,\bar{\eta}] \propto \int \mathcal{D}[\psi,\bar{\psi}] e^{i\int d^4 x \left[\bar{\psi}(i\slashed{\partial})\psi + \frac{g^2}{2}(\bar{\psi}\psi)^2 + \bar{\eta}\psi+\bar{    \psi}\eta \right]}~,
	\ee
{\it i.e.} the generating functional for the Gross-Neveu model given by the action~(\ref{GNaction}). Thus, the correlation functions calculated using either~(\ref{GNaction}) or~(\ref{GNamod}) should be the same, and both models are actually equivalent. In this context, the scalar field $\phi$ represents a fermion condensate $\propto g^2<\bar{\psi}\psi>$ which is formed if the coupling is strong enough and, in this case, breaks the discrete chiral symmetry, allowing a mass term for $\psi$ to appear, as we see below.

\subsection{Dynamical mass generation: the effective potential approach}

In this section we consider the Gross-Neveu model in $3+1$ dimensions given by~(\ref{GNaction}) (or equivalently~(\ref{GNamod})), and, by using the effective potential approach, we show how a mass term for the fermion field is dynamically generated.

The advantage of working with~(\ref{GNamod}) instead of~(\ref{GNaction}) is that the former action presents a Yukawa term, similar to the one that appears when the Higgs field couples to a fermion in the standard model. Therefore, from our experience with the Higgs mechanism (see section~(\ref{sec:EWSB})), we know that when the scalar field acquires a non-trivial vev, it generates a mass for the fermion via the Yukawa term. Thus, we can verify the generation of a mass term ($m_d = g \phi_0$) for the fermion field in our theory by first calculating the potential for $\phi$ and then its {\it vev} $\phi_0$. 

To obtain the effective potential for $\phi$, instead of integrating out the scalar field and going back to the original GN action, we integrate out the fermion fields in~(\ref{Zsc}), and find
	\be\label{spot} 
		V(\phi) = \frac{\phi^2}{2}+ i ~\text{Tr} \int \frac{d^4 p}{(2\pi)^4} \ln(\slashed{p}-g\phi)~.
	\ee  
The minimum, $\phi_0$, of this potential is a solution of $\left(dV(\phi)/d\phi\right)_{\phi_0}=0$. Differentiating~(\ref{spot}) with respect to $\phi$, we obtain, in Euclidean space-time (after a Wick rotation),
	\be\label{minint} 
		\phi_0 = 4g^2\phi_0 \int \frac{d^4 p}{(2\pi)^4} \frac{1}{p^2+g^2 \phi_0^2}.
	\ee 
From the equation above, it is obvious that $\phi_0 = 0 $, {\it i.e.} no dynamical mass generation, is the trivial solution. However, since we are interested in dynamical mass generation, we need to look for non-trivial solutions: $ \phi_0 \neq 0$. Because the integral in~(\ref{minint}) is divergent, let us use a Lorentz-invariant cut off $\Lambda^2 =p^2=\omega^2+\vec{p}^2$ to regularise it. After integration, eq.~(\ref{minint}) becomes
	\be 
		\frac{2(2\pi)^2}{g^2\Lambda^2} = 1- \left(\frac{g\phi_0}{\Lambda}\right)^2\ln\left(1+\frac{\Lambda^2}{(g\phi_0)^2}\right)~.
	\ee 
Assuming now that $(m_d/\Lambda)<1$, where $m_d=g\phi_0$ is the dynamical mass, we see that the right hand side of the equation above allows for values between $(1-\ln 2)$ and $1$ only, which leads to the following constraint on the coupling constant
	\be\label{critcoup} 
		 g_{crit} \leq g \leq \frac{g_{crit}}{\sqrt{1-\ln 2}}~~~\mbox{with}~~~g_{crit}=\frac{\sqrt{2}(2\pi)}{\Lambda}~.
	\ee 
	
Therefore, a non-trivial solution for the fermion mass is only possible if the coupling constant $g$, governing the strength of the four-fermion interaction, is equal to or greater than the critical coupling ($g_{crit}$) given in~(\ref{critcoup}), but, at the same time, $g$ must not be greater than $\approx 1.8~g_{crit}$. Because dynamical mass generation takes place for a very limited range of values for $g$, we recognise this as a fine-tuning problem. Furthermore, it is worth noticing that the upper-limit for $g$ in~(\ref{critcoup}) goes to $g_{crit}$ when $m_d/\Lambda\to 0$, which means that the smaller $m_d$ with respect to $\Lambda$, the more fine-tuned $g$ should be in order to allow for dynamical mass generation to take place.

\chapter{Global and local Lorentz violation}\label{chap3}
\mbox{}

Lorentz symmetry, whether global or local, is one of the pillars of two of the most successful physical theories of all times: the standard model of particle physics and general relativity. This fact is, by itself, strong enough to motivate a serious investigation of whether or not Lorentz symmetry is indeed an exact symmetry of nature. However, a more commonly discussed motivation for the study of Lorentz invariance violating theories comes from the fact that, although a fully consistent description of the quantum effects of gravity has not yet been found, many approaches for quantum gravity favour Lorentz invariance violation; string theory~\cite{string}, space-time foam models~\cite{foam}, brane-world scenarios~\cite{brane}, and non-commutative geometry~\cite{non-comm} are some well known examples. 

In this chapter, exploring the possibility that Lorentz symmetry is not an exact symmetry of nature, but only an approximate one in the low-energy limit, we work with two different ways of introducing Lorentz violation in quantum field theory.\footnote{Although, in this chapter, we restrict ourselves to two main frameworks for Lorentz violation, it is worth noting that many other interesting possibilities have been proposed, {\it e.g.} varying speed of light (VSL) theories \cite{VSL}. For a review of Lorentz violation see, for example, \cite{Liberatireview} and references therein.} In the first section we consider an effective field theory in which global Lorentz symmetry is spontaneously broken by non-trivial vacuum expectation values of tensor fields. Then, in the following section, we study explicit violation of Lorentz symmetry due to the anisotropic scaling between space and time, focusing mainly on the local violation of Lorentz invariance in a modified quantum theory of gravity.

\section{The Standard Model Extension}\label{SME}
\mbox{}

In this section we study global Lorentz violation that takes place due to the presence of constant tensor fields which break Lorentz symmetry spontaneously. For this task, we consider the standard model extension (SME) which began to be developed by Kostelecky $\it et~al$ in the late $`90s$~\cite{SME1,SME2}. 

The SME is a low-energy effective theory which extends the standard model of particle physics by adding all possible Lorentz- (and CPT-) violating terms that could arise from spontaneous symmetry breaking at a fundamental level, but leaves the $SU(3)\times SU(2) \times U(1)$ gauge structure of the SM unmodified. In general, the SME can be divided in two parts: the minimal SME~\cite{SME1,SME2}, containing operators up to dimension 4 only; and the non-minimal SME~\cite{nmSME1,nmSME2,nmSME3} which contains non-renormalisable operators (of dimension greater than 4).

An important and subtle point to understand is how exactly the new terms break Lorentz invariance. Initially, we need to consider the two kinds of Lorentz transformations (LTs) describing boosts and rotations: observer and particle Lorentz transformations. Observer transformations are the most common, or conventional, kind of LTs. They can be seen as coordinate changes relating observations of a given physical system made by distinct inertial frames. On the other hand, the particle transformations are those which relate the properties of particles (or fields) with different momenta or spin orientations within a specific inertial frame.

Although observer and particle LTs can in general be used interchangeably, with one being simply the inverse of the other, this is not true in the presence of a tensor background field. Tensor background fields transform accordingly under observer transformations, but they are insensitive to particle transformations under which they ``transform'' as scalar fields instead. Thus, because the Lorentz-violating terms in the SME are expected to come from a fundamental Lorentz-covariant theory when tensor fields acquire a non-vanishing vev, they must be invariant under {\it observer} LTs, while breaking {\it particle} LTs. In summary, observer LTs, which are defined as coordinate changes, are preserved in the SME, whereas particle LTs, which involve boosts and rotations on the physical system (particles or localised fields), but not on the background fields, are not, thus manifesting the Lorentz invariance violation in the SME.

In the SME, every new term is made of a dimension $d$ operator contracted with a $(4-d)$-dimensional coefficient for Lorentz violation. When higher-order operators are taken into account, as in the non-minimal SME, an infinite number of new terms appear in the theory. Nonetheless, because Lorentz violation is expected to be fairly small (otherwise it should have already been observed in nature) the LIV terms are treated perturbatively and, as a consequence, operators with a lower mass dimension are expected to dominate in the IR. In this way, it is natural that renormalisable LIV operators of dimension $3$ and $4$ hold a special place in the SME. However, in some specific cases, additional symmetries of a more fundamental model can forbid the appearance of lower dimensional operators; {\it e.g.} in supersymmetric theories, where LIV operators of renormalisable mass dimension are forbidden, implying, therefore, that the leading order LIV contributions should come from higher-dimensional operators with dimension greater than $4$~\cite{SUSYLV}.
 
Fortunately, there exists a plethora of experimental and observational measurements which can be used to test and constrain the coefficients for Lorentz violation. In~\cite{bounds}, it is possible to find a comprehensive list of experimental- and observational-based bounds on LIV coefficients of the matter, photon, neutrino and gravity sectors of the SME.


\subsection{Neutrino sector}\label{sec:SMEneut}
 
In this section we describe the neutrino sector of the SME. First, we present the most general action for such a sector containing both minimal~\cite{mSMEneutrino1,mSMEneutrino2} and non-minimal~\cite{nmSME2} operators. Then, in order to obtain the expressions for oscillation probabilities, we present the effective Hamiltonian. Finally, we consider a specific model for neutrino oscillations based on the SME, and discuss some of its essential features and results.

We start by considering three flavour neutrinos ($ \nu_e, \nu_\mu, \nu_\tau $) and combining them with their respective charge conjugate field ($\nu_a^C = C \bar{\nu}_a^T$) in the $6\times 1$ multiplet below, as in~\cite{nmSME2},
	\be\label{N}
		N = \begin{pmatrix} \nu_a\\ \nu_a^C
															\end{pmatrix}~,~~ \mbox{with}~~	a= e, \mu, \tau~.
	\ee	
The most general quadratic SME action for the free neutrinos in~(\ref{N}), allowing for operators of any mass dimension, can be written as
	\be\label{SMEneutaction}
		S_{\nu}^{SME} = \int d^4 x \left\{\frac{1}{2}\bar{N}(i\slashed{\partial}-M+\hat{Q})N+\mbox{h.c.}\right\}~,
	\ee 
with
	\bea\label{MQ}
		M &=& m +i m_{5}\gamma_5~,\\
		\hat{\mathcal{Q}} &=& \sum_I \hat{\mathcal{Q}}^I \gamma_I = \hat{\mathcal{S}}+i\hat{\mathcal{P}}\gamma_5+\hat{\mathcal{V}}^\mu\gamma_\mu+\hat{\mathcal{A}}^\mu\gamma_5 \gamma_\mu+\frac{1}{2}\hat{\mathcal{T}}^{\mu\nu}\sigma_{\mu\nu}~,\nn
	\eea	
where $m$, $m_5$ and $\hat{\mathcal{Q}}^I$ are $6 \times 6$ hermitian matrices. In~(\ref{MQ}), $\hat{\mathcal{Q}}$ was expanded in the basis of Dirac matrices and contains scalar $\hat{\mathcal{S}}$, pseudo-scalar $\hat{\mathcal{P}}$, vector $\hat{\mathcal{V}}^\mu$, axial vector $\hat{\mathcal{A}}^\mu$ and tensor~$\hat{\mathcal{T}}^{\mu\nu}$ contributions. Moreover, the $\hat{\mathcal{Q}}^I$ operators are derivative-dependent and can be expanded as a sum of operators of mass dimension $d$ 
	\be\label{Qexp}
		\hat{\mathcal{Q}}^I = \sum_{d=3}^{\infty} i^{(d-2)} \mathcal{Q}^{(d)I \alpha_1\alpha_2\cdots\alpha_{d-3}}\partial_{\alpha_1}\partial_{\alpha_2}\cdots \partial_{\alpha_{d-3}}~,
	\ee
where $\mathcal{Q}^{(d)I}$ is the dimension $4-d$ coefficient governing the corresponding $d$-dimensional operator. In general, the following terms are allowed
	\bea\label{Qdec}
		\hat{\mathcal{S}} &=& \hat{m}+i\hat{e}^\mu \partial_\mu~,~~ \hat{\mathcal{P}} = \hat{m}_{5}+i\hat{f}^\mu \partial_\mu~,~~ \hat{\mathcal{V}}^\mu=\hat{a}^\mu +i\hat{c}^{\mu\nu} \partial_\nu~,\\
		\hat{\mathcal{A}}^\mu &=& \hat{b}^\mu+i\hat{d}^{\mu\nu} \partial_\nu~,~~ \hat{\mathcal{T}}^{\mu\nu} = \hat{H}^{\mu\nu} +i \hat{g}^{\mu\nu\lambda} \partial_\lambda~.\nonumber
	\eea
Again all terms with a hat on can be expanded as in~(\ref{Qexp}). 

In addition to breaking Lorentz invariance, operators with an odd number of indices ($ \hat{a}^\mu, \hat{b}^\mu, \hat{e}^\nu ,\hat{f}^\nu$ and  $\hat{g}^{\kappa\lambda\nu}$ ) are also not invariant under the discrete CPT transformation. It is easy to see that, since all Lorentz indices appear properly contracted among themselves, all terms in~(\ref{SMEneutaction}) are invariant under observer Lorentz transformations. However, as mentioned above, due to the presence of non-vanishing tensor vevs, represented by the coefficients for Lorentz violation, this effective model is not invariant under particle LTs. 

As we are interested in calculating oscillation probabilities, it is more convenient to work directly with the effective Hamiltonian associated with~(\ref{SMEneutaction}). In addition to this, at this point, since the neutrinos observed in nature are chiral (left-handed) fields, we should project the formalism above onto left-handed fields. Then, following~\cite{nmSME2}, the effective Hamiltonian for the neutrino sector is found to be given by
	\be\label{effH} 
		h_{eff} = (h_{eff})_0 + \delta h~,
	\ee
where the dominant contribution, $(h_{eff})_0$, is the usual Lorentz-symmetric Hamiltonian written in terms of $3 \times 3$ block matrices as
	\be\label{effH0} 
		(h_{eff})_0 = |\vec{p}| \begin{pmatrix}   1 & 0 \\ 0 & 1\end{pmatrix} +\frac{1}{2 |\vec{p}|}\begin{pmatrix}   m_l m_l^\dagger & 0 \\ 0 &  m_l^\dagger m_l\end{pmatrix}~,
	\ee
with $m_l$ being the effective left-handed mass matrix. The LIV perturbative corrections come from
	\be\label{effHLV} 
		\delta h = \frac{1}{|\vec{p}|}  \begin{pmatrix}   \hat{a}_{eff}-\hat{c}_{eff} & -\hat{g}_{eff}+\hat{H}_{eff} \\ -\hat{g}_{eff}^\dagger+\hat{H}_{eff}^\dagger & -\hat{a}_{eff}^T-\hat{c}_{eff}^T\end{pmatrix} ~.
	\ee
The effective terms above, $\hat{a}_{eff},\hat{c}_{eff}, \cdots$, represent combinations of the LIV operators shown in~(\ref{Qdec}). For brevity, the exact form of such terms are omitted here, since they are rather involved, but they can be found in~\cite{nmSME2}.

\subsubsection{Neutrino oscillations}

In order to investigate neutrino oscillations in the SME, we can make use of the time evolution operator, as in section~\ref{sec:NOsc}, to calculate the oscillation probabilities. For neutrino oscillations, the diagonal terms present in $(h_{eff})_0$ do not play any role and can be consistently omitted. By doing this, the time-evolution operator relevant to oscillations becomes $S(t) = \exp{(-i h_{osc}t)}$, with
	\be\label{hosc}
		h_{osc} = \frac{1}{|\vec{p}|} \begin{pmatrix} \frac{1}{2}  m_l m_l^\dagger +\hat{a}_{eff}-\hat{c}_{eff} & \hat{H}_{eff}-\hat{g}_{eff} \\ \hat{H}_{eff}^\dagger-\hat{g}_{eff}^\dagger & \frac{1}{2} m_l^\dagger m_l-\hat{a}_{eff}^T-\hat{c}_{eff}^T  \end{pmatrix}~.
	\ee
Thus, oscillation probabilities can be calculated following the standard procedure (see section~\ref{sec:NOsc}) which leads to $\mathcal{P}_{\nu_a\to\nu_b}(t) = |(U^\dagger e^{-i E_{eff} t} U)_{ab}|^2$, where $U$ is the unitary matrix which diagonalises the effective Hamiltonian~(\ref{hosc}).

The effective Hamiltonian for oscillations~(\ref{hosc}) represents a system with an infinite number of degrees of freedom per space-time point, when taking into account all possible terms of any dimension in the SME. In order to make it more manageable, so that one can effectively calculate oscillation probabilities, it is necessary to restrict the amount of parameters in the model. This can be performed in different ways, one of which is by finding a more fundamental theory which naturally reduces the number of LIV operators due to its symmetries, and hopefully agrees with current measurements. Another way, as in the example below, is to choose the parameters so that the resulting model reproduces most or all experimental data available.
	

	\begin{itemize}
		\item Neutrino oscillations in the SME: the puma model
	\end{itemize}

In the last decade, many models based on the minimal~\cite{mSMEneutrino2,SMEneutosc2,SMEneutosc3,SMEneutosc4} and non-minimal~\cite{puma1,puma2} SME have been considered as alternative ways of describing neutrino oscillations. Despite the partial success of models containing one mass parameter at most, it is clear now that these ``simple'' LIV models cannot reproduce all the current data. Nonetheless, they teach us much about the power of the SME in describing physical phenomena. In this section, we choose one of these models, the so-called ``puma model''~\cite{puma1,puma2}, and present some of its interesting features and results.

The puma model Hamiltonian is a very particular case of the general expression given in~(\ref{hosc}) which only contains three parameters. Assuming three flavours of left-handed active neutrinos, the Hamiltonian can be written in the flavour basis as the following texture 
	\bea\label{pumaH}	
		h_{osc}^{puma} &=& A\begin{pmatrix} 1 & 1 & 1\\ 1 & 1 & 1\\ 1 & 1 & 1\end{pmatrix}+B\begin{pmatrix} 1 & 1 & 1\\ 1 & 0 & 0\\ 1 & 0 & 0  \end{pmatrix}+C \begin{pmatrix} 1 & 0 & 0\\ 0 & 0 & 0\\ 0 & 0 & 0  \end{pmatrix}~,\\
 	 \mbox{with}~~   A(E) &=& \frac{m^2}{2E}~,~~B(E) = \mathring{a} E^2~,~~ C(E) = \mathring{c} E^5	~.\nn
	\eea
As explained in~\cite{puma1,puma2}, this specific effective model~(\ref{pumaH}) was obtained through a systematic search for a SME model describing the established neutrino oscillations data (at that time), by considering models with no more than three parameters and a simple analytical form. 

It can be seen in~(\ref{pumaH}) that in the IR limit the Hamiltonian is dominated by the first matrix, which depends on the mass parameter. The other two terms, proportional to $\mathring{a}$ and $\mathring{c}$, which are associated with operators of the non-minimal SME of dimension 5 and 8, respectively, break Lorentz invariance ($\mathring{a}$ breaks CPT as well), and have a non-standard energy dependence which leads them to dominate over the mass term in the UV.

Since the Hamiltonian above only takes into account neutrinos, (\ref{pumaH}) represents the upper left $3\times 3$ block of~(\ref{hosc}) for the present case. Then, the effective Hamiltonian describing anti-neutrinos $\bar{h}_{osc}^{puma}$, which is associated with the lower right block of $(\ref{hosc})$, can be obtained by performing a CPT transformation in $h_{osc}^{puma}$. In the present case, as the only CPT-odd term is the one related to $\mathring{a}$, finding $\bar{h}_{osc}^{puma}$ is a straightforward task: one only needs to change the sign of the respective coefficient.

Another unusual feature of this model is found when diagonalising the Hamiltonian~(\ref{pumaH}): one of the eigenvalues vanishes. The other two eigenvalues are 
	\be\label{eig}
		\lambda_{\pm}  = \frac{1}{2} \left[3 A + B + C \pm \sqrt{(A-B-C)^2+8(A+B)^2}  \right]~. 
	\ee 
As usual, the oscillation probability is found in terms of the unitary matrix $U^{puma}$ which diagonalises the Hamiltonian above	
	\be\label{U}
	U^{puma} =\begin{pmatrix} \frac{\lambda_+-2A}{N_+} & \frac{A+B}{N_+} & \frac{A+B}{N_+} \\ \frac{\lambda_- -2A}{N_-} & \frac{A+B}{N_-} & \frac{A+B}{N_-} \\ 0 & -\frac{1}{\sqrt{2}} & \frac{1}{\sqrt{2}}\end{pmatrix}~,
	\ee
with $N_\pm = \sqrt{(\lambda_\pm-2A)^2+2(A+B)^2}$. 	

The unitary matrix~(\ref{U}) assumes, in the IR region, the so-called ``tribimaximal form''\footnote{The term ``tribimaximal'' follows from the fact that, in this configuration, one of the massive neutrinos is described by a maximal (uniform) mixture of the three flavour neutrinos (``trimaximal mixing''), while another massive neutrino is described by a maximal mixture of only two of the three flavour neutrinos (``bimaximal mixing''). }~\cite{tribi}; therefore, in such a region, the theoretical results found with the puma model can be made compatible with those of the $3\nu SM$\footnote{The $3\nu SM$ model is the standard Lorentz-symmetric extension of the minimal SM including three massive left-handed neutrinos~\cite{PDG}; it is characterised by six parameters: two mass-squared differences, three mixing angles, and one CP-violating phase (for Dirac fields).} extension, for which the mixing is described by the PMNS matrix~(\ref{PMNS}), when, among other things, the mixing angle $\theta_{13}$ vanishes. When the energy increases, the eigenvalue $\lambda_-$ becomes proportional to $\mathring{a}^2/\mathring{c}E$, behaving thus as a mass term, even though the original mass term in the model becomes negligible in the UV. 

In the $3\nu SM$, each mixing angle, not being energy-dependent, assumes a unique value which can be chosen accordingly to the data. Consequently, they represent degrees of freedom of the respective model. In the puma model, on the other hand, the mixing angles are energy dependent, changing with the energy in a way determined by the texture~(\ref{pumaH}). Therefore, in this case, the values for the mixing angles cannot be chosen to match the data as in the $3\nu SM$, because their values are intrinsic properties of the model~(\ref{pumaH}).

By comparing the expressions for oscillation probabilities found using both the puma model and the $3\nu SM$~(\ref{aoscp3}), and taking into account the experimental results, we can determine the values of the three parameters of the former model. For example, the survivor probability for reactor anti-neutrinos, in the low energy limit, in the puma model and in the $3\nu SM$ are respectively given by:
	\bea\label{survprob}
		\mathcal{P}_{\bar{\nu}_e\to\bar{\nu}_e}^{puma} \approx 1-\frac{8}{9} \sin^2\left(\frac{3 m^2}{4 E}\right)~~~\\
		\mathcal{P}_{\bar{\nu}_e\to\bar{\nu}_e}^{3\nu SM} \approx 1-\sin^2(2 \theta_{12}) \sin^2\left(\frac{\Delta m^2_\odot L}{4 E}\right)~~~.
	\eea 
As discussed above, the value for the mixing angles in the puma model are determined by the texture and, in this case, comparing both expressions above, we see that $\sin^2(2 \theta_{12})_{eff} = 8/9$, which agrees with the experimental results. Moreover, comparing the mass terms, we see that $m^2 =\Delta m_\odot^2/3$ which matches the data, if we choose $m^2$ according to experimental results for $\Delta m_\odot^2$. The other degrees of freedom of the model $\mathring{a}$ and $\mathring{c}$ can be determined by similar procedures when looking at different experiments~\cite{puma1,puma2}. As a result, the three parameters of the puma model can be chosen as
	\be
		m^2 = 2.6 \times 10^{-23} GeV^2~,~~\mathring{a} = -2.5\times 10^{-19} GeV^{-2}~, ~~ \mathring{c} = 1.0 \times 10^{-16} GeV^{-4}~.
	\ee	
With this choice, it was shown in~\cite{puma1,puma2} that by the time the puma model was proposed, it was compatible with most experimental results on neutrino oscillations\footnote{At that time, $\theta_{13}$ was believed to be very small ($\approx 0$).}, except for some anomalies that also could not be explained by the $3\nu SM$ model. This is no longer the case, however, as we explain in what follows. The mixing angle $\theta_{13}$ has been recently measured to an unprecedented degree of accuracy~\cite{T131,T132} and, to the surprise of many, its value, $\theta_{13} \simeq \pi/20$ \cite{PDG}, is not as small as once thought. On the other hand, the puma model, due to its tribimaximal IR structure, as already mentioned, requires that $\theta_{13}$ should vanish. Thus, the present version of the puma model is no longer phenomenologically accurate.

In summary, the puma model reveals the power of the SME, since it only needs three parameters to reproduce most experimental data.

\section{Lifshitz-type theories}\label{sec:Lifshitz}
\mbox{}

In this section we present a different approach to Lorentz violation. Instead of treating Lorentz violation as a result of spontaneous symmetry breaking caused by non-trivial vevs of tensor fields as in the SME, we consider explicit Lorentz violation based on the assumption that space and time satisfy the Lifshitz, or anisotropic, scaling, as defined below.

Lifshitz-type theories are characterised by their invariance under the anisotropic scaling between space and time
	\be\label{Lsc}
		\vec{x} \to b \vec{x}~~~\mbox{and}~~~t\to b^z t~,
	\ee
where $z$ is the critical exponent. When $z=1$, space and time scale isotropically as in special relativity, whereas for $z\neq 1$ space and time are treated differently; consequently, although still invariant under Galilean transformations, theories invariant under~(\ref{Lsc}) with $z\neq 1$ are no longer Lorentz-symmetric. This anisotropic scaling implies that the dimensions of space and time coordinates are given by $[x_i] =-1$ while $[t]=-z$.

The study of models with anisotropic scaling, especially in the gravitational context, is motivated by the desire for a consistent way of constructing UV-complete theories. Whereas covariant higher-order derivative modifications of GR present an improved UV behaviour~\cite{Stelle}, they generally contain ghost excitations (Ostrogradky's ghosts~\cite{Ostrog, Ostrog2}), which lead to violation of unitarity, as a result of the presence of higher-order time derivatives. The Lifshitz scaling allows one to build theories with higher-order space derivatives only, keeping the number of time derivatives to its minimum, which present an improved UV behaviour without introducing Ostrogradky's ghosts. 	

In the remainder of this section, we will explore the consequences of the Lifshitz scaling~(\ref{Lsc}) for quantum field theories.\footnote{For a comprehensive introduction to the subject, see~\cite{Lifreview}.} Firstly, the main features of such an idea will be investigated through the study of a simple Lifshitz-type model involving only scalar fields. We will then focus on how the Lifshitz scaling is used in the formulation of a modified theory of quantum gravity, the so-called Horava-Lifshitz gravity, and present three different versions.

\subsection{Lifshitz scalar field theory}

We start this section by deriving the action for a Lifshitz-type model containing only scalar fields in $d+1$ space-time dimensions
	\be
		S_{sc} =\int dt d^d x~\mathcal{L}_{sc}~.
	\ee
When considering the anisotropic scaling between space and time~(\ref{Lsc}), we have that $[dt d^d x] = -z-d$. As a consequence, only terms of dimension up to $z+d$ need to be considered for the Lagrangian $\mathcal{L}_{sc}$ of our theory. In order to avoid pathologies associated with higher-order time derivatives, we start defining the dimension of $\phi$ by requiring that $[(\partial_t \phi)^2]=z+d$. Since $[\partial_t]=z$, we find that
	\be
		[\phi]=\frac{d-z}{2}~.
	\ee
The scalar field $\phi$ is dimensionless when $d=z$ and, in such a case, the theory is at least power-counting renormalisable.

From now on, since we are interested in the renormalisability of the theory, let us take $z=d$, which leads to
	\be
		[dt d^d x] = -2z~,~~~[(\partial_t \phi)^2]= 2z~~\mbox{and}~~[\partial_i\phi\partial^i\phi]=2~.
	\ee
As shown above, because the scalar field is dimensionless as a result of $z=d$, the usual term containing only two spatial derivatives is of dimension 2 only. Therefore, terms involving up to 2z spatial derivatives can be consistently added to the Lagrangian, and for $z=d>1$, it means that higher-order spatial derivatives are allowed in the theory, e.g. for $z=2(3,4...)$ we can add terms with up to 4(6,8...) spatial derivatives. 

Thus, in general, for a free scalar field in $d+1$ dimensions with $z=d$, we can write the following action
	\be\label{scac}
		S_{sc} = \frac{1}{2}\int dt d^d x \left[ \dot{\phi}^2 - \phi\left(\sum_{i=1}^z \Lambda_{i}^{2(z-i)} (-\Delta)^i\right)\phi-m^{2z} \phi^2\right]~,
	\ee
with $\dot{\phi} =\partial_t \phi$, $\Delta = -\partial_k \partial^k =\vec{\partial}\cdot\vec{\partial}$ is the Laplacian and $[\Lambda_{i}] =1$.	

The dispersion relation for such a free scalar is then
	\be\label{dr}
		\omega^2 = m^{2z} + \sum_{i=1}^z \Lambda_{i}^{2(z-i)}\vec{p}^{2i}~.
	\ee
In the UV region the dispersion relation is clearly not Lorentz symmetric: $\omega^2 \simeq \vec{p}^{2z}$. However, in the IR, if $\Lambda_1\neq 0$, one can rescale the whole expression to obtain
	\be\label{irdr} 
		\omega'^2 \simeq m'^2 + \vec{p}^2, ~~~\mbox{where}~~~ \omega' = \frac{\omega}{\Lambda_1^{z-1}}~~~\mbox{and}~~~ m' = \frac{m^z}{\Lambda_1^{z-1}}~.
	\ee 
Therefore, in the IR region, Lorentz symmetry is approximately recovered, showing that in this limit $z\to 1$. Nonetheless, as the energy increases, the right hand side of~(\ref{irdr}) starts to receive non-negligible contributions from the higher-order operators, breaking Lorentz invariance.

Let us now introduce self-interactions to the system. As the scalar field is dimensionless, the model allows us to add any self-interaction terms of the form $\alpha_n \phi^n$, where $[\alpha_n] = 2z$. For simplicity, we only consider the following term with $n=4$, and take $d=z=3$ (which is sufficient for our purposes here);
\be\label{phi4int} 
	\mathcal{L}_{int} =\alpha_4 \phi^4~.
\ee 
To understand the basic advantages of such theories, compared to Lorentz-symmetric ones, let us roughly estimate the one-loop correction to the scalar field propagator. When the only interaction term is the one in~(\ref{phi4int}), the only one-loop diagram that need to be taken into account is that which contains a single fermion propagator, and its correction, in Euclidean space-time, is proportional to
	\be
		\int \frac{ d \omega d^3 p }{(2\pi)^4} \frac{1}{\omega^2 + m^{6} + \Lambda_1^4 \vec{p}^2 +\Lambda_2^2 \vec{p}^4 + \vec{p}^6} \propto \int dp \frac{p^2}{\sqrt{m^6 + \Lambda_1^4 p^2 +\Lambda_2^2 p^4 + p^6}}~,
	\ee 
with $p^2 = \vec{p}^2$. As can be seen, instead of being quadratically divergent as in the Lorentz symmetric case, the integral above is actually only logarithmically divergent due to the higher-order spatial derivatives in~(\ref{scac}). We used this example to present a general feature of Lifshitz-type theories: loop integrals appearing in these theories present an improved UV behaviour, being less divergent than their corresponding Lorentz-symmetric correction, or even convergent.

In addition to improving the convergence of loop integrals, Lifshitz-type models usually allow for a larger set of renormalisable interactions. For example, a self-interaction term involving $\phi^6$ in the Lorentz-symmetric case would necessarily appear with a coupling constant $\alpha'_6$, where $[\alpha'_6]=-2$, because for $d=3$ and $z=1$, we have $[\phi]=1$. Therefore, such a term is irrelevant. On the other hand, if we consider the same interaction term in the Lifshitz-type model above, the coupling constant $\alpha_6$ will have dimension of 6, thus becoming relevant. In this way, another advantage of Lifshitz-type theories is that the class of renormalisable interactions is, in general, extended. 

Some important works have been dedicated to the above-mentioned ideas regarding improved convergence of loop integrals and, consequently, renormalisabity of LIV theories containing higher space derivatives~\cite{Anselmi1,Anselmi2,Anselmi3,Anselmi4,Visser}. Detailed aspects of such theories have been studied, and technical machinery developed, in work by Anselmi {\it et al}~\cite{Anselmi1,Anselmi2,Anselmi3,Anselmi4}. However, instead of Lifshitz (or anisotropic) scaling, another nomenclature, with expressions such as ``weighted scale invariant theories'' and ``weighted power counting'', is adopted.

As one might expect, however, Lifshitz-type theories do not bring advantages without presenting new challenges . While such models are clearly Lorentz-violating in the UV, in the IR we expect Lorentz symmetry to be approximately recovered in order to agree with the current experimental bounds on Lorentz violation. While in a free theory we have the freedom to rescale the parameters, as in~(\ref{irdr}), to recover Lorentz symmetry, when different particles interact, this procedure cannot be performed for all particles at the same time. As a result, different particles will experience different limiting speeds of propagation. Calculating these speed differences and comparing them with the experimental bounds on Lorentz invariance is a useful way to constrain some of the parameters of the relevant Lorentz-violating theory and, in some cases, even rule out certain Lifshitz-type models \cite{deviation,ABH}.

\subsection{Horava-Lifshitz gravity}\label{subsec:HL}
\mbox{}

With the concepts developed in the previous section in mind, it is natural to try using Lifshitz scaling as a tool in the quest for a renormalisable quantum theory of gravity. This task was originally performed by Horava in $2009$~\cite{Horava} when he developed what is now known as Horava-Lifshitz (HL) gravity. In this section we present the original HL theory with $d=z=3$, its different versions, together with their main features and problems. The metric signature in this section on HL gravity is $(-,+,+,+)$.\footnote{To avoid any confusion, we emphasise, as initially stated, that throughout this thesis every time gravity is taken into account, we use the mostly plus metric signature $(-,+,+,+)$. On the other hand, when gravitational fields are not considered, we use the mostly minus metric signature $(+,-,-,-)$. This follows the conventions adopted in the respective fields.}

The anisotropic scaling between space and time~(\ref{Lsc}) can be applied to gravity if we give up on one of the most important building blocks of GR, invariance under 4-dimensional diffeomorphisms. Instead, as space and time must be treated differently, the theory will be invariant under a reduced set of symmetries. It is not necessary to look very far for a consistent way to describe a gravity model in which space and time are treated differently, the Arnowitt-Deser-Misner (ADM) decomposition of the metric provides the necessary formalism:
	\be\label{ADMdec}
			g_{\mu\nu} d x^\mu d x^\nu = -c^2 N^2 d t^2 + g_{ij}(dx^i+N^idt)(dx^j+N^jdt)~,
	\ee
where $N$, $N_i$ and $g_{ij}$ are the ADM fields; the lapse function, shift vector and spatial metric, respectively. For convenience, we kept the speed of light $c$ in this decomposition so that we can more easily find the dimensions of the ADM fields. A natural choice is to postulate that the spatial metric is dimensionless $[g_{ij}]=0$ which implies that $[g_{ij} dx^i dx^j]=-2$. Then, since the terms $N^2 c^2 dt^2$ and $g_{ij} N^i dx^j dt$ have the same dimension as $g_{ij} dx^i dx^j$, {\it i.e.} $-2$, we conclude that, assuming the Lifshitz scaling with $z=3$, $[c]=2$, $[N]=0$ and $N^i=2$.

Because of~(\ref{Lsc}), time plays a privileged role in this construction, and therefore space-time is seen as a foliation, with its leaves being spatial hypersurfaces of constant time. As the usual GR diffeomorphisms do not protect such a space-time structure, a theory of gravity built upon the Lifshitz scaling~(\ref{Lsc}) is invariant under the following reduced symmetry:
	\be \label{fpdiff}
		\delta t = f(t)~~~\mbox{and}~~~\delta x^{i} = \xi^i(t,x)~,
	\ee 
known as foliation-preserving diffeomorphisms. We will also refer to~(\ref{fpdiff}) as 3-d diffeomorphisms, although foliation-preserving diffeomorphisms represent a combination of time-dependent space (3-d) diffeomorphisms, with space-independent time reparametrisations. 

In order to build an action which is invariant under foliation-preserving diffeomorphisms, we first need to know how the fields present in the theory, the ADM fields in our case, transform under~(\ref{fpdiff}). If the ADM fields are functions of space and time, they transform as
	\bea\label{ADMtransf}
		\delta g_{ij}&=&\partial_{i}\xi_j+\partial_{j}\xi_i+\xi^{k}\partial_{k}g_{ij}+f\dot{g}_{ij}~, \\
		\delta N_{i}&=&\partial_{i}\xi^k N_k+\xi^k\partial_k N_{i}+\dot{\xi^{j}}g_{ij}+\dot{f}N_i+f \dot{N}_{i}~, \nn \\
		\delta N&=&\xi^k\partial_kN+\dot{f}N+f \dot{N}~,\nn
	\eea 
where the overdot represents a derivative with respect to time ($t$).
While GR contains two propagating degrees of freedom (a massless spin-2 excitation, the graviton), a modified gravity model, invariant under foliation-preserving diffeomorphisms, allows for a new degree of freedom, known as the scalar graviton. The study of the scalar graviton, specially in the IR region, is of crucial importance in understanding whether Horava-Lifshitz gravity can be taken seriously as a phenomenologically viable alternative to GR. In the subsections below, we present three different versions of HL gravity. Initially, we consider the original model and show that the respective scalar graviton presents some undesirable features in the IR. Then, with the aim of finding a new version of the theory in which the additional degree of freedom either presents a more ``acceptable'' behaviour or does not exist, we consider first the non-projectable, and then the covariant versions of HL gravity. 

\subsubsection{The original model: Projectable version with detailed balance}

In the original work by Horava~\cite{Horava}, in addition to imposing invariance under foliation-preserving diffeomorphisms, it was assumed that the lapse function was a function of time only: $N=N(t)$. When this additional condition is imposed, we have what is known as the ``projectable version'' of HL gravity. 

The most general kinetic term for this theory is constructed by requiring it to be quadratic in $\dot{g}_{ij}$. This can be achieved by considering the extrinsic curvature $K_{ij}$, which is invariant under foliation-preserving diffeomorphisms, 
	\be\label{Kij}
		K_{ij} = \frac{1}{2 N}(\dot{g}_{ij}-\nabla_i N_j - \nabla_j N_i)~~~\mbox{with}~~~\nabla_i N_j = \partial_i N_j - \Gamma^{k}_{ij} N_k~,
	\ee 
where $\nabla_i$ and $ \Gamma^{k}_{ij}$ are the covariant derivative and Christoffel symbol, respectively, defined with respect to the spatial metric $g_{ij}$ on the constant time hypersurfaces.
Thus, in terms of~(\ref{Kij}), we can write the following kinetic term for the theory
	\be\label{SK} 
		S_{K} = \frac{2}{\kappa^2} \int dt d^3x \sqrt{g} N \left(K_{ij}K^{ij} -\lambda K^2\right)~,
	\ee
where $K=g^{ij}K_{ij}$, and $\kappa$ and $\lambda$ are coupling constants. Such a kinetic term~(\ref{SK}) is quite general and valid for all versions of HL gravity that we to study here, the only ``difference'' being that in some cases the lapse function in~(\ref{SK}), which appears in the action because it is a part of the volume element $dV =dt d^3 x \sqrt{g} N$, will be allowed to be a function of space and time, while in the present version, as already discussed, $N$ is a function of time only.

It is worth emphasising at this point that $\kappa^2$, which would have mass dimension of $(-2)$ in theories without anisotropic scaling such as GR, where $k^2\propto G_N$, is here dimensionless and therefore represents a huge improvement in terms of the renormalisability of the theory. Another important point to note is that, due to the reduced symmetry, the kinetic term in~(\ref{SK}) contains a new coupling constant $\lambda$ which in GR is automatically set to $1$ according to the invariance under full 4-dimensional diffeomorphisms. Therefore, one might expect that $\lambda\to 1$ if the present theory fully describes GR in the IR. However, as we will see later, this is probably not the case.

Let us consider now the potential term $S_V$. If we wanted to find the most general $S_V$, containing all possible terms of dimension equal to or less than 6, a huge number of new terms and coupling constants would appear in the theory. In order to restrict the number of terms in the potential of the theory, Horava imposed an extra symmetry. Following~\cite{Horava}, the only terms allowed in the potential $S_V$ are those which can be put in the following form,
	\be\label{SpVdef} 
		S_V^{proj} = \frac{\kappa^2}{8} \int dt d^3 x \sqrt{g} N ~E^{ij} \mathcal{G}_{ijkl} E^{kl}~,~~\mbox{with}~~\sqrt{g}E^{ij} = \frac{\delta W[g_{kl}]}{\delta g_{ij}}~,
	\ee 
where $W$ represents some action, and $\mathcal{G}_{ijkl}$ is the inverse of the De Witt metric and 
	\be
		\mathcal{G}^{ijkl} =\frac{1}{2}\left(g^{ik}g^{jl}+ g^{il}g^{jk}\right) - \lambda g^{ij}g^{kl}~.
	\ee
A potential of this form is said to satisfy the ``detailed balance condition''. 

Let us consider that $W$ in~(\ref{SpVdef}) is 
	\be\label{W}
		W = W_{CS} + \mu\int d^3 x \sqrt{g}(R-2\Lambda_W)~,
	\ee
where the first term, $W_{CS}$, is the gravitational Chern-Simons term as given in~\cite{Horava}, while the second is the Einstein-Hilbert action, with $R$ the Ricci scalar associated with $g_{ij}$. The coupling constants $\omega$, $\mu$ and $\Lambda_W$ are of dimension $0,1$ and $2$, respectively.
Thus, the most general potential term, within the detailed balance class, reads
	\bea\label{SpV}
		S_V^{proj} &=& \frac{\kappa^2}{2}\int dt d^3x \sqrt{g}N\left[\frac{\mu^2}{4(1-3\lambda)}\left(\frac{1-4\lambda}{4} R^2+\Lambda_W R - 3\Lambda_W^2\right)\right.\\
		&&\left.-\frac{1}{\omega^4} C_{ij}C^{ij}+\frac{\mu}{\omega^2}\epsilon^{ijk} R_{il}\nabla_j R^l_k-\frac{\mu^2}{4} R_{ij}R^{ij}\right]~,\nn
	\eea
where $R_{ij}$ is the Ricci tensor associated with $g_{ij}$, $\epsilon^{ijk}$ is the Levi-Civita symbol and	
	\be 
		C^{ij} = \epsilon^{ikl}\nabla_k\left(R^j_l-\frac{1}{4}R \delta^j_l \right)
	\ee
is the Cotton tensor.	

An interesting connection between the Ashtekar formulation of relativity and HL gravity is pointed out in \cite{MagueijoHL}, where it is shown that HL gravity can emerge as a specific case of the former, when in the presence of a fermion aether which spontaneously breaks the 4-dimensional diffeomorphism invariance.\\

\underline{IR spectrum}\\

We now study the spectrum of the original version of HL gravity in the IR region, particularly the scalar graviton, in order to see whether or not it is compatible with GR.

Putting together~(\ref{SK}) and ~(\ref{SpV}), we have the full action for the original version of HL gravity (or projectable version of HL gravity with detailed balance). In the IR limit, dimension 4 and 6 operators from the potential~(\ref{SpV}) can be consistently neglected as, in this limit, dimension 2 operators are dominant and therefore the theory flows towards $z=1$. Additionally, in order to be able to compare the resulting IR theory with GR, it is useful to rescale the time coordinate $t$ as
	\be
		x^0 = c' t~,~~ \mbox{with}~~ c' = \frac{\kappa^2\mu}{4}\sqrt{\frac{\Lambda_W}{(1-3\lambda)}}~,
	\ee
where $c'$ ($[c']=2$) is the emergent speed of light and $x^0$ has the same dimension as the time coordinate in isotropic models ($z=1$), {\it i.e.} $[x^0]=-1$.	
We can then write the IR version of the present model as
	\be\label{SpIR}
		S^{proj}_{IR} = M_P^2 \int dx^0 d^3x \sqrt{g} N \left[K_{ij} K^{ij} -\lambda K^2 +R-2\Lambda \right]~,
	\ee
where
	\be 
		\Lambda =\frac{3}{2}\Lambda_W ~~~\mbox{and}~~~ M_P^2 = \frac{1}{16\pi G_N}~~\mbox{with}~~ G_N = \frac{\kappa^2}{32\pi c'}~,
	\ee
which only differs from GR written in the ADM formalism because, in our case, $\lambda$ is not constrained to be equal to 1.

To better understand how the theory works in the IR region, let us expand the ADM fields $g_{ij}, N_i, N$ in terms of small fluctuations around the flat background (with $\Lambda =0$), and look for the spectrum of the theory,
	\be\label{fluct} 
		g_{ij} = \delta_{ij} + h_{ij},~~~~ N_i = n_i~~\mbox{and}~~N=1+n~,
	\ee
where the tensor and vector fluctuations $h_{ij}$ and $n_i$ can be further decomposed into different spin components
	\bea\label{spindec}
		h_{ij} &=& H_{ij}+ (\partial_i W_j+ \partial_j W_i ) + \left(\partial_{i}\partial_{j}-\frac{\delta_{ij}}{3}\partial^2\right)B +\frac{\delta_{ij}}{3}h~,\\
		n_i &=& n_i^T + \partial_i \rho~,
	\eea
with $H_{ij}$ being a transverse-traceless tensor ($\partial^i H_{ij} = 0$ and $H_{ii} =0$), $W_i$ and $n_i^T$ are transverse vectors ($\partial^i W_i = \partial^i n_i^T =0$); and $B,~h$ and $\rho$ are scalar fields. Moreover, because the flat space metric, from eq.(\ref{fluct}), is $\delta_{ij}$, we can lower all the spatial indices, for simplicity, i.e. $h^{ij}\to h_{ij}$.

At this point, it is useful to make use of the gauge freedom in~(\ref{ADMtransf}) to eliminate some non-propagating degrees of freedom from the theory. A natural choice, which fixes most of the gauge symmetry, is made by taking
	\be\label{gaugechoice}
		B = 0 ~~~\mbox{and}~~~ W_i=0~,
	\ee	
and leads to the following quadratic action	
	\bea\label{SpIR2}
		S_{IR}^{proj(2)} &=& M_P^2 \int dt d^3 x \left[ \frac{1}{4}H_{ij}(\partial^2-\partial_t^2)H_{ij} -\frac{1}{2}n_i^T \partial^2 n_i^T-(\lambda-1)\rho (\partial^2)^2\rho\right.\nn\\
		&&+\left.\frac{(3\lambda-1)}{12}h \partial_t^2h -\frac{1}{18}h \partial^2h+\frac{(3\lambda-1)}{3}\rho \partial^2 \dot{h} \right]~,
	\eea
where $\partial^2=\partial_i\partial_i$

All dependence on $n(t)$ disappears in this expansion when considering integration by parts and $\partial_i n(t) =0$. Finally, to arrive at an action containing only propagating degrees of freedom, we make use of the momentum constraint associated with the shift vector. By varying the action with respect to the auxiliary fields ($\rho$ and $n_i^T$), we find the following ``equations of motion'' or constraints:
	\bea\label{pconst}
		\partial^2 \rho &=& \frac{X}{6} \dot{h}~~~\mbox{with}~~~X=\frac{3\lambda-1}{\lambda-1}~,\\
		\partial^2 n_i^T &=& 0~,
	\eea
Then, we substitute~(\ref{pconst}) back into the action~(\ref{SpIR2}) to find
	\be\label{SpIR2f}
		S_{IR}^{proj(2)} = M_P^2 \int dt d^3 x \left[ \frac{1}{4}H_{ij}(\partial^2-\partial_t^2)H_{ij} +\frac{1}{18}h\left(-\partial^2-\frac{3\lambda-1}{\lambda-1}\partial_t^2\right)h \right]~,
	\ee
which now depends only on the spin-2 tensor field $H_{ij}$ and the scalar graviton $h$, the only propagating degrees of freedom of the theory.

From~(\ref{SpIR2f}) we note that the spin-2 degree of freedom has the usual dispersion relation in the IR
	\be\label{Hdr}
		\omega^2 = \vec{p}^2.
	\ee
On the other hand, the scalar graviton does not seem to be well-behaved, as we describe below. In order to guarantee the stability of the theory we might expect that
	\be\label{stcond}
		X=\frac{3\lambda-1}{\lambda-1} >0~.
	\ee
This implies that either $\lambda>1$ or $\lambda<1/3$. However, if we write the dispersion relation for $h$, we obtain
		\be\label{hdrp}
			\omega^2 = -X^{-1}\vec{p}^2~.
		\ee
As a result, we either have a ``stable'' theory with $h$ being a ghost field (when $X>0$) or the theory is unstable ($X<0$). In both cases, the projectable version of HL gravity presents a problematic IR limit due to the extra degree of freedom.

Another problem that emerges in this region is the strong coupling of the scalar graviton~\cite{StrongCpHL1,StrongCpHL2}. Ideally, when trying to recover GR in the IR region, we would expect that as $\lambda\to 1$, the extra degree of freedom $h$ decouples from the theory. However, what happens is exactly the opposite, as $\lambda\to 1$, $h$ becomes strongly coupled at a low energy scale, compromising the phenomenological viability of the present version.

In the subsequent sections, we present alternative versions of HL gravity which aim to solve the IR limit problems of the original proposal.

\subsubsection{Non-projectable version} 

In the non-projectable version of HL gravity, the lapse function is allowed to be a function of space and time $N=N(t,x^i)$, and the detailed balance condition is not imposed. Therefore, this version is much more general than the original. In such a case, a new term appears which is invariant under the foliation-preserving diffeomorphisms and, consequently, must be included in all possible forms in the potential of the theory:
	\be\label{aterm}
		a_i = \partial_i \ln N~.
	\ee
In order to avoid confusion with other works, we consider the non-projectable version of HL gravity as the one in which, in addition to having $N=N(t,x^i)$, terms constructed from~(\ref{aterm}) are also taken into account. This version of HL gravity is sometimes known as the ``healthy extension''~\cite{non-proj}.

The kinetic term, as mentioned earlier, is given by~(\ref{SK}) with the lapse function being a function not only of time, but also space.

The potential term, on the other hand, allows for many more coupling constants than in~(\ref{SpV}), since we do not impose the detailed balance condition. Furthermore, even more coupling constants appear as we include all the new possible terms up to dimension 6 containing the new term~(\ref{aterm}). For simplicity, we present here the potential term for the non-projectable version of HL gravity, contributing to a quadratic action only,  
	\bea\label{SnpV}
		S^{np}_V &=&  \int dt d^3 x \sqrt{g} N \left\{ \xi R+\alpha a_i a^i+ f_1 R_{ij}R^{ij} + f_2 R^2 + f_3 R \nabla_i a^i + f_4 a_i \Delta a^i\right.\nonumber\\
		&&\left.+ s_1 (\nabla_i R_{jk})^2 + s_2 (\nabla_i R)^2 + s_3 (\Delta R \nabla_i a^i) + s_4 (a_i \Delta^2 a^i) \right\}~,
	\eea
where the coupling constants have the following dimensions: $[\xi]=[\alpha]=4$, $[f_i]=2$ and $[s_i]=0$. Consequently, the operators of dimension 6, those related to the $s_i$ couplings, dominate in the UV where $z=3$.\\

\underline{IR spectrum}\\

Let us now investigate what happens in the IR region for the non-projectable HL gravity~\cite{non-proj} described by~(\ref{SK}) and~(\ref{SnpV}).

Rescaling the time component, as in the previous section, so that the rescaled time variable $x^0$ has dimension -1 as in GR, we find that the dominant contribution in the IR is given by
	\be\label{SnpIR}
		S^{np}_{IR} = S^{proj}_{IR} + M_P^2 \int d x^0 d^3 x \sqrt{g}N \alpha a_i a^i~,
	\ee  
where $S^{proj}_{IR}$ is the IR action for the projectable case given by~(\ref{SpIR}). 

Expanding now the action above up to quadratic order in the ADM field fluctuations as in~(\ref{fluct}), and using the spin decomposition given by~(\ref{spindec}) with the gauge choice~(\ref{gaugechoice}), we find
	\bea\label{SnpIR2}
		S_{IR}^{np(2)} &=& M_P^2 \int dt d^3 x \left[ \frac{1}{4}H_{ij}(\partial^2-\partial_t^2)H_{ij} -\frac{1}{2}n_i^T \partial^2 n_i^T-(\lambda-1)\rho (\partial^2)^2\rho\right.\\
		&&+\left.\frac{(3\lambda-1)}{12}h \partial_t^2h -\frac{1}{18}h \partial^2h-\alpha n \partial^2 n+\frac{(3\lambda-1)}{3}\rho \partial^2 \dot{h}- \frac{2}{3}n \partial^2 h \right]~,\nonumber
	\eea
To remove the auxiliary fields of the action above, we use the momentum and Hamiltonian constraints, which consist in substituting back into the action the linearised ``equations of motion'' for the lapse function and shift vector. Since the terms containing the shift vector were not modified when moving from the projectable version to the non-projectable one, the constraints associated with $n_i^T$ and $\rho$ are the same as~(\ref{pconst}). The new terms in the action~(\ref{SnpIR}) compare to that of the projectable case are all related to the lapse function. Varying~(\ref{SnpIR}) with respect to $n$, we find the following constraint:
	\be\label{nconst}
		-\alpha\partial^2 n=\frac{1}{3} h~.
  \ee 

Finally, substituting the constraints~(\ref{pconst}) and (\ref{nconst}) into the action~(\ref{SnpIR}), we have
		\be\label{SnpIR2f}
			S^{np(2)}_{IR}  = M_P^2\int dt d^3x \left\{ \frac{1}{4} H_{ij} \left( \partial^2- \partial_t^2 \right) H_{ij}+ \frac{1}{18} h \left[ \left(\frac{2-\alpha}{\alpha}\right)\partial^2-X\partial_t^2\right] h\right\} \nonumber~.
		\ee
The behaviour of $H_{ij}$ in the IR is the same as in the projectable case~(\ref{Hdr}); that is, we do not need to worry about it. In addition, when considering the scalar graviton, we see that, as in the projectable case, one needs $X>0$ in order to have a stable theory. However, as can be seen from the spatial derivative term in~(\ref{SnpIR2f}), the scalar graviton behaves differently due to the addition of new terms which depend on the lapse function. The dispersion relation for $h$ is now given by
	\be\label{hdrnp}
		\omega^2 = \frac{2-\alpha}{\alpha X}  \vec{p}^2~.
	\ee 
In this way, together with the condition $X>0$, the dispersion relation~(\ref{hdrnp}) tells us that the scalar graviton in the non-projectable version of HL gravity is well-behaved if either one of the conditions below is satisfied:
	\bea
		&&\lambda<1/3~~\mbox{and}~~0<\alpha<2~~~\mbox{or}\\
		&&\lambda>1~~\mbox{and}~~0<\alpha<2~.\nn
	\eea 
Finally, we mention here another feature that makes this version more viable than the original one. By requiring that the scale suppressing the higher-order operators is low enough, the non-projectable version of HL gravity, in contrast with the original~\cite{StrongCpHL1,StrongCpHL2}, is free from strong coupling when $\lambda\to 1$~\cite{StrongCnpHL}.

Before presenting the covariant version of HL gravity, we briefly mention an interesting relation between the non-projectable version of HL gravity and Einstein-aether theory. It has been shown in~\cite{Einstein-aether} that the Einstein-aether theory, described by GR coupled to a unit timelike tensor field, reduces to the IR limit of non-projectable HL gravity when the timelike vector is taken to be hypersurface orthogonal. Thus, a number of results obtained when considering Einstein-aether theory can be shown to also be valid in the case of the non-projectable version of HL gravity, in particular, those related to spherically symmetric solutions~\cite{Einstein-aether}.

\subsubsection{Covariant version}

The covariant version of HL gravity~\cite{covariant} is the only version studied here in which the extra degree of freedom is eliminated from the theory. Thus, in this version, only two degrees of freedom, as in GR, propagate. This is achieved by extending the symmetries of the model with the introduction of a new gauge symmetry. In this way, the resulting theory contains the same number of symmetries as in GR, {\it i.e.} $d+1$ per space-time point. In the following paragraphs we briefly describe some important points behind such an idea.\footnote{For a detailed explanation about the construction of the action, see the original paper~\cite{covariant}.}

The possibility of extending the symmetries of HL gravity as a way to eliminate the scalar graviton was first proposed by Horava in~\cite{membranes}, when it was realised that at $\lambda =1$ in the linearised approximation around flat space-time, the projectable version of HL presented an additional symmetry, acting only in the shift vector as $\delta_\theta N_i = \partial_i \theta$~ with $\theta=\theta(x_i)$ ($\delta_\theta N =0$ and $\delta_\theta g_{ij}=0$). It was then noticed that this symmetry could be promoted to a space-time dependent gauge symmetry of the full theory (not only of the linearised approximation), leading thus to a new version of HL with ``nonrelativistic general covariance''. The term ``nonrelativistic general covariance'' comes from the fact that, as already mentioned, this version has the same number of symmetries per space-time point as GR (thus, the ``general covariance''), but it is not relativistic. 

It was later shown in~\cite{covariant} that in order to impose on HL gravity the invariance under the new $U(1)$ symmetry, it is necessary to include two auxiliary fields $A$ and $\nu$, which under the $U(1)$ symmetry transform as
	\be\label{Anutrasnf}
		\delta_\theta A =  \dot{\theta} -N^i \nabla_i \theta~~~\mbox{and}~~~\delta_\theta \nu = \theta~,
	\ee
where $\theta$ has been promoted to be a function of space and time $\theta(t,x_i)$, and $N_i$ now transforms as
	\be
		\delta_\theta N_i = N\nabla_i \theta~,
	\ee
Thus, when the new fields are included together with the new symmetry, the action of the theory can be written as
	\be\label{covaction} 
		S^{cov} = \frac{2}{\kappa^2} \int d t d^3 x \sqrt{g} \left\{ N\left[K_{ij}K^{ij}-\lambda K^2-V+\nu\Theta^{ij}(2K_{ij}+\nabla_i\nabla_j \nu)\right]-A(R-2\Omega)\right\}~,
	\ee
where $\Omega$ is a new coupling constant with the same dimension as the cosmological constant $[\Omega]=[\Lambda]=2$, $[\nu]=1$, $[A]=4$, and
	\be
		\Theta^{ij} = R^{ij}-\frac{1}{2}R g^{ij}+\Omega g^{ij}~.
	\ee
Because neither $N$ nor $g_{ij}$ varies when the new $U(1)$ gauge transformation is performed, the potential $V$ is automatically invariant under the new symmetry. Therefore, $V$ can be either~(\ref{SpV}), if projectability and detailed balance are imposed,~(\ref{SnpV}), as in the non-projectable case, or it can assume other forms if additional constraints are imposed. For simplicity, let us assume that the potential in~(\ref{covaction}) is given by~(\ref{SpV}). Moreover, the variation of the action~(\ref{covaction}) with respect to the field $A$ generates a new constraint, $R-2\Omega=0$, which, as we see below, makes clear how the scalar graviton disappears.\\

\underline{IR spectrum}\\

As we did for the previous versions of HL gravity, we now investigate the spectrum of the covariant version in the IR region by expanding it around a flat background as in~(\ref{fluct}), which requires that $\Lambda=\Omega=0$.\footnote{In~\cite{covariant}, the absence of the scalar graviton in the IR of the theory was investigated for a more general case.} In the IR, the dominant term in the potential~(\ref{SpV}) is $\propto R$. However, as observed above, the constraint coming from the auxiliary field $A$ implies that $R=0$ when $\Omega=0$, which, at linear order, when taking into account our foliation-preserving diffeomorphisms gauge choice~(\ref{spindec}), with $B=0$ and $W_i=0$ is equivalent to
	\be\label{Aconst}
		\partial^2 h =0~.
	\ee
The expression~(\ref{Aconst}), together with the appropriate boundary conditions $h(\infty)=h(-\infty)=0$, imply that $h=0$. Therefore, the scalar graviton, which was problematic in the other versions of HL gravity, is not a propagating degree of freedom of the covariant version. 

Finally, it is worth emphasising that although the covariant version of HL gravity contains the same number of propagating degrees of freedom as GR, this feature is achieved by the inclusion of an extra symmetry in the theory. This new symmetry, however, does not come from any fundamental principle, and is only enforced as a way of eliminating an undesirable degree of freedom from the theory. Therefore, unless a relevant physical motivation for this new symmetry emerges, one must bear in mind that the covariant version of HL gravity is lacking an understanding of one of its basic elements.
\chapter{Lorentz-violating regulator gauge fields as the origin of dynamical flavour oscillations}\label{chap4}
\mbox{}

This chapter is based upon the paper~\cite{ALM1}.

\section{Introduction}

In this chapter, considering what has been discussed in chapter~\ref{chap1} about the need of an alternative mechanism to generate neutrino masses and oscillations, we propose a model where these features arise dynamically from the flavour-mixing interaction of two massless bare fermions with an Abelian gauge field, which has a LIV propagator.

Lorentz symmetry violation in our model comes from the introduction of higher-order space derivatives, which are suppressed by a large mass scale $M$. As shown in \cite{JA1,JA2}, such a mass scale allows the dynamical generation of fermion masses, which can be obtained when using the Schwinger-Dyson approach, discussed in section \ref{sec:SD}. Furthermore, the inclusion of the operator associated with the mass scale $M$ leads to finite gap equations, regulating the model. Other studies using a similar model were done in \cite{AM1} regarding the generation of fermion mass hierarchies.

An important point is related to the structure of the dynamical fermion mass \cite{JA1,JA2}
	\be\label{mdyn}
		m_{dyn}\simeq M \, \exp(-a/e^2)~, 
	\ee
where $a$ and $e$ are a positive constant and the coupling constant, respectively. The non-analytical form of (\ref{mdyn}) is well-known in the studies of magnetic catalysis \cite{ExtMagLad1,mag2,ExtMagLad2,mag4,mag5,mag6} and, as shown in section~\ref{sec:examples}, can only be derived from a non-perturbative approach, such as the Schwinger-Dyson one, used in \cite{JA1,JA2} and here. From~(\ref{mdyn}), we observe that it is possible to take the simultaneous limits
	\be\label{melim}
		M\to\infty \quad {\rm and} \quad e \to 0 ~,
	\ee
in such a way that the dynamical mass (\ref{mdyn}) remains finite, corresponding to a physical fermion mass.

When the limit (\ref{melim}) is taken, the non-physical gauge field decouples from the theory; hence, the gauge dependence of the
dynamical mass is avoided (although this problem can be understood perturbatively in the framework of the pinch technique~\cite{PT}, as explained in \cite{NM}). We give emphasis here to an essential feature of the mechanism described in this chapter: although LIV operators are suppressed by a large mass scale, corresponding to negligible effects at the classical level, quantum corrections completely change this picture, leading to finite effects. In this case, the finite effects are the dynamical generation of fermion masses and oscillations, which are present even after setting the LIV-suppressing mass scale $M$ to infinity. Note that the order of the steps taken is crucial: quantization is done for finite mass $M$ and coupling $e$, and only after that the simultaneous limits (\ref{melim}) are taken.

At this point, we note that the use of LIV operators as UV regulators of quantum field theories has also been considered in \cite{Visser}, but from a rather different perspective. Here, we aim at discussing the dynamically generated mass for fermions and the induced oscillations among different fermion flavours, using the coupling of the fermions to a LIV regulator gauge field.

The dynamical generation of flavour oscillations in the context of Lifshitz theories was studied in \cite{Anselmi5,Anselmi6}. A detailed analysis of this mechanism for two Lifshitz fermions coupled through a renormalisable four-fermion interaction can be found in \cite{ABH}.

This chapter is organised in the following way. Next, in section \ref{sec:2}, we introduce the model and consider the corresponding Schwinger-Dyson equations for the fermion propagators which must be satisfied by the solutions, {\it i.e.} the dynamical masses. Taking into account the constraints derived in the previous section, we calculate the dynamical masses in the relevant cases in section \ref{sec:3}. In section \ref{sec:4} we discuss the recovery of Lorentz invariance when taking the limit (\ref{melim}), for which the LIV gauge field decouples from fermions, and we demonstrate that relativistic dispersion relations for fermions are indeed recovered. Finally, in section \ref{sec:5} we present extensions of the Dirac fermion case to Majorana fermions, as appropriate for neutrinos either in the standard model or in seesaw-type extensions thereof, involving sterile neutrinos. Technical aspects of our work are given in Appendix A and Appendix B.

\section{Dynamical fermion mass matrix \label{sec:2}}

\subsection{The field theory model}

The LIV model we consider here, which is an extension of the model studied in~\cite{JA1,JA2}, is given by
	\be\label{Lag}
		\mathcal{L}= -\frac{1}{4} F_{\mu\nu}(1-\frac{\Delta}{M^2})F^{\mu\nu}+\overline{\Psi}(i \slashed{\partial}-\tau \slashed{A})\Psi,
	\ee
where $F_{\mu\nu}$ is the Abelian field strength for the gauge field $A^\mu$, and $\Delta=-\partial_i\partial^i$ is the Laplacian. The Lagrangian above is not invariant under Lorentz transformations due to the presence, in the gauge sector, of the LIV derivative operator $\Delta$ suppressed by the mass scale $M$, which can be thought of as the Planck mass, and eventually will be set to infinity. $\Psi$ is a massless fermion doublet, containing two fermions with flavours $l_1$ and $l_2$,
	\be
		\Psi = \begin{pmatrix} 
					 \psi_{l_1} \\
           \psi_{l_2}
           \end{pmatrix}~,
	\ee
and $\tau$ is the flavour mixing matrix written in terms of the gauge couplings $(e_1,e_2,\epsilon)$ as 
	\be\label{tau}
		\tau = \begin{pmatrix}e_1 & -i\epsilon \\i\epsilon & e_2\end{pmatrix}=
		\frac{e_1 + e_2}{2}{\bf1} +\frac{e_1 - e_2}{2} \sigma_3 +\epsilon \sigma_2~,
	\ee
where $\sigma_i$ are the usual Pauli matrices, and ${\bf 1}$ is the $2\times 2$ identity matrix. The fermions $\psi_{l_1}$ and $\psi_{l_2}$ in (\ref{Lag}) represent Dirac fields, but the structure of the gap equations that will be derived below remains the same whether the fermions are Dirac or Majorana, hence the corresponding dynamical masses are independent of the nature of fermions. 

We mention here that the Lagrangian (\ref{Lag}) can be derived from a stringy space-time foam model, as shown in \cite{NM}. Also, such a space-time foam model was already used to study decoherence in flavour oscillations, both in flat space-time and in a Friedman-Robertson-Walker metric \cite{AFMP1,AFMP2}.

The bare propagator for the gauge field is 
	\be\label{gaugeprop}
		D_{\mu\nu} = -\frac{i}{1 +\vec{p}^2/M^2}\left(\frac{\eta_{\mu\nu}}{\omega^2-\vec{p}^2}+\zeta \frac{p_\mu p_\nu}{(\omega^2-\vec{p}^2		 )^2}\right)~,
	\ee
where $\zeta$ is a gauge fixing parameter. Although $\zeta$ appears in the final expression for the dynamical masses, it does not play 
a role in the simultaneous limits 
	\begin{equation}\label{metlim}
		M \to \infty \quad {\rm and} \quad e_1,    e_2,   \epsilon \to 0~, 
	\end{equation}
that leave the dynamical masses finite, as we discuss further on. 
 
We note that the flavour mixing interaction $\ol\Psi\tau\slashed{A}\Psi$, in addition to generating fermion masses, can be associated with the generation of a gauge boson mass, thus playing the role of an alternative to the Higgs mechanism. This possibility was explained in \cite{dynmassA1,dynmassA2}, while in \cite{AM2} it was extended to a LIV model. In our case, however, we disregard the possibility of dynamical mass generation for the gauge boson, since, as we shall demonstrate below, the flavour mixing coupling $\epsilon$ vanishes necessarily for consistency of the model when dynamical generation of fermion oscillations take place. 

From~(\ref{Lag}), we have that the fermion propagator is the usual, Lorentz-symmetric, one: $S=i/\slashed{p}$, where $p_\mu=(\omega,\vec p)$.  Let us now assume the dynamical generation of the following fermion mass matrix
	\be\label{mm}
		{\bf M}=\begin{pmatrix} m_1 & \mu \\ \mu & m_2\end{pmatrix}
		=\frac{m_1 + m_2}{2}{\bf1} + \frac{m_1 - m_2}{2} \sigma_3 + \mu \sigma_1 ~,
	\ee
which has as eigenvalues
	\be\label{eigenmasses}
		\lambda_{\pm} = \frac{m_1+m_2}{2} \pm \frac{\sqrt{(m_1-m_2)^2+4\mu^2}}{2}~.
	\ee
The generation of these dynamical masses $m_1,m_2,\mu$ by quantum corrections will be checked in the next sections, using the Schwinger-Dyson approach.
	
Taking into account the dynamically generated mass matrix~(\ref{mm}), but neglecting other quantum corrections, the dressed fermion propagator $G$, obtained by solving the equation
	\be 
		G(\slashed p-{\bf M})=i{\bf 1}~,
	\ee 
is then
	\bea\label{gfermi}
		G&=& i\frac{p^2 + \slashed{p}( m_1 + m_2) + m_1 m_2 -\mu^2}{(p^2 - m_1^2)(p^2 - m_2^2)-2 \mu^2(p^2 + m_1 m_2)+\mu^4}\\
		&&\times \left[(\slashed{p} - \frac{m_1 + m_2}{2}) {\bf1} + \frac{m_1 - m_2}{2} \sigma_3 +\mu \sigma_1 \right]~.\nonumber
	\eea

Based on our discussion on mixing and mass matrices in chapter \ref{chap1}, we note that because the mass matrix (\ref{mm}) contains, in general, non-vanishing off-diagonal elements, the flavour eigenstates $|\psi_\alpha \rangle $, with $\alpha=l_1, l_2$, are not the same as the mass eigenstates $|\psi_\pm \rangle$.\footnote{For convenience, since we are working with two fermions only, we define the mass states and eigenvalues as $\psi_\pm$ and $\lambda_\pm$ instead of $\psi_i$ and $m_i$, with $i=1,2$, as in chapter \ref{chap1}.} The mixing matrix relating mass and flavour states is then parametrised by a mixing angle $\theta$, as in eq.(\ref{mm2}):
	\begin{equation}\label{mixed}
		\begin{pmatrix} \psi_{l_1} \\ \psi_{l_2} \end{pmatrix} = \begin{pmatrix} {\rm cos}\theta \quad {\rm sin}\theta \\ -{\rm sin} \theta 		\quad {\rm cos}\theta \end{pmatrix}   
		\begin{pmatrix} \psi_+ \\ \psi_- \end{pmatrix}~.
	\end{equation}
Moreover, provided that the mixing angle $\theta$ is not trivial and that the energy eigenvalues $E_\pm = \sqrt{p^2 + \lambda_\pm^2}$\footnote{We shall check in section \ref{sec:4} that the relativistic dispersion relations for the fermions are indeed obtained in the Lorentz Invariant Limit (\ref{metlim}), after (finite) dynamical mass generation.} are different, neutrino oscillations take place.

Then, according to section \ref{sec:NOsc}, if at time $t=0$ a flavour neutrino $\psi_{\alpha}(t=0)$ is emitted, the probability of obtaining (under Hamiltonian evolution) a neutrino with a different flavour at a later time $t>0$,{\it i.e.} $\psi_{\beta}(t)$, is given by (\ref{nosc2}) 
	\begin{equation}\label{osc}
		{\mathcal P}_{\alpha\to \beta} (t) \approx{\rm sin}^2   2		\theta   {\rm sin}^2 \Big[\frac{(\lambda_{+}^2 - \lambda_{-}^2) L}{4 E}  \Big]~
	\end{equation}
and the survival probability ${\mathcal P}_{\alpha\to \alpha} = 1 - {\mathcal P}_{\alpha\to \beta}$, where $\lambda_{\pm}$ are the masses associated with the mass states $\psi_{\pm}$, given by (\ref{eigenmasses}). Therefore, we must stress that, as it can be seen in (\ref{osc}), non-trivial mixing, $\theta \ne 0$, is a necessary, but not sufficient, condition for oscillatory behaviour among flavours (different energy levels $E_+ \ne E_- $ ($\lambda_{+}^2 \ne \lambda_{-}^2$) are also required). Our aim is to identify the cases where masses and mixing are generated dynamically, as a result of the coupling of the fermions with the LIV gauge bosons.

\subsection{Schwinger-Dyson gap equations}

The Schwinger-Dyson equation for the fermion propagator, which has already been derived in section \ref{sec:SD}, is not modified by the presence of the LIV term in the Lagrangian (\ref{Lag}). Taking the simplest approximation, {\it i.e.} neglecting corrections to the wave functions, the vertices and the gauge propagator, the Schwinger-Dyson equation (\ref{SDE}), in matrix form, for our model reads  
	\begin{eqnarray}\label{SD}
		G^{-1}-S^{-1} = \int_p~ \tau\gamma^\mu~ D_{\mu\nu} ~G ~\tau\gamma^{\nu}~.
	\end{eqnarray}
In a Lorentz symmetric case, this loop integral would diverge, but the presence of the LIV term $\vec p^2/M^2$ in the denominator of the gauge propagator (\ref{gaugeprop}) makes it finite. The matrix equation (\ref{SD}), as shown in Appendix A, leads to the following four gap equations, which must be satisfied by the three masses $m_1,m_2,\mu$,
	\bea\label{4equs}
		\frac{m_1}{4+\zeta} &=& (e_1^2 m_1 + \epsilon^2 m_2) I_1+ (\mu^2-m_1 m_2)(e_1^2 m_2 +\epsilon^2 m_1) I_2\\
		\frac{m_2}{4+\zeta} &=& (e_2^2 m_2 + \epsilon^2 m_1) I_1 + (\mu^2-m_1 m_2)(e_2^2 m_1 +\epsilon^2 m_2) I_2\nn\\
		\frac{\mu}{4+\zeta} &=& \mu(e_1 e_2-\epsilon^2)[I_1 - (\mu^2-m_1 m_2)I_2]\nn\\
		0 &=&\epsilon(e_1 m_1 + e_2 m_2) I_1+\epsilon(\mu^2-m_1 m_2) (e_1 m_2 +e_2 m_1) I_2~,\nonumber
	\eea
where 
	\bea\label{I1I2J}
		I_1&=& \frac{J(A_+^2)-J(A_-^2)}{A_+^2-A_-^2}  \\
		I_2&=& \frac{1}{A_+^2-A_-^2} \left[\frac{J(A_+^2)}{A_+^2}-\frac{J(A_-^2)}{A_-^2}\right]\nonumber~,
	\eea
and 
	\bea\label{Apmdef}
		J(A_\pm^2)&=&\frac{1}{4 \pi^3}\int_0^{\infty}{dp} \frac{\vec{p}^2}{1 + \vec{p}^2/M^2} \int_{-\infty}^{\infty}{d\omega} 
		\left( \frac{1}{\omega^2+\vec{p}^2}-\frac{1}{\omega^2+\vec{p}^2+A_{\pm}^2}\right)\nn\\
		A_\pm^2&=&\frac{m_1^2+m_2^2+2\mu^2}{2} \pm \frac{\sqrt{(m_1^2-m_2^2)^2+4\mu^2(m_1+m_2)^2}}{2}~.
	\eea
Performing the integrals in $J(A_\pm^2)$ over frequency and momentum, and expanding the results for $M>>m_1,m_2,\mu$, we find
	\bea\label{I1I2}
		I_1&\simeq&\frac{1}{16\pi^2} \frac{1}{A_+^2 - A_-^2} \left[A_-^2\ln\left(\frac{A_{-}^2}{M^2}\right)- A_+^2\ln\left(\frac{A_{+}^2}{		M^2}\right)\right]\nn\\
		I_2&\simeq& \frac{1}{16\pi^2}  \frac{1}{A_+^2 - A_-^2} \ln{\left(\frac{A_{-}^2}{A_{+}^2} \right)}~.
	\eea 
The four equations (\ref{4equs}) must be satisfied by only three unknowns $m_1,m_2,\mu$. Consequently, they do not have obvious solutions. In what follows we investigate different solutions, paying attention to the fact that the solutions allowing for the generation of flavour oscillations must have $\mu\ne0$.

\subsection{Constraints}

By manipulating the first two equations in (\ref{4equs}) and assuming that $e_1^2e_2^2\ne\epsilon^4$, we find the following relations
	\bea\label{I1I2bis}
		I_1&=&\frac{1}{4+\zeta}~\frac{e_2^2m_1^2-e_1^2m_2^2}{(e_1^2e_2^2-\epsilon^4)(m_1^2-m_2^2)}\\
		(\mu^2-m_1m_2)I_2&=&\frac{1}{4+\zeta}\frac{m_1m_2(e_1^2-e_2^2)+\epsilon^2(m_2^2-m_1^2)}{(e_1^2e_2^2-\epsilon^4)(m_1^2-m_2^2)}~.		\nonumber
	\eea
Although the denominators above vanish when $m_1^2=m_2^2$, no singularity arises, because, as we will see in the next section, when $m_1^2=m_2^2$ we necessarily have that $e_1=e_2$, causing the numerator to vanish as well.

Similarly, the third and forth equations lead to the following constraints
	\bea\label{constraints}
		\mu(m_1+m_2)(e_2m_1+e_1m_2)(e_1-e_2)&=&0\\
		\epsilon(e_2m_1+e_1m_2)&=&0~.\nonumber
	\eea
In the next section, we make use of these constraints to look for the different solutions to our problem.

\section{Solutions of the gap equations - dynamical fermion masses and mixing \label{sec:3}}

We investigate now the different solutions to the gap equations (\ref{4equs}). The simplest solution to these equations is given by $m_1=m_2=\mu=0$, implying that no fermion mass is generated dynamically, and, consequently, such a solution is of no interest to us here.  
In what follows, starting from the constraints (\ref{constraints}), we focus on the cases where fermion masses are generated.

\subsection{The case $\epsilon=0$, $\mu = 0$}

This case is a straightforward generalisation of the original model presented in \cite{JA1,JA2}, which involved one fermion, to the two 
fermion-flavour case with no mixing at all. It can be divided into two situations: i) $m_1 \ne 0$ and $m_2 \ne 0$ and  ii) $m_1=0$ or $m_2=0$. As we shall discuss in section \ref{sec:5}, these may be relevant for Majorana neutrinos in the standard model and extensions thereof, involving right-handed neutrinos. 

\subsubsection{i) $m_1\ne 0$ and $m_2 \ne 0$ \label{nusm}}

In this first situation, the mass matrix (\ref{mm}) is diagonal; therefore, its eigenvalues (\ref{eigenmasses}) are the diagonal elements $m_1$ and $m_2$. Because $\epsilon=0$ and $\mu=0$, the constraints (\ref{constraints}) are trivially satisfied, and the last two equations in (\ref{4equs}) vanish. We are then left with the first two equations in (\ref{4equs}). Although, at first sight, they seem to mix $m_1$ and $m_2$, these two equations are completely independent of each other. This can be verified by taking both equations (with $\epsilon=0$ and $\mu=0$) in (\ref{4equs}) and substituting the definitions (\ref{I1I2J}) for $I_1$ and $I_2$ in terms of the integrals $J$, leading to
	\bea\label{2equs}
		\frac{m_1}{4+\zeta} &=& e_1^2 m_1[ I_1 - m_2^2 I_2]= e_1^2 \frac{J(m_1^2)}{m_1}\\
		\frac{m_2}{4+\zeta} &=& e_2^2 m_2[ I_1 - m_1^2 I_2]=e_2^2 \frac{J(m_2^2)}{m_2}\nn~.
	\eea
These equations represent two decoupled Schwinger-Dyson equations for the fermion propagators of two decoupled fermions interacting with a LIV gauge field. Solving $J(m_i^2)$ for $M\gg m_i$, we have
	\be
		J(m_i^2) \approx -\frac{1}{16\pi^2} m_i^2 \ln\left(\frac{m_i^2}{M^2}\right)~,
	\ee
so that (\ref{2equs}) becomes
	\be
		\frac{m_i}{4+\zeta} = -\frac{e_i^2}{16\pi^2} m_i \ln\left(\frac{m_i^2}{M^2}\right)~.
	\ee
Finally, because $m_i\neq 0$, we obtain the same solution as in \cite{JA1,JA2}, {\it i.e.}
	\be\label{2mass}
		m_i=M\exp\left(\frac{-8\pi^2}{(4+\zeta)e_i^2}\right)~~~,~i=1,2~.
	\ee 
In conclusion, the dynamically generated mass matrix is diagonal with masses $m_i$ among the two flavours. 
Hence, as expected, no mixing (\ref{mixed}) or oscillations (\ref{osc}) take place between the flavours $\psi_{\alpha}$, $\alpha=l_1,l_2,$. 

\subsubsection{ ii) $m_1 = 0$ or $m_2=0$}  \label{sec:3nu}

In this case, it is clear from (\ref{2equs}) that there is also a consistent solution, with either $m_1 =0$ 
with $m_2 \ne 0$ or $m_2 =0$ and $m_1 \ne 0$ (since $J(m_i^2)/m_i \to 0$ as $m_i\to 0$), and that the two cases are completely symmetric. For reasons that will become clear in our discussion on Majorana neutrinos in section \ref{sec:5}, we concentrate here in the former case, i.e. $m_1 =0$. 

When $m_1 =0$, both sides of the first equation in (\ref{2equs}) vanish, not bringing any important information. The second equation, on the other hand, can be solved as in the previous section, yielding 
	\be\label{apm}
		 m_2 \simeq  M ~\exp\left(-\frac{8  \pi^2 }{(4 + \zeta)  e_2^2}\right) ~.
	\ee
	
We emphasise that in this case the mass eigenvalues are $m_1 = 0$ and $m_2 \ne 0$ given by (\ref{apm}), the mixing angle $\theta$ vanishes, and thus there are no oscillations between the two flavours.

\subsection{The case $m_1=m_2=0$ and $\mu\ne0$}\label{sec:case2}

In this case, where the mass matrix (\ref{mm}) presents vanishing diagonal elements, the eigen masses are
	\be
		\lambda_\pm=\pm\mu~,
	\ee
and the mass eigenstates are 
	\be\label{eigenstates}
		\psi_\pm=\frac{1}{\sqrt2}(\psi_{l_2}\pm\psi_{l_1})~,
	\ee
such that the mixing angle (\ref{mixed}) is $\theta =-\pi/4$, in our conventions.

This case does not include a mass hierarchy, hence there are no oscillations among the fermion flavours either, since, according to (\ref{osc}), the oscillation probability $\mathcal{P}_{l_1 \to l_2}$ vanishes as a result of $\lambda_+^2=\lambda_-^2=\mu^2$. 

Among the four gap equations (\ref{4equs}), only the third is not trivial, leading to
	\bea
		\frac{1}{4+\zeta}=(e_1e_2-\epsilon^2)(I_1-\mu^2I_2)~.
	\eea
Since $A_\pm^2=\mu^2$, the expressions (\ref{I1I2}) give
	\bea
		I_1&\simeq&\frac{-1}{16\pi^2}\left(1+\ln\left(\frac{\mu^2}{M^2}\right)\right)\\
		I_2&\simeq&\frac{-1}{16\pi^2}\frac{1}{\mu^2}~,
	\eea
and we obtain
	\be
		\ln\left(\frac{\mu^2}{M^2}\right)=\frac{-16\pi^2}{(4+\zeta)(e_1e_2-\epsilon^2)}~.
	\ee
We note that this expression has a meaning only for $e_1e_2>\epsilon^2$, otherwise $\mu^2>M^2$. Thus, with $e_1e_2>\epsilon^2$, we finally obtain that $\mu$ is dynamically generated 
	\be
		\mu\simeq M\exp\left(\frac{-8\pi^2}{(4+\zeta)(e_1e_2-\epsilon^2)}\right)~.
	\ee

\subsection{The case $e_2m_1+e_1m_2=0$ and $m_1^2\ne m_2^2$}

In this situation, the first equation (\ref{I1I2bis}) leads to $I_1=0$. The expression (\ref{I1I2}) for $I_1$ leads to
	\be
		A_+^2=A_-^2=\exp(-1)M^2~.
	\ee
This solution, however, must be disregarded, otherwise the dynamical masses would then be necessarily of the order $M$, which will eventually be taken to infinity.

\subsection{The case $m_1=-m_2\ne0$ \label{sec:3.4}}

It can be seen from eqs.(\ref{4equs}) that when $m_1=-m_2\equiv m$, we necessarily have $e_1=e_2$, such that both constraints (\ref{constraints}) are satisfied. Additionally, eqs.(\ref{4equs}) are equivalent to
	\be
		\frac{1}{4+\zeta}=(e^2-\epsilon^2)[I_1-(\mu^2+m^2)I_2]~,
	\ee
and $A_\pm^2=m^2+\mu^2$, such that we find
	\be
		m^2+\mu^2=M^2\exp\left(\frac{-16\pi^2}{(4+\zeta)(e^2-\epsilon^2)}\right)~,
	\ee
which has a meaning only if $e^2>\epsilon^2$. This condition for the coupling constants allows one to take the limit $\epsilon \to 0$ without affecting the mass eigenvalues or mixing angles (see below). This is important because, as already mentioned, if $\epsilon\neq 0$, a mass term for the vector boson may be dynamically generated~\cite{dynmassA1,dynmassA2,AM2}, thereby spoiling its nature as a regulator field.  

The eigen masses are 
	\be\label{lpm}
		\lambda_\pm=\pm\sqrt{m^2+\mu^2}~,
	\ee
and we stress here that we cannot determine $m$ and $\mu$ independently. The mass eigenstates, related to the eigen masses above, can be written as
	\be
		\psi_\pm=\frac{1}{N_\pm}\left(\psi_{l_1}+\frac{\mu}{m\pm\sqrt{m^2+\mu^2}}\psi_{l_2}\right)~,
	\ee
where
	\be
		N_\pm^2=\frac{2m^2+2\mu^2\pm2m\sqrt{m^2+\mu^2}}{2m^2+\mu^2\pm 2m\sqrt{m^2+\mu^2}}~,
	\ee
and the mixing angle $\theta$ (\ref{mixed}) is given by 
	\be
		\tan\theta=\frac{-\mu}{m+\sqrt{m^2+\mu^2}}~.
	\ee
In order to fix the mixing angle, one would need an additional ingredient, since the present model gives us $m^2+\mu^2$, but not $\mu$ alone.

Again, as in section \ref{sec:case2}, because there is no mass hierarchy, $\lambda_+^2=\lambda_-^2$, oscillations (\ref{osc}) among fermion flavours cannot take place.

\subsection{The case $m_1=m_2\ne0$: dynamical flavour oscillations} \label{sec:3.5}

For $m_1=m_2$, we find from eqs.(\ref{4equs}) that necessarily $e_1=e_2$, $\epsilon=0$ and $\mu^2=m^2$, satisfying therefore the constraints~(\ref{constraints}). In this case, we obtain 
	\be\label{I1final}
		\mu^2=m_1m_2=m^2~~~\mbox{and}~~~I_1=\frac{1}{(4+\zeta)e^2}~,
	\ee
where $e=e_1=e_2$. Using then the solution (\ref{I1I2}) for the integral $I_1$ with $A_-^2=0$ and $A_+^2=4m^2$, (\ref{I1final}) becomes
	\be
		-\frac{1}{16\pi^2}\ln\left(\frac{4m^2}{M^2}\right)=\frac{1}{(4+\zeta)e^2}~,
	\ee
thus the dynamical mass is 
	\be\label{dynmass}
		m=\frac{M}{2}\exp\left(-\frac{8\pi^2}{(4+\zeta)e^2}\right)~
	\ee
which, as expected, is not perturbative in $e$. 

In this situation, the mass eigenvalues are
	\be
		\lambda_+=2m=M\exp\left(-\frac{8\pi^2}{(4+\zeta)e^2}\right)~~,~~\lambda_-=0~,
	\ee
and the corresponding mass eigenstates are the same as the ones in eq.(\ref{eigenstates}). Finally, the mixing angle (\ref{mixed}) is $\theta = \mp  \pi/4$, depending on the sign of $\mu=\pm   m$, respectively.

In this case, one of the fermions is massless, and the other massive, with mass $2m$, {\it i.e.} twice the solution found in~(\ref{dynmass}). Therefore, there is a mass hierarchy ($\lambda_+^2-\lambda_-^2\neq 0$), and, since the mixing angle is non-trivial, oscillations (\ref{osc}) among the fermion flavours take place. 

We note that because of the constraints (\ref{constraints}), this is the only case in the present model (\ref{Lag}) where 
oscillations among fermion flavours are allowed. As we have seen above, the flavour-mixing gauge couplings $\epsilon$ must vanish, so one does not have to worry about dynamical generation of gauge boson masses, and thus the latter play the role of regulator fields.

\subsection{Energetics arguments}

Among the different possibilities to generate masses dynamically, it is natural to question the preference for the system to exhibit finite masses, rather than no dynamical mass at all. With this in mind, we focus here on an energetics argument to support the choice of non-vanishing dynamical masses~\cite{AM2}.

Our argument is based upon the Feynman-Hellmann theorem~\cite{feynman, feynman2}. This theorem states that if there is a ground state $|\Psi_\chi\rangle $ of a system described by a Hamiltonian $\widehat{H}$ that depends on a parameter $\chi$, then 
	\begin{equation}\label{fhth}
		\frac{\partial \mathcal{E}}{\partial \chi} = \langle \Psi_\chi | \frac{\partial \widehat{H}}{\partial \chi} | \Psi_\chi \rangle~,
	\end{equation} 
where $\mathcal{E}$ is the energy associated with the ground state $| \Psi_\chi \rangle$. In the present situation, let us choose the parameter $\chi=M^{-2}$, such that
	\begin{equation}\label{fhthours}
		\frac{\partial \mathcal{E}}{\partial \chi } = +\frac{1}{4}~ {_M }\langle 0 | \int d^4 x^E \left(F_{\mu\nu} \Delta F^{\mu\nu} 
		\right)_E | 0 \rangle_M ~, \quad \chi = M^{-2} ~.
	\end{equation}
where we used the fact that the Hamiltonian of the system can be identified with minus the effective Euclidean action, hence the index $E$. As the Lorentz-violating nature of the vacuum $|0\rangle_M$ suggests, in general, the non vanishing of the right-hand side, we have that the vacuum energy must depend on the mass scale M. Furthermore, using the cyclic Bianchi identity for the gauge bosons field strengths, 
	\begin{equation}\label{bianchi}
		\partial_{\left[\mu\right.} F_{\left. \nu\rho\right]} = 0~,
	\end{equation}
where $[ \dots ]$ denotes anti-symmetrisation of the appropriate indices, we obtain
	\be
		\frac{\partial \mathcal{E}}{\partial \chi } =- \frac{1}{4}~ {_M }\langle 0 | \int d^4 x^E \left(F_{\mu\nu}\partial_i[\partial^\mu F^{\nu i		}+\partial^\nu F^{i\mu}]\right)_E | 0 \rangle_M ~.
	\ee
If we now integrate by parts assuming that the fields decay away at space-time infinity, eq.(\ref{fhthours}) may be written as
	\be\label{fhthours2}
		\frac{\partial \mathcal{E}}{\partial \chi } = +\frac{1}{2} ~{_M} \langle 0 | \int d^4 x^E \left(\partial^\mu F_{\mu\nu} \partial_i F^{\nu 		i}\right)_E | 0 \rangle_M~.
	\ee
Writing the equations of motion for the vector fields from the Lagrangian (\ref{Lag}), neglecting the operator $\Delta/M^2$, we obtain:
	\be\label{squarecharge}
		\frac{\partial \mathcal{E}}{\partial \chi } = +\frac{1}{2} ~{ _M} \langle 0 | \int d^4 x^E \left((J^0)^2+\vec J\cdot\vec J -J_k\partial_0		F^{k0}\right)_E|0\rangle_M~,
	\ee
where the current is $J^{\mu} = \overline{\Psi} \gamma^\mu \tau \Psi$.

In the framework of the present LIV model, one might face a situation where non-trivial condensates of the covariant square of the stationary four-current $J^\mu$ are observed in the (rotationally invariant) vacuum. For such stationary currents, where $\partial_0 F^{k0}=0$, eq.(\ref{squarecharge}) becomes
	\begin{eqnarray}\label{monot}
		\frac{\partial \mathcal{E}}{\partial \chi } = \frac{1}{2}~_M\langle 0 | \int d^4 x^E \left( J^\mu J_\mu \right)_E |0\rangle_M \ge 0~.
	\end{eqnarray} 
Therefore, the vacuum energy $\mathcal{E}$ is a monotonically decreasing function of $M^2$, which tends to its minimum in the limit we are interested in, {\it i.e.} the Lorentz symmetric limit: $M\to\infty$.

The argument given above in favour of the stability of the Lorentz invariant limit (\ref{melim}) can be also used in favour of the dynamical fermion mass generation, as follows. In a finite $M < \infty$ situation, the gauge coupling $e$ is seen as an independent quantity from $M$, and thus, in view of (\ref{mdyn}), the LIV mass scale is proportional to the fermion mass $m > 0$ (absolute value if $m < 0$). In this sense, from (\ref{monot}), we have
	\be\label{mono2} 
		\frac{\partial   \mathcal{E}}{\partial   m} = \frac{\partial   M}{\partial m}   \frac{\partial \chi}{\partial M}  \frac{\partial \mathcal{E}}{		\partial \chi} = - \frac{1}{m  M^2}   _M\langle 0 | \int d^4 x^E \left( J^\mu J_\mu \right)_E |0\rangle_M \le 0~.
	\ee
Thus, the vacuum energy $\mathcal{E}$ for any finite value of $M$ is also a monotonically decreasing function of the fermion mass. In the Lorentz-symmetric limit (\ref{melim}), $\mathcal{E}$ exhibits a plateaux ($\partial \mathcal{E}/\partial m =0$), as far as its dependence on the finite $m > 0$ is concerned, but its value is lower than in the case where $m=0$.  

We must stress, however, that the arguments presented above rely on the formation of condensates for the covariant square of the current. Such a property is at present a conjecture, and its proof goes far beyond our considerations in this work.

\section{Lorentz symmetric limit \label{sec:4}}

Lorentz invariance can be finally recovered by taking the simultaneous limits
	\be\label{limit}
		M\to\infty~~~~\mbox{and}~~~~e_1,~e_2,~\epsilon\to0~,
	\ee
in such a way that the dynamical masses remain finite. Such a procedure is independent of the gauge parameter $\zeta$, and the resulting fermion mass can be set to any desired value. Furthermore, in the limit above, the gauge field decouples from fermions, and the only finite effect from Lorentz violation in the original model is the presence of finite dynamical masses for fermions. 

We now demonstrate the validity of this statement by showing that the fermion dispersion relations are relativistic in the limit (\ref{limit}). We focus here for concreteness on the solution which allows for oscillations, described in subsection \ref{sec:3.5}, with $\mu=+m$, but clearly the same conclusion holds for all the other solutions. Because possible one-loop infrared (IR) divergences may appear as a result of the fact that one of the fermions is massless, we consider $m_1=m_2=m$ and $m-\mu=m\delta$ with $\delta<<1$. As will be seen, however, after the limit (\ref{limit}) is taken, the fermion self energy will not depend on $\delta$, such that the limit $\delta\to 0$ will not introduce any IR divergence. 

We calculate in Appendix B the one-loop fermion self energy, using the Feynman gauge since the limit (\ref{limit}) is gauge independent. To lowest order in momentum, we obtain
	\be
		\Sigma=\begin{pmatrix} Z_{diag}^0 & Z_{off}^0 \\ Z_{off}^0  & Z_{diag}^0 \end{pmatrix}\omega\gamma^0
		-\begin{pmatrix} Z_{diag}^1 & Z_{off}^1 \\ Z_{off}^1  & Z_{diag}^1 \end{pmatrix}\vec p\cdot\vec\gamma-{\bf M}~,
	\ee
where $(\omega,\vec p)$ is the external 4-momentum and
	\bea
		Z_{diag}^0&=& \frac{e^2}{8\pi^2}\left(\frac{1}{4}-\frac{1}{2} \ln{2} + \frac{1}{2} \ln{\delta} + \ln\left(\frac{m}{M}\right)\right)\\
		Z_{diag}^1&=&\frac{e^2}{8\pi^2}\left(-\frac{1}{12}-\frac{1}{2} \ln{2} + \frac{1}{2} \ln{\delta} + \ln\left(\frac{m}{M}\right)\right		)\nn\\
		Z_{off}^0&=&Z_{off}^1=\frac{e^2}{16\pi^2}(\ln{2} - \ln{\delta})~.\nonumber
	\eea
As expected, because of Lorentz-symmetry violation, we have that $Z_{diag}^0\ne Z_{diag}^1$. However, since
	\be
		e^2\ln\left(\frac{m}{M}\right)=-2\pi^2~,
	\ee
the limit (\ref{limit}) leads to
	\be
		\Sigma~~\to~~-\frac{1}{4}(\omega\gamma^0-\vec p\cdot\vec\gamma){\bf 1}-{\bf M}_r~,
	\ee
where ${\bf M}_r$ is the corresponding ``renormalised'' mass matrix.	

Therefore, since time and space derivatives are dressed with the same corrections in the limit (\ref{limit}), the one-loop dispersion relations are relativistic. Such corrections can be absorbed in a fermion field redefinition, and we are left with two free relativistic fermion flavours oscillating.

\section{Extension to Majorana neutrinos \label{sec:5}}

So far we have considered Dirac fermions only. However, in order to present the above-described dynamical mass generation scenario as a viable alternative to standard seesaw mechanisms for neutrinos and explain neutrino oscillations as a dynamical phenomenon, we now extend our previous considerations to the case where the fermions are Majorana (as most likely is the case realised in nature). 

We do this in what follows by considering two separate cases. In the first case, fermions correspond to Majorana mass eigenstates obtained from  the left-handed flavour neutrino fields of the standard model. The second case, on the other hand, involves sterile right-handed neutrinos as in seesaw extensions of the standard model. 

We shall discuss a connection of our previous findings to both types of neutrino masses. This connection relies on the fact that, as shown in section \ref{sec:MMT}, Majorana fermions are mass eigenstates, involving both chiralities. We start by linking our dynamical mass generation scenario described in the first case of section \ref{nusm} to the standard model left-handed neutrinos. We then connect what has been discussed in the last case of section \ref{sec:3nu} to a dynamical seesaw model, involving right-handed Majorana neutrinos that exist in extensions of the standard model. Since in our scenarios the values of the masses can be fixed according to phenomenology, we can assume that any other possible mass contributions to neutrinos (\emph{e.g.} due to a Higgs mechanism in conventional seesaw models) are sub-dominant. The advantage of our dynamical mass generation approach is that it can be directly applied to left-handed standard model neutrinos, without the need of introducing right-handed ones (although there may be other reasons to introduce the latter), as well as it can provide a mechanism for generation of heavy sterile neutrino masses.

\subsection{Majorana masses for left-handed neutrinos}

We consider here the coupling of a doublet of Majorana fields, which, as seen in section \ref{sec:MMT}, are mass eigenstates, to the regulator U(1) gauge field $A_\mu$ in the case discussed in the first part of section \ref{nusm}. Because a Majorana field contains both chiralities, a straightforward extension of the Dirac case to the current situation is possible. 

In this way, we are able to generate dynamically different mass eigenvalues for the two species, without mixing, as given by (\ref{2mass}), {\it i.e.}  
	\be
		m_i=M\exp\left(\frac{-8\pi^2}{(4+\zeta)e_i^2}\right)~~~,~i=1,2~.
	\ee 
This is therefore a consistent way of discussing the dynamical appearance of a Majorana mass for left-handed neutrinos of the standard model, without the need of right-handed neutrino fields. 

Non-trivial mixing of flavour neutrinos, {\it i.e.} those fields coupled to the physical $SU(2)_L$ gauge fields of the standard model, and, consequently, flavour oscillations can then be obtained in the case where the mass eigenvalues $m_1$ and $m_2$ are different. Furthermore, in order to recover Lorentz invariance, we need to take simultaneously $e_1, e_2 \to 0$ in such a way that their ratio is fixed to the phenomenologically desired value. It is important once again to point out that in this approach we started from Majorana mass eigenstates coupled to the regulator gauge fields, with no mixing. The latter is obtained by expressing the Majorana mass eigenstates, with the help of the respective mixing matrix, in terms of the flavour neutrino eigenstates.

\subsection{Extensions of the standard model with right-handed neutrinos}

In this section we link our results on dynamical mass generation with the seesaw mechanism.
We begin by developing some of the concepts discussed in section \ref{sec:Seesaw} for the case of one generation only: one active and one sterile neutrino, and, after that, we show how our results can be used in the context of the seesaw mechanism.

\subsubsection{The seesaw mechanism with two neutrino fields}

As discussed in section \ref{sec:Seesaw}, when right-handed (sterile) neutrino components, $N_R$, are also present in the model, one can define a Dirac and Majorana mass term as in (\ref{66}) and (\ref{dmmt}). Considering the one generation case, this term reads
	\be\label{nMD}
		{\mathcal L}^{D+M}_m = -\frac{1}{2} {\overline n}_L   M^{D+M}   (n_L)^c ~,
	\ee
with
	\be\label{mmss}
		n_L = \begin{pmatrix} \nu_L \\ (N_R)^c \end{pmatrix}~~~\mbox{and}~~~M^{D+M} = \begin{pmatrix} 		 m^L \quad m^D \\ m^D \quad m^R \end{pmatrix}~.
	\ee
Additionally, for our toy purposes, we assume no CP violation in the lepton sector~\cite{BilenkyBook}, such that the elements of the mass matrix above are real numbers.

The matrix $M^{D+M}$ can be diagonalised by a Hermitean matrix $U$:
$$ M^{D+M} =  U {\mathfrak m}' U^T = O  {\mathfrak m}' \eta O^T $$
with~\cite{BilenkyBook}:
	\be\label{Umix}
		U = O  {\eta}^{1/2}~, \qquad O = \begin{pmatrix} {\rm cos}  \theta \quad {\rm sin}   \theta \\   	-{\rm sin}  \theta \quad {\rm cos}  \theta \end{pmatrix} ~.
	\ee
The orthogonal matrix $O$ diagonalises the mass matrix $M^{D+M}$ to ${\mathfrak m}'$, with eigenvalues
	\be\label{frakmp}
		m^\prime_{1,2} = \frac{1}{2} (m^R + m^L) \mp \frac{1}{2}   \sqrt{ (m^R - m^L)^2 + 4 		(m^D)^2}~,
	\ee
and we make use of the matrix $\eta$, with eigenvalues $\eta_i = \pm 1 $, to obtain positive mass eigenvalues: $\tilde{m}_i = \eta_i m_i'$. According to (\ref{frakmp}), $m_2'$ is always positive, and we choose $\eta_2=1$. On the other hand, the other mass eigenvalue can be negative; hence, for $m'_1<0$, we choose $\eta_1=-1$, such that we are left with two real and positive neutrino masses.

The mixing angle $\theta$ can be found in terms of the masses
	\be\label{mangle}
		{\rm tan}2\theta = \frac{2 m^D}{m^R - m^L}~.
	\ee
Finally, the Majorana fields, involving both chiralities, are defined in terms of $U$ as
	\be\label{finalmaj}
		\nu^M = U^\dagger   n_L + (U^\dagger   n_L)^c = \begin{pmatrix} \nu_1 \\ \nu_2 \end{pmatrix} ~, \quad \nu_i^c = \nu_i,   \quad i=1,2~.
	\ee

Thus, the original left-handed flavour neutrinos, appearing in the Lagrangian (\ref{nMD}), are related to these mass eigenstates as follows:
	\begin{eqnarray}\label{etanu}
		\nu_L &=& {\rm cos}  \theta \sqrt{\eta_1} \nu_{1  L} + {\rm sin}  \theta \sqrt{\eta_2}   \nu_{2   L}  \nonumber \\
		(N_R)^c &=& - {\rm sin}  \theta   \sqrt{\eta_1}   \nu_{1  L} + {\rm cos}   \theta \sqrt{		\eta_2}   \nu_{2  L}
	\end{eqnarray} 
In the standard seesaw scenarios~\cite{seesaw1,seesaw2,seesaw3}, as already mentioned in section \ref{sec:Seesaw}, there is no mass term for the left-handed fields ($m^L = 0$), and the Majorana mass associated to the RH neutrino is assumed much heavier than the Dirac masses ($m^R \gg m^D$). The Dirac mass is generated via the usual Higgs mechanism by Yukawa coupling terms of the form (\ref{YukNeut})
	\be\label{yuk}
		y \overline{\psi}_L \phi^C N_R + {\rm h.c.},  
	\ee
where $y$ is the Yukawa coupling, and $\phi^c = i \sigma_2 \phi^\star $ is the dual of the Higgs doublet. In this limit, from (\ref{frakmp}) and (\ref{mangle}), the mass eigenstates and mixing angle are given by
	\bea
		\tilde{m}_1 &\simeq& \frac{(m^D)^2}{m^R} \ll m^D ~~\mbox{and}~~ \tilde{m}_2 \simeq m^R \gg m^D~,\\
		\theta &\simeq& \frac{m^D}{m^R} \ll 1~, \nn
	\eea
with $\eta_1 = -1$ and $\eta_2 = 1$, hence from (\ref{etanu}) we obtain 
	\bea\label{seesaw} 
		\nu_L \simeq  i  \nu_{1 L} ~~~\mbox{and}~~~	(\nu_{R})^c \simeq \nu_{2 L}~. 
	\eea

\subsubsection{The generation of mass for heavy sterile neutrinos}

Having presented above the basics behind the seesaw mechanism in the simplest case of only two neutrinos, our purpose from now on is to adopt the previous procedure and generate dynamically masses for the Majorana fields by coupling them to LIV gauge fields. 

We start by rewriting the initial Lagrangian (\ref{Lag}) in terms of $N_R = \frac{1}{2} \Big(1 + \gamma_5 \Big)  N$ and $\nu_L = \frac{1}{2} \Big(1 - \gamma_5 \Big)  \nu$, a RH and a LH flavour neutrino, respectively, as
	\bea\label{Lag3}
		\mathcal{L}&=& -\frac{1}{4} F_{\mu\nu}(1-\frac{\Delta}{M^2})F^{\mu\nu}+ 
		\bar{N}(i \slashed{\partial}-e_1 \slashed{A})  \frac{1}{2} \Big(1 + \gamma_5 \Big)  N \\
		&&~~~~~~~~~~~~~~~~~+\bar{\nu}(i \slashed{\partial}- e_2 \slashed{A})  \frac{1}{2} \Big(1 - 				\gamma_5 \Big)  \nu~,\nonumber
	\eea
where, due to the opposite chiralities of the two spinor fields, the off diagonal flavour mixing gauge couplings $\epsilon$ are irrelevant because the corresponding terms vanish identically.
Here, $N, \nu$ are non-chiral spinors, which may be taken to be Majorana. This would be a simple version of the minimal (non supersymmetric) extension of the standard model of ref. \cite{numsm1,numsm2}, termed $\nu$MSM. In this case, instead of avoiding sterile neutrinos, we use the dynamical mass generation mechanism presented here to give masses to them.

According to our general discussion on combined Dirac and Majorana masses in section \ref{sec:Seesaw} and above, we may express the Lagrangian (\ref{Lag3}) in terms of Majorana fields, in such a way that $\nu$ and $N$ form a Majorana field doublet $\nu^M$ (\ref{finalmaj}), which then couples to the vector fields. Nevertheless, as discussed above eq.~(\ref{seesaw}), dynamically generated mixing of the two should involve a small mixing angle in phenomenologically realistic situations.

Comparing the mass matrix (\ref{mmss}) with that of our original model (\ref{mm}), we have the following correspondence: $m^L=m_1$, $m^R=m_2$ and $m^D=\mu$. Thus, as in the seesaw mechanism, we want a solution where $m^L=m_1=0$. Considering the solutions previously studied in section \ref{sec:3}, we see that the only compatible solution is the one in the subsection \ref{sec:3nu}. In such a case, the masses $m_1=m^L$ and $m_2=m^R$ can be identified with the dynamically generated mass eigenvalues 
	\bea\label{dgm}
		m^L&=&0\\
		m^R&=&M\exp\left(\frac{-8\pi^2}{(4+\zeta)e_2^2}\right)\nonumber~.
	\eea
As a result, the dynamically generated masses above correspond to a seesaw type mass matrix (\ref{nMD}) of the form:
	\be\label{nMD2}
		M^{D+M} = \begin{pmatrix} 0 \quad 0 \\ 0 \quad m_R \end{pmatrix}
	\ee
for the Majorana neutrinos. Therefore, in this situation, there is no non-trivial Dirac mass $\mu$, since the latter vanishes in the dynamical solution, as explained in subsection \ref{sec:3nu}. 

Although our dynamical solution does not provide a Dirac mass term, we expect it to come from the standard mechanism, {\it i.e.} via Yukawa couplings (\ref{yuk}) with the Higgs field. In this scenario, therefore, it is the heavy sterile neutrino mass that can be generated dynamically, due to the coupling with the LIV gauge sector. As previously discussed, since the finite mass in the Lorentz Invariant limit (\ref{metlim}) is arbitrary, it is possible to make it much heavier than the Higgs-generated Dirac mass, leading to naturally light active neutrinos via the seesaw mechanism. We discuss this idea in some detail below.

We consider the Schwinger-Dyson equations in the background of a Higgs field.\footnote{Any contribution of the fluctuations of the Higgs to the Schwinger-Dyson equations will be suppressed  by the Higgs mass and, consequently, will be ignored in the leading order approximation adopted here.} As usual, when the scalar field acquires a vev $\langle \phi \rangle = v$, it gives rise to a Dirac mass term of the form $yv$, where $y$ is the corresponding Yukawa coupling. Then, the new ``bare'' fermion propagator $S$ must contain a Dirac-mass term proportional to the Higgs-induced $\mu_0 = y v$, and the dressed fermion propagator $G$ will have a form similar to that in (\ref{gfermi}), but with the replacement of $\mu$ by the sum  $\mu + \mu_0$, where $\mu$ corresponds to any Dirac mass term generated dynamically. Taking into account these modifications to the Schwinger-Dyson equations, the expressions (\ref{I1I2bis}) become
	\bea\label{sdmn}
		I_1 &=& \frac{1}{4+\zeta} \frac{e_2^2 m_1^2 - e_1^2 m_2^2}{ (e_1^2 e_2^2-\epsilon^4)(m_1^2-m_2^2)}\\
		((\mu +\mu_0)^2 - m_1 m_2) I_2 &=& \frac{1}{4+\zeta} \frac{m_1 m_2(e_1^2-e_2^2)+\epsilon^2 (m_2^2-m_1^2)}{(e_1^2 e_2^2-\epsilon^4)(m_1^2-m_2^2)}~,\nn
	\eea
while the constraints (\ref{constraints}) are replaced by
	\bea\label{constraints2}
		(m_1+m_2)  \Big[ \mu   (e_2m_1+e_1m_2)(e_1-e_2) & - & \mu_0   \Big( m_1   (\epsilon^2 + e_2^2 		) - m_2   (\epsilon^2 + e_1^2) \Big) \Big] = 0 \nonumber \\
	\epsilon(e_2m_1+e_1m_2)&=&0~.
	\eea
The integrals $I_i$, $i=1,2$ are now given by the same expressions in (\ref{I1I2}), but with $\mu$ replaced by $\mu+ \mu_0$. 

For consistency with our considerations above, we seek solutions of (\ref{sdmn}) with $m_2 \ne 0$ and $m_1 = \mu = \epsilon  = 0$. According to the constraints above, such a specific solution requires that $e_1 = 0$. Furthermore, we consider the case where $\mu_0 \ll m_2 $, which is consistent with light active neutrinos. Thus, to leading order in $x \equiv \frac{\mu_0}{m_2} \ll 1$, we find
	\bea
		I_1 &\simeq& \frac{1}{16  \pi^2}\Big(-{\rm ln}(\frac{m_2^2}{M^2})-2x^2 + {\mathcal O}(x^4)\Big)~,\\
		\mu_0^2   I_2 &\simeq& \frac{1}{16  \pi^2}   \Big( 4 x^2   {\rm ln} x  + {\mathcal O}(x^4) \Big)~,\nn
	\eea
which, except for the fact that the present mass matrix has bare $\mu_0 = y v$ Dirac terms, leads to the same solution as in (\ref{dgm}): 
	\be\label{nMD3}
		M^{D+M} = \begin{pmatrix} 0 \quad yv \\ yv \quad m_2 \end{pmatrix}~, ~~     yv \ll m_2=m^R~,
	\ee
where $m_2=m^R$ is given by (\ref{dgm}). Therefore, this shows that the dynamical mass generation scenario described here provides a novel way for generating heavy right-handed neutrino masses when applied to extensions of the standard model containing such states, as in the model presented in~\cite{numsm1,numsm2}.

\section*{Appendix A: gap equations}\label{App1}

We present here the main steps to obtain (\ref{4equs}) from the Schwinger-Dyson equation (\ref{SD}) which is rewritten below:
\begin{eqnarray}\label{SDap}
G^{-1}-S^{-1} = \int_p ~ \tau\gamma^\mu ~D_{\mu\nu}~G ~\tau\gamma^{\nu}~.
\end{eqnarray}

We start by commuting the first $\tau$ and $\gamma^\mu$ in (\ref{SDap}), so that in the middle of the integrand we have the following matrix product 
\begin{eqnarray}\label{tgt}
\tau G \tau &=& X\begin{pmatrix} e_1 & -i\epsilon \\ i\epsilon & e_2 \end{pmatrix} \begin{pmatrix} \slashed{p}-m_2 & \mu \\
\mu & \slashed{p}-m_1 \end{pmatrix}\begin{pmatrix} e_1 & -i\epsilon \\ i\epsilon & e_2 \end{pmatrix}\\
&=& X \begin{pmatrix}
         e_1^2(\slashed{p}-m_2)+\epsilon^2(\slashed{p}-m_1) & -Y+\mu(e_1 e_2 - \epsilon^2) \\
        Y+\mu(e_1 e_2 - \epsilon^2) & \epsilon^2(\slashed{p}-m_2)+e_2^2(\slashed{p}-m_1)
         \end{pmatrix}~, \nonumber
\end{eqnarray}
where
\begin{eqnarray}
X &=& i\frac{p^2 + \slashed{p}( m_1 + m_2) + m_1 m_2 -\mu^2}{(p^2 - m_1^2)(p^2 - m_2^2)-2 \mu^2(p^2 + m_1 m_2)+\mu^4}\\
  &=& i\frac{p^2 + \slashed{p}( m_1 + m_2) + m_1 m_2 -\mu^2}{(p^2 - A_{-}^2)(p^2-A_{+}^2)}~;\nonumber\\
Y &=& i \epsilon[ e_1 (\slashed{p}-m_2)+ e_2 (\slashed{p}-m_1)]~,\nonumber
\end{eqnarray}
with $A_{\pm}^2$ defined as in (\ref{Apmdef}). Identifying each matrix element in the Schwinger-Dyson equation (\ref{SDap}) individually, we obtain for the $M_{11}$ element
\begin{eqnarray}
i m_1 &=& \int_p D_{\mu\nu}~\gamma^\mu X [e_1^2(\slashed{p}-m_2)+\epsilon^2(\slashed{p}-m_1)] \gamma^{\nu}\\
     &=&\int_p \frac{(4+\zeta)}{(1+ \vec{p}^2/M^2) } \frac{p^2(e_1^2 m_1 + \epsilon^2 m_2) 
     + (\mu^2-m_1 m_2)(e_1^2 m_2 +\epsilon^2 m_1)}{p^2(p^2 - A_{-}^2)(p^2-A_{+}^2)}~.\nonumber
\end{eqnarray} 
The previous expression can be written as
\begin{eqnarray}\label{m1}
\frac{m_1}{4+\zeta} =(e_1^2 m_1 + \epsilon^2 m_2) I_1 + (\mu^2-m_1 m_2)(e_1^2 m_2 +\epsilon^2 m_1)I_2,
\end{eqnarray}
where
\begin{eqnarray}\label{I1I2mink}
I_1 &=& -i \int_p \frac{1}{1+\vec{p}^2/M^2} \frac{1}{(p^2 - A_{-}^2)(p^2-A_{+}^2)}\\
I_2 &=& -i \int_p \frac{1}{(1+\vec{p}^2/M^2) } \frac{1}{p^2(p^2 - A_{-}^2)(p^2-A_{+}^2)}~.\nonumber
\end{eqnarray} 
After performing a Wick rotation $p_0 \rightarrow i\omega$, the integrals above become 
\begin{eqnarray}\label{i1i2in}
I_1 &=& \frac{1}{4\pi^3} \int_0^{\infty} \frac{\vec{p}^2d\vec p}{1 + \vec{p}^2/M^2} \int_{-\infty}^{\infty}
\frac{d\omega}{(\omega^2+ \vec{p}^2 + A_{+}^{2})(\omega^2+ \vec{p}^2 + A_{-}^{2})}\\
I_2&=&\frac{-1}{4\pi^3} \int_0^{\infty}\frac{\vec{p}^2d\vec p}{1 + \vec{p}^2/M^2} \int_{-\infty}^{\infty} 
\frac{d\omega}{(\omega^2+\vec{p}^2)(\omega^2+ \vec{p}^2 + A_{+}^{2})(\omega^2+ \vec{p}^2 + A_{-}^{2})}~.\nonumber
\end{eqnarray}
The part of the integrand of $I_1$ which depends on $\omega$ only can be written as
\bea\label{I1t}
&&\frac{1}{A_{+}^2-A_{-}^2}\left[\left(\frac{1}{\omega^2+\vec{p}^2}-\frac{1}{(\omega^2+ \vec{p}^2 + A_{+}^{2})}\right)\right.\nn\\
&&~~~\left.~~~~~~~~~~~~~~-\left(\frac{1}{\omega^2+\vec{p}^2}-\frac{1}{(\omega^2+ \vec{p}^2 + A_{-}^{2})}\right) \right]~,
\eea
and, similarly, for the part of the integrand of $I_2$ which depends on $\omega$ only, we have
\bea\label{I2t}
&&\frac{1}{A_{+}^2-A_{-}^2}\left[\frac{1}{A_{-}^2}\left(\frac{1}{\omega^2+\vec{p}^2}-\frac{1}{(\omega^2+ \vec{p}^2 + A_{-}^{2})}\right)\right.\nn\\
&&\left.~~~~~~~~~~~~~~~~~~~~~~~~-\frac{1}{A_{+}^2}\left(\frac{1}{\omega^2+\vec{p}^2}-\frac{1}{(\omega^2+ \vec{p}^2 + A_{+}^{2})}\right) \right]
\eea
Thus, substituting (\ref{I1t}) and (\ref{I2t}) into (\ref{i1i2in}), we obtain the first equation of (\ref{4equs}). Furthermore, due to the symmetry of our model, the second equation of (\ref{4equs}) can be easily obtained from the first one by exchanging $m_1$ and $m_2$. Finally, the left-hand side of (\ref{SDap}) is symmetric, with non-diagonal elements given by $i \mu$, therefore, the non-diagonal elements of the right-hand side must also be equal; however, looking at (\ref{tgt}), we realise that it is only possible if the terms related to $Y$ vanish. Thus, the non-diagonal elements lead to the following equations
\begin{eqnarray}
i \mu &=&\int_p \frac{(4+\zeta)}{(1+ \vec{p}^2/M^2) } \mu(e_1 e_2 -\epsilon^2) \frac{p^2  + m_1 m_2 -\mu^2}{p^2(p^2 - A_{-}^2)(p^2-A_{+}^2)}~,\\
0 &=& \int_p D_{\mu\nu}~\gamma^\mu X Y \gamma^{\nu}\nonumber\\
&=& \epsilon \int_p \frac{(4+\zeta)}{(1+ \vec{p}^2/M^2) } \frac{p^2(e_1 m_1 + e_2 m_2) 
+ (\mu^2-m_1 m_2)(e_1 m_2 +e_2 m_1)}{p^2(p^2 - A_{-}^2)(p^2-A_{+}^2)}~,\nonumber
\end{eqnarray}
where using eq.(\ref{I1I2mink}), we find the last two equations in (\ref{4equs}).

\section*{Appendix B: one-loop fermion self energy}\label{App2}

In this appendix we choose the Feynman gauge to calculate the fermion wave function renormalisation for the case \{$e_1=e_2$ and $\epsilon=0$\}. In order to avoid IR divergences obtained in the one-loop calculation for $m_1=m_2=\mu$, where one of the eigen masses vanishes, we initially consider that $m_1=m_2=m\ne\mu$. Thus, the fermion propagator is given by
\begin{eqnarray}\label{fprop}
G(p)=i\frac{p^2+2m\slashed{p}+m^2-\mu^2}{[p^2-(m+\mu)^2][p^2-(m-\mu)^2]}\begin{pmatrix} \slashed{p}-m  & \mu \\ \mu & \slashed{p}-m \end{pmatrix}.
\end{eqnarray}
The fermion wave function renormalisation is obtained by differentiating the fermion self-energy with respect to the external momentum and then setting it to zero. Since the fermion propagator (\ref{fprop}) has two independent flavour components, we consider the one-loop diagonal self energy $\Sigma^{(1)}_{diag}$ 
and the one-loop off-diagonal part $\Sigma^{(1)}_{off}$, {\it i.e.}
\begin{eqnarray}
&&\Sigma^{(1)}_{diag} (\omega,\vec{p})\\ 
&=& \frac{-i e^2}{(2\pi)^4} \int\frac{d^4 k}{1+\vec{k}^2/M^2} \left\{\frac{\gamma^\mu \gamma_\mu [(p-k)^2 - (m^2-\mu^2)]m}
{k^2 [(p-k)^2-(m+\mu)^2][(p-k)^2-(m-\mu)^2]}\right.\nn\\
&&~~~~~~~~~~~+\left.\frac{\gamma^\mu(\slashed{p}-\slashed{k})\gamma_\mu[(p-k)^2-(m^2+\mu^2)] }{k^2 [(p-k)^2-(m+\mu)^2][(p-k)^2-(m-\mu)^2]}\right\}\nn\\
&&\Sigma^{(1)}_{off}(\omega, \vec{p}) \nn\\
&=& \frac{-i e^2}{(2\pi)^4} \int\frac{d^4 k}{1+\vec{k}^2/M^2} \frac{ \gamma^\mu\gamma_\mu [(p-k)^2+m^2-\mu^2]\mu 
+2 m \mu\gamma^\mu(\slashed{p}-\slashed{k})\gamma_\mu }{ k^2 [(p-k)^2-(m+\mu)^2][(p-k)^2-(m-\mu)^2]}~.\nonumber
\end{eqnarray}
Differentiating now these terms with respect to $p_\rho$ and then setting the external frequency and momentum to zero, we find
\begin{eqnarray}
\frac{\partial \Sigma^{(1)}_{diag}}{\partial p_\rho}\big|_{p=0} 
&=& \frac{i e^2}{8 \pi^4} \int\frac{d^4 k}{1+\vec{k}^2/M^2} \left\{ \frac{k^2 \gamma^\rho - (m^2 +\mu^2) \gamma^\rho 
+ 2k^\rho \slashed{k}}{k^2[k^2-(m+\mu)^2][k^2-(m-\mu)^2]} \right.\\
&& - \left.\frac{ 4 k^\rho \slashed{k} k^4 -8 k^\rho \slashed{k} k^2 (m^2+\mu^2) +4k^\rho \slashed{k} 
(m^2+\mu^2)^2}{k^2[k^2-(m+\mu)^2]^2[k^2-(m-\mu)^2]^2}\right\}\nn\\
\frac{\partial \Sigma^{(1)}_{off}}{\partial p_\rho} \big|_{p=0} 
&=& - \frac{i \mu m e^2}{4 \pi^4}\int\frac{d^4 k}{1+\vec{k}^2/M^2} \left\{ \frac{-\gamma^\rho}{ k^2 [k^2-(m+\mu)^2][k^2-(m-\mu)^2]} \right.\nn\\
&&  + \left.\frac{ 4 k^\rho \slashed{k}k^2 - 4 k^\rho \slashed{k}(m^2+\mu^2) }{k^2[k^2-(m+\mu)^2]^2[k^2-(m-\mu)^2]^2}\right\}~.\nn
\end{eqnarray}
We then write
\bea
\Sigma^{(1)}_{diag}&=&-m+Z_{diag}^0\omega\gamma^0-Z_{diag}^1\vec p\cdot\vec\gamma\nn\\
\Sigma^{(1)}_{off}&=&-\mu+Z_{off}^0\omega\gamma^0-Z_{off}^1\vec p\cdot\vec\gamma~,
\eea
and since we are actually interested in the limit $\mu\to m$, we choose $m-\mu=m\delta$, with $\delta<<1$ and approximate $m + \mu \approx 2 m$. Writing the expressions above in terms of new variables $x=\sqrt{ \vec{k}^2 }/m$, $y = k_0/m$, $\lambda = m/M\ll 1$ and after a Wick rotation, we obtain
\begin{eqnarray}
Z_{diag}^0 &=& \frac{e^2}{2 \pi^3}\int_{0}^{\infty} \frac{x^2 dx}{1+\lambda^2 x^2} \int_{-\infty}^{\infty} dy
\left[ \frac{-(x^2+y^2)-2y^2-2}{(x^2+y^2)(x^2+y^2+4)(x^2+y^2+\delta^2)}\right. \nn\\
&& +\left. \frac{4 y^2 (x^2+y^2)^2 +16y^2(x^2+y^2)+16y^2}{(x^2+y^2)(x^2+y^2+4)^2(x^2+y^2+\delta^2)^2}\right]\\
Z_{off}^0 &=& \frac{e^2}{\pi^3}\int_{0}^{\infty} \frac{x^2 dx}{1+\lambda^2 x^2}  \int_{-\infty}^{\infty} dy
\left[ \frac{1}{(x^2+y^2)(x^2+y^2+4)(x^2+y^2+\delta^2)}\right.\nn\\
&& -\left. \frac{4 y^2 (x^2+y^2) + 8 y^2}{(x^2+y^2)(x^2+y^2+4)^2(x^2+y^2+\delta^2)^2}\right]~,\nonumber
\end{eqnarray}
and 
\begin{eqnarray}
Z^1_{diag} &=& \frac{e^2}{2 \pi^3}\int_{0}^{\infty} \frac{x^2 dx}{1+\lambda^2 x^2} \int_{-\infty}^{\infty} dy
\left[ \frac{-(x^2+y^2)-2x^2/3-2}{(x^2+y^2)(x^2+y^2+4)(x^2+y^2+\delta^2)}\right. \nn\\
&& +\left. \frac{4}{3}\frac{ x^2 (x^2+y^2)^2 +4x^2(x^2+y^2)+4x^2}{(x^2+y^2)(x^2+y^2+4)^2(x^2+y^2+\delta^2)^2}\right]\\
Z^1_{off} &=& \frac{e^2}{\pi^3}\int_{0}^{\infty} \frac{x^2 dx}{1+\lambda^2 x^2}  \int_{-\infty}^{\infty} dy
\left[ \frac{1}{(x^2+y^2)(x^2+y^2+4)(x^2+y^2+\delta^2)}\right.\nn\\
&& -\left. \frac{4}{3}\frac{x^2 (x^2+y^2) + 2 x^2}{(x^2+y^2)(x^2+y^2+4)^2(x^2+y^2+\delta^2)^2}\right]~.\nonumber
\end{eqnarray}
Finally, we solve the integrals above to find 
\begin{eqnarray}
Z^0_{diag} &=& \frac{e^2}{8 \pi^2}\left(\frac{1}{4}-\frac{1}{2} \ln{2} + \frac{1}{2} \ln{\delta} + \ln{\lambda}   \right) \\
Z^0_{off} &=& \frac{e^2}{16 \pi^2}\left(\ln{2} - \ln{\delta} \right)~,  \nonumber
\end{eqnarray}
and
\begin{eqnarray}
Z^1_{diag} &=&\frac{e^2}{8 \pi^2}\left(-\frac{1}{12}-\frac{1}{2} \ln{2} + \frac{1}{2} \ln{\delta} + \ln{\lambda}   \right)\\
Z^1_{off} &=& \frac{e^2}{16 \pi^2}\left(\ln{2} - \ln{\delta} \right)~.  \nonumber
\end{eqnarray}

\chapter{Quasi-relativistic fermions and flavour oscillations}\label{chap5}
\mbox{}

This chapter is based upon the paper~\cite{ALM2}.

\section{Introduction}

In this chapter we investigate another possibility to generate neutrino masses and oscillations from a LIV model. The model studied here contains higher-order space derivatives suppressed by a mass scale $M$, but keeps the number of time derivatives to its minimum, in order not to generate ghosts. Nonetheless, the kinematics are different from those in Lifshitz-type models (see \cite{Lifreview} for a review), since, in both the infrared and the ultraviolet, the dispersion relation for fermions is almost relativistic, differing from relativistic kinematics in an intermediate energy regime only, characterised by the mass scale $M$. Because of this behaviour, we say that the fermions in our model are ``quasi-relativistic''.

The quasi-relativistic fermions in our model are coupled via a four-fermion interaction, which, as we will see, allows for the generation of fermion masses, however small is the coupling strength governed by $g^2$. This is clearly in contrast with what happens in Lorentz symmetric theories with four-fermion interactions, where a critical coupling is naturally defined by the gap equation, as discussed in section \ref{sec:4f}. The originality of our model consists in generating fermion masses and flavour oscillations from quantum corrections, and not tree-level processes. These corrections imply finite effects in the IR even when the Lorentz-symmetric limit, consisting in taking $M\to \infty$ and $g\to 0$, simultaneously, is taken.

A particular consequence of our model is the analytic properties of the mass solution, as a function of the coupling constant. This feature is unusual in the case of Lorentz-symmetric theories, since a fermion mass cannot be generated by quantum corrections only, from a perturbative expansion in the Standard Model. In the case discussed here, however, although a non-perturbative approach is used to calculate the dynamical mass, an expansion of the result in the coupling constant could also be obtained by a one-loop calculation. 

In the next section we show the main properties of our model for the massive and massless single flavour case. We demonstrate that the possibility to generate masses for any coupling strength $g$ is crucial to recover Lorentz symmetry through the limit $M\to\infty$. Indeed, the fermion mass we find is proportional to $g^2M$, such that it can be kept fixed if we take the limits $M\to\infty$ and $g\to0$ simultaneously, in such a way that $g^2M\to$ constant. Also, in this limit, the four-fermion interaction vanishes, leaving us with a free relativistic fermion whose mass has been generated by quantum corrections. 

Section 3 generalizes the previous analysis to the case of two fermion flavours. We then show that in the relativistic limit $M\to\infty$, we are left with two massive free fermions, with a flavour-mixing mass matrix generated dynamically. This is similar to what was described in the previous chapter, where fermions interact with a LIV Abelian gauge field, which plays the role of regulator and eventually decouples from fermions in the Lorentz symmetric limit. 

In section 4 we extend our results to Majorana fermions by including sterile right-handed neutrinos in a seesaw-type extension. 

Finally, we present an Appendix to provide some technical details on the effective action for the auxiliary field which is introduced in section \ref{sec:mm}.

\section{Single-flavour case}

\subsection{Massive model and classical properties}

The LIV model we study here is defined by the Lagrangian 
{\footnotesize
  \be\label{model}
		\mathcal{L}_1 = \bar{\psi} \left[i\partial_0\gamma^0\left(1-\frac{a}{M^2}\Delta\right) -i\vec\partial\cdot\vec\gamma\left(1-i\frac{b}{M}\vec\partial\cdot\vec\gamma-\frac{c}{M^2}\Delta\right) -m_0\right]\psi 
		+ \frac{g^2}{M^2} (\ol\psi\psi)^2,
  \ee}
where $g^2$ is a dimensionless coupling, such that the mass scale $M$ is used to control both the LIV operators and the strength of the four-fermion interaction. We are mainly interested in the case with non-vanishing coefficients ($a, b, c$) which, as we explain below, leads to a quasi-relativistic dispersion relation for the fermion field.

The quadratic part of the Lagrangian above can be obtained from the standard model extension \cite{nmSME2}. The general Lagrangian describing the fermion sector of the SME can be easily obtained from equation (\ref{SMEneutaction}) by replacing the field $N$ with a standard Dirac field $\psi$
  \be
		{\cal L}^{SME}_f=\ol\psi\left(i\slashed\partial-M+\mathcal{Q}\right)\psi~+h.c.,
  \ee
where, as explained in section (\ref{sec:SMEneut}), $M$ and $\hat{Q}$ can be written as
  \bea
	M &=& m +i m_{5}\gamma_5~,\\
		\hat{\mathcal{Q}} &=& \sum_I \hat{\mathcal{Q}}^I \gamma_I = \hat{\mathcal{S}}+i\hat{\mathcal{P}}\gamma_5+\hat{\mathcal{V}}^\mu\gamma_\mu+\hat{\mathcal{A}}^\mu\gamma_5 \gamma_\mu+\frac{1}{2}\hat{\mathcal{T}}^{\mu\nu}\sigma_{\mu\nu}~,\nn
		\eea
with $\hat{\mathcal{Q}}$ representing derivative-dependent operators which can be further expanded as in (\ref{Qexp}). Thus, the Lagrangian (\ref{model}) is obtained by considering the specific case
  \be
		\hat{\mathcal{S}}= \frac{b}{M}\Delta~,~~\hat{\mathcal{V}}_0=-i\frac{a}{M^2}\Delta\partial_0~,~~\hat{\vec{\mathcal{V}}}=-i\frac{c}{M^2}\Delta\vec\partial~,~~M=m_0,~~\hat{\mathcal{P}}=\hat{\mathcal{A_\mu}}=\hat{\mathcal{T}}_{\mu\nu}=0~,
  \ee
where $a,b,c$ are dimensionless constants ($a>0$ and $c>0$), such that
  \be
		\hat{\mathcal{Q}}=-i\partial_0\gamma^0\frac{a}{M^2}\Delta+i\vec\partial\cdot\vec\gamma \left(i\frac{b}{M}\vec\partial\cdot\vec\gamma+\frac{c}{M^2}		\Delta\right)~.
  \ee
This choice is also motivated by a gravitational microscopic model \cite{mavroLV}. \\

In what follows, we will naturally assume that $m<<M$. From the Lagrangian (\ref{model}), we find that the fermion dispersion relation is
  \be\label{disprel}
		\omega^2=m^2\left(\frac{1+bp^2/(Mm)}{1+ap^2/M^2}\right)^2+p^2\left(\frac{1+cp^2/M^2}{1+ap^2/M^2}\right)^2~,
  \ee
which, in spite of being similar to a re-summation of higher-order powers of the momentum $p$, comes from the local Lagrangian (\ref{model}), containing a finite number of space derivatives. In the IR region the dispersion relation above is approximately relativistic: $\omega^2\simeq m^2 + \vec{p}^2$ for any values of $a, b, c$ (as far as $a,b,c\ll M$), but it is clearly modified at higher energies. Nonetheless, in the limit $M\to \infty$, at fixed $p$ and $m$, the Lorentz symmetric dispersion relation is recovered at all scales, as expected. 

Few specific cases are worth mentioning:\\

$\bullet$ {\it $a=b=0$ and $c\ne0$:} In this case the dispersion relation becomes $\omega^2\simeq (1/M^4) p^6$ in the UV and is therefore equivalent to a $z=3$ Lifshitz theory, where time has been rescaled by $M^2$ (see section~\ref{sec:Lifshitz} ). As we wish to avoid such a deviation from relativistic kinematics in the UV, we do not study this situation here;\\

$\bullet$ {\it $a\ne0$ and $b=c=0$:} We also discard this possibility here because it leads to a non-physical situation, since energy behaves as a decreasing function of momentum when $p^2>M^2$;\\

{$\bullet$ \it $a\ne0$, $b\ne0$ and $c=0$:} If only $c$ vanishes, the energy goes to a constant value in the UV: $\omega^2\simeq (bM/a)^2$. This possibility is also of no interest, as it leads to group velocity which goes to 0 when $p\to\infty$;\\

$\bullet$ {\it $a\ne0$ and $c\ne0$:} In this situation, the dispersion relation in the UV regime is $\omega^2\simeq (cp/a)^2$ which is relativistic for $a=c$. Therefore, the dispersion relation is not relativistic in the intermediate regime $p\sim M$ only. This is the interesting case on which we focus from now on.\\

By imposing $\omega$ to be an increasing function of $p$, we have that the different constants in the model (\ref{model}) must satisfy
  \be
		2b^2+4c\ge a+2ab\, m/M~,
  \ee
and without loss of generality\footnote{A more general study would keep free parameters $a,b,c$, but our aim is to give emphasis on the mechanism of mass generation, for which the choice $a=c=1$ is enough.} we shall choose $a=c=1$. With this choice the product of the group and phase velocities is then 
  \be\label{vpvg}
		v_pv_g=\frac{\omega}{p}\frac{d\omega}{dp}=1+\frac{2}{M^2}\frac{m+bp^2/M}{(1+p^2/M^2)^3}(bM-m)~,
  \ee
which shows that, for a typical SM mass $m$ and a typical Planckian mass $M$, the upper bound for Lorentz symmetry violation \cite{bounds}
  \be\label{upperbound}
		|v_pv_g-1|\lesssim10^{-16} ~,
  \ee
is satisfied for $p\lesssim 10^{-8}M$ (if $b$ is of order 1). Although it is far outside the current range of energies available experimentally in the laboratory, it approaches the Greisen-Zatsepin-Kuzmin cut off limit of high energy cosmic rays.

Finally, the bare propagator $S$ for the model (\ref{model}), with our choice $a=c=1$, is 
  \be\label{prop}
		S=i~\frac{(\omega \gamma^0 -\vec{p}\cdot\vec{\gamma})(1+p^2/M^2)+m+bp^2/M}{(\omega^2-p^2)(1+p^2/M^2)^2-(m+bp^2/M)^2}~,
  \ee
and we note that, for non-vanishing $b$, its trace, even in the massless case, is different from zero, which will be important 
for the analytical properties of the mass generated, as explained below.

\subsection{Massless model and mass generation} \label{sec:mm}

We investigate here the possibility to generate mass due to the four-fermion interaction, in the situation where the bare mass $m_0$ vanishes. We make use of the usual approach which consists in introducing a Yukawa coupling of fermions to an auxiliary field $\phi$, then integrate over fermions and look for a non-trivial minimum for the effective potential $V(\phi)$, which leads to a mass term in the original Yukawa interaction, as previously discussed in section \ref{sec:4f}. By following these steps, we neglect fluctuations of the auxiliary field about its vev, but these can be omitted in the limit $g^2\to0$, which will be taken in the process to recover Lorentz symmetry (see next subsection).

Thus, we consider the intermediate Lagrangian
  \be\label{intermediate}
		\mathcal{L}_1'= \bar{\psi} \left[i(\partial_0\gamma^0-\vec\partial\cdot\vec\gamma)\left(1-\frac{\Delta}{M^2}\right)
		+b\frac{\Delta}{M}\right]\psi -\frac{M^2}{4}\phi^2-g\phi\ol\psi\psi~,
  \ee
for which the integration over $\phi$ leads back to the original model (\ref{model}) with $a=c=1$. 
The Lagrangian does not contain a kinetic term for the scalar field at the tree level, and the large mass of such a field is an important feature of this approach to dynamical mass generation because it suppresses possible fluctuations of $\phi$ about its vev $\phi_1$, such that $g\phi\simeq g\phi_1$ can be identified with the fermion mass. As a consequence, to calculate the effective potential $V(\phi)$ and its minimum $\phi_1$, it is sufficient to consider a homogeneous configuration for $\phi$. Nevertheless, the field $\phi$ can be physically seen as a scalar collective excitation of the original fermionic fundamental degrees of freedom, whose kinetic term is generated by integrating out fermions, if one allows $\phi$ to depend on space-time coordinates. In the Appendix A, we derive this kinetic term and show that it vanishes in the Lorentz-symmetric limit considered in the next subsection, being therefore consistent with the assumption that $\phi$ is frozen to its vev $\phi_1$ in this limit.

From the Lagrangian (\ref{intermediate}), we integrate over fermions for a homogeneous field $\phi$, leading to the effective potential
  \be\label{Veff1}
		V_1(\phi)= \frac{M^2}{4}\phi^2 + i~ tr \int \frac{d^4 p}{(2 \pi)^4} 
		\ln\left[(\omega\gamma^0-\vec p\cdot\vec\gamma)(1+p^2/M^2)-bp^2/M-g\phi\right]~.
  \ee
In order to obtain the gap equation we minimise the potential: $(dV_1/d\phi)_{\phi_1}=0$, which, after a Wick rotation, gives
  \be\label{minimize}
		\frac{M^2}{2}\phi_1=\frac{g}{\pi^3}\int p^2 dp \int d\omega \left[\frac{ (g\phi_1 + bp^2/M)}{(\omega^2+p^2)(1+p^2/M^2)^2+(g\phi_1+		bp^2/M)^2} \right]~,
  \ee
and leads to the mass $m=g\phi_1$. Performing the integral over frequency, we then find
  \be\label{gap}
		\mu\frac{\pi^2}{2g^2}=\int \frac{x^2dx~(\mu+bx^2)}{(1+x^2)\sqrt{x^2(1+x^2)^2+(\mu+bx^2)^2}}~,
  \ee
where $x=p/M$ and $\mu = m/M$. 

We note that, unlike the case in conventional studies of dynamical mass generation, for $b\neq0$, $\mu=0$ is not a solution of the gap equation (\ref{gap}). Furthermore, if we consider $b=0$, although the UV behaviour of the integral above improves as it will become convergent, the existence of a non-vanishing mass requires the coupling constant $g$ to be larger than some critical coupling, as in the conventional case (we note that $b=0$ also coincides with a subluminal product $v_pv_g$ in eq.(\ref{vpvg})). We disregard this possibility, since we eventually will take $g^2\to0$ for the Lorentz-symmetric limit. 

Therefore, from now on, we take $b=1$ and regularize the gap equation (\ref{gap}) by $M$, such that the domain of integration in the gap equation is $0\leq x\leq 1$. Solving the integral (\ref{gap}) with $\mu\ll 1$, we then find
  \be\label{mdyn1}
		\mu=\frac{m}{M}=\frac{\alpha g^2}{1-2g^2/(5\pi^2)}+{\cal O}(\mu^2)~,
  \ee
where
  \be\label{alpha}
		\alpha=\frac{\ln(1+2/\sqrt5)-\arctan(1/2)}{\pi^2}\simeq~0.018~.
  \ee
The solution can be further simplified by taking into account that $g^2<<1$:
  \be\label{mdyn2}
		m\simeq \alpha g^2M~.
  \ee
We checked that the solution (\ref{mdyn1}) indeed corresponds to a minimum of the effective potential (\ref{Veff1}). 

An interesting point is that the solution (\ref{mdyn1}) is analytic in the coupling constant $g^2$, unlike other standard cases, such as the Lifshitz 4-fermion interaction in \cite{Lif4fermion,ABH}, where a dynamical mass has the typical non-analytic form 
  \be\label{non-analytic}
		m_{dyn}^{Lif}\simeq M\exp(-a/g^2)~, 
  \ee
where $a$ is a constant. We mention here, however, that the expression (\ref{mdyn1}) consists of a re-summation in powers of $g^2$ and goes beyond a one-loop calculation. Nevertheless, the approximate result (\ref{mdyn2}) can also be obtained from the usual one-loop correction to the fermion mass. This special features is due to the LIV propagator (\ref{prop}) whose, as already pointed out, trace does not vanish, even in the massless case. We are therefore in the unusual situation where a fermion mass generated dynamically can be derived using a perturbative expansion, whereas a mass of the form (\ref{non-analytic}) can only be obtained from a non-perturbative approach.

For completeness, we give the expressions for the mass generated in two other cases:\\
  \bea
		m&\simeq& b g^2 M \frac{2 \ln 2 -1}{2 \pi^2} ~~~~\mbox{for}~~0\ne b<<1\\
		m&\simeq& g^2 M \left(\frac{4-\pi}{2\pi^2} +\mathcal{O}(1/b^2)\right)~~~~\mbox{for}~~b>>1~.\nonumber
  \eea 
In the first case, the limit $b\to 0$ continuously leads to the vanishing solution of the gap equation (\ref{gap}) when $b=0$, as in a first order phase transition, and the non-trivial solution (involving a critical coupling) is not recovered. In addition, the situation $b<<1$ leads to a suppression of LIV effects in the dispersion relation (\ref{disprel}), and thus seems more relevant than the case $b>>1$.

\subsection{Lorentz symmetric limit} 

An important point, which differs from other models involving four-fermion interactions, such as the one discussed in section~\ref{sec:4f}, is that mass generation takes place here for any coupling strength, and no critical coupling exists below which this non-perturbative process does not occur. This interesting feature allows us to consider the Lorentz symmetric limit of the model, $M\to\infty$, in such a way that the dynamical mass (\ref{mdyn1}) remains finite, provided that $g$ depends on $M$ as
  \be\label{gpropto}
g(M)\sim\sqrt\frac{m}{\alpha M}~~,~~~~\mbox{when}~~M\to\infty~,
  \ee
where $m$ is fixed.

After taking this limit, where the product $Mg^2$ tends to a finite value, we are left with a free relativistic massive fermion, for which the mass has been generated by quantum corrections. A similar limit is considered in our previous work \cite{ALM1}, where the dynamical mass has the form (\ref{non-analytic}) though, as explained in the previous chapter.

Because our model (\ref{model}) breaks Lorentz invariance, space and time derivatives are dressed differently by quantum corrections. This fact can lead to problems when several species interact\footnote{If only one kind of particle self-interacts, then it is always possible to rescale momentum in order to absorb quantum corrections in such a way that the dressed dispersion relation remains relativistic in the IR.} \cite{deviation}. Indeed, it has been shown, in different Lifshitz models for example, that loop corrections to classical dispersion relations can lead to worse deviations from Relativity than the classical dispersion relation \cite{JA1,JA2,ABH,deviation,AB}. Nevertheless, since a consistent Lorentz symmetric limit  implies $g^2\to0$ according to (\ref{gpropto}), loop corrections to the kinetic terms in the model (\ref{model}) eventually vanish in this limit, such that the classical upper bound (\ref{upperbound}) for Lorentz violation remains satisfied.

Let us now illustrate this point by calculating the superficial degree of divergence $D$ of an $L$-loop graph $\Sigma^{(L)}$ contributing to the self energy. Each loop integral measure and each propagator (\ref{prop}) carry mass dimension 4 and -1, respectively. Because integrals are regularised by $M$, corrections of the form $p^2/M^2$ are at most equal to 1, so they do not play a role for the superficial degree of divergence. Hence $D=4L-I$, with $I$ being the number of internal propagators. Moreover, $L=I-V+1$ by momentum conservation, where $V$ is the number of vertices, and, since each vertex has four legs, and each internal propagator relates two vertices, we have $4V=E+2I$, where $E$ is the number of external propagators. As a consequence, we have, as for the usual NJL model,
  \be
D=2L+2-\frac{E}{2}~,~~\mbox{and}~~~V=L-1+\frac{E}{2}~.
  \ee
In our case, however, each vertex comes with a factor $g^2/M^2$, hence for the self energy ($E=2$) we have 
  \be
\Sigma^{(L)}\propto\left(\frac{g^2}{M^2}\right)^VM^{D}=Mg^{2L}~.
  \ee
Finally, taking into account the limit (\ref{gpropto}), we obtain
  \be\label{sigmaL}
\Sigma^{(L)}\propto\frac{m^L}{M^{L-1}}~.
  \ee
Because one-loop self energy is independent of the external momentum, the first non-trivial loop corrections to the kinetic terms occur at two loops ($L=2$). Therefore, in the case of interest, {\it i.e.} $L\ge2$, with a fixed mass $m$, the property (\ref{sigmaL}) shows that the loop correction $\Sigma^{(L)}$ vanishes when $M\to\infty$: quantum corrections to the kinetic terms vanish in the Lorentz symmetric limit (\ref{gpropto}).\\

Let us make here an additional remark about the present model in the context of the structure of the standard model. One could think that the introduction of LIV terms in the neutrino sector may cause unwanted consequences to emerge for the charged leptons. However, it is important to note that the higher-order derivative terms added in the neutrino sector are not invariant under $U(1)$ gauge transformations, unless one introduces new interactions which are not renormalisable, hence these LIV terms are not allowed in the charged lepton sector of the SM. Thus, our LIV model does not directly imply new physics for charged leptons. Moreover, although some can still expect radiative corrections to the charged leptons as a result of their interactions with neutrinos through weak gauge bosons, such new corrections are suppressed by the mass scale $M$, which eventually is taken to infinity. Therefore, we emphasise that the only observable effect of the present Lorentz violating model is the generation of neutrino masses and oscillations.

\section{Two-flavour case and oscillations}

Let us now extend the model studied above to the case of a massless fermion doublet $\Psi$, which self-interacts according to the flavour-mixing coupling matrix $\tau$, 
  \be\label{modelbis}
\mathcal{L}_2 = \bar{\Psi} \left[i(\partial_0\gamma^0-\vec\partial\cdot\vec\gamma)\left(1-\frac{\Delta}{M^2}\right)
+\frac{\Delta}{M}\right] \Psi + \frac{1}{M^2} (\ol\Psi\tau\Psi)^2,
  \ee
where 
  \be
\Psi=  \begin{pmatrix} 
       \psi_1\\
       \psi_2
       \end{pmatrix}~,
~~~~\mbox{and}~~~~
\tau=  \begin{pmatrix} 
       g_1 & g_3\\
       g_3 & g_2
       \end{pmatrix}~.
  \ee
In what follows we show how flavour oscillations can be generated dynamically.

\subsection{Minimization of the effective potential}

The original Lagrangian (\ref{modelbis}) is equivalent to the following Lagrangian involving an auxiliary field
  \be\label{lag2}
		\mathcal{L}_2' = \bar{\Psi}\left[i(\partial_0\gamma^0-\vec\partial\cdot\vec\gamma)\left(1-\frac{\Delta}{M^2}\right)
		+\frac{\Delta}{M}\right]\Psi-\frac{M^2}{4}\phi^2-\phi\ol\Psi\tau\Psi~.
  \ee
Before integrating over fermions, we first find the eigen values, in flavour space, of the operator
{\footnotesize
  \be
		{\cal O}=  \begin{pmatrix} 
       (\omega\gamma^0-\vec p\cdot\vec\gamma)(1+\frac{p^2}{M^2})-\frac{p^2}{M}-g_1\phi & -g_3\phi\\
       -g_3\phi & (\omega\gamma^0-\vec p\cdot\vec\gamma)(1+\frac{p^2}{M^2})-\frac{p^2}{M}-g_2\phi
       \end{pmatrix}~,
  \ee}
which are
  \be\label{Oeigen}
		\lambda_\pm=(\omega\gamma^0-\vec p\cdot\vec\gamma)(1+p^2/M^2)-p^2/M-h_\pm\phi~,
  \ee
where the eigen values $h_\pm$ of the coupling matrix $\tau$ are given by
  \be
		h_\pm=\frac{1}{2}(g_1+g_2)\pm\frac{1}{2}\sqrt{(g_1-g_2)^2+4g_3^2}~.
  \ee
The effective potential for the auxiliary field can now be written in terms of $\lambda_{\pm}$ as
  \be
		V_2=\frac{M^2}{4}\phi^2+i ~tr\int \frac{d^4 p}{(2 \pi)^4} (\ln\lambda_++\ln\lambda_-)~,
  \ee
and its minimization $(dV_2/d\phi)_{\phi_2}=0$ leads to 
{\footnotesize 
 \be
		\frac{M^2}{2}\phi_2=\sum_{s=+,-}\frac{h_s}{\pi^3}\int p^2 dp \int d\omega \left[\frac{ (h_s\phi_2 + p^2/M)}{(\omega^2+p^2)(1+p^2/M^2)^2+(h_s\phi_2+p^2/M)^2} \right]~.
  \ee}
Then, integrating over frequencies, we find the following gap equation, regularized by the mass scale $M$,
  \be\label{gap3}
		\kappa\frac{\pi^2}{2} = \sum_{s =+,-}h_s\int_0^1 \frac{x^2dx~(h_s\kappa+x^2)}{(1+x^2)\sqrt{x^2(1+x^2)^2+(h_s\kappa+x^2)^2}} ~,
  \ee
where $x = p/M $ and $\kappa = \phi_2/M$. Solving the integral above for $\kappa\ll 1$ gives
  \be
		\kappa = \frac{\alpha(h_+ + h_- )}{1-2(h_+^2 + h_-^2)/(5\pi^2)}+{\cal O}(\kappa^2)~,
  \ee
where $\alpha$ is given by eq.(\ref{alpha}). Finally, taking into account  that $h_\pm<<1$, we obtain 
  \be\label{min2}
		\kappa \simeq \alpha(g_1 + g_2 )~.
  \ee
This non-vanishing value for the minimum of the potential allows for the generation of a mass matrix, as we discuss below.

\subsection{Flavour oscillations}

From the previous results, we find that the mass matrix $\mathcal{M} = \kappa M \tau$ below is generated
  \be
		\mathcal{M}=\alpha(g_1+g_2)M  \begin{pmatrix} g_1 & g_3 \\ g_3 & g_2\end{pmatrix}~,
  \ee
and the mass eigen values $m_\pm = \kappa M h_\pm$ and the mixing angle $\theta$ are given by
  \bea\label{md12}
		m_\pm &=& \frac{\alpha}{2} M\left[(g_1+g_2)^2\pm\sqrt{(g_1^2-g_2^2)^2+4g_3^2(g_1+g_2)^2}\right]\nn\\
		\tan\theta&=&\frac{g_1-g_2}{2g_3}+\sqrt{1+\left(\frac{g_1-g_2}{2g_3}\right)^2}~.
  \eea
Using the expressions above, we can write the dimensionless couplings $g_i$ in terms of the mass eigenvalues, mixing angle and mass scale $M$, as
  \bea\label{g1g2g3}
		g_1 &=&\frac{\mu_++\mu_-+(\mu_+-\mu_-)\cos(2\theta)}{2\sqrt{\alpha(\mu_++\mu_-)}}~,\\
		g_2 &=&\frac{\mu_++\mu_--(\mu_+-\mu_-)\cos(2\theta)}{2\sqrt{\alpha(\mu_++\mu_-)}}~,\nn\\
		g_3 &=&\frac{\mu_--\mu_+}{2\sqrt{\alpha(\mu_++\mu_-)}}\sin(2\theta)~,\nonumber
  \eea
where 
  \be 
		\mu_\pm = \frac{m_\pm}{M}~.
  \ee
In this way, the couplings $g_i$ can be written in the form
  \be\label{gi}
		g_i=\frac{a_i}{\sqrt M}~~,~i=1,2,3~,
  \ee
where the constants $a_i$ are completely determined by the experimental values for $m_\pm$ and $\theta$. 

From the last expression, which shows the explicit dependence of the coupling constants $g_i$ on the mass scale $M$, we see that the couplings $g_i$ go to zero in the Lorentz-symmetric limit, {\it i.e.} when $M\to \infty$. Thus, by taking such a limit, we are left with two relativistic free fermions, for which flavour oscillations have been generated. Therefore, by considering the coupling constants (\ref{g1g2g3}), any set of values for $m_\pm$ and $\theta$ can be described by the Lorentz-symmetric limit of our model.

Let us discuss now the oscillation probability for our model. For the two flavour case, in the usual, Lorentz invariant, situation where $E_+-E_- \approx (m_1^2-m_2^2)/2E$, the oscillation probability is is given by (\ref{nosc2}).

In our case, however, with the dispersion relation given by~(\ref{disprel}) and considering, as usual, $m_\pm^2/p^2\ll 1$
and $m_\pm/M<<1$, we find
  \be\label{endiff}
		(E_+ - E_-) t = \frac{(m_+^2 - m_-^2) L}{2 E} + (m_+ - m_-) \frac{E L}{M}+ \mathcal{O}(m_\pm^2/M^2)~.
  \ee
Then, the corresponding oscillation probability is
  \bea\label{proba}
		\mathcal{P}(\nu_{  \beta_1} \to \nu_{  \beta_2} ) &=& \sin^2(2 \theta) \sin^2\left[\frac{ (m_+^2 - m_-^2) L}{4 E} + 
		(m_+ - m_-)\frac{E L}{2 M}+ ...\right]~, \\
		&\simeq& \frac{\sin^2 [A ~(g_1 + g_2)^3 \sqrt{(g_2-g_1)^2 + 4g_3^2}~]}{1 + (g_2 - g_1)^2/(4 g_3^2)},
		~~~~\mbox{with}~A = \frac{\alpha^2 M^2 L}{4 E}~.\nn
  \eea
where, on account of (\ref{md12}) and (\ref{g1g2g3}), ${\rm sin}^{-2}(2\theta)$ is just the denominator of the right-hand-side of (\ref{proba}). According to eq.(\ref{gi}), the argument of the sine function in (\ref{proba}) goes to a finite limit when $M\to\infty$, since it is proportional to the finite $M^2g_i^4$, $i=1,2,3$, and so does the denominator in the last expression of (\ref{proba}). Moreover, we note that, by looking at the right-hand side of the first line of (\ref{proba}), the first term on the argument of the sine is the usual relativistic expression, while the second term is the first contribution coming from the Lorentz-violating features of our model. This second term, however, goes to zero when $M\to \infty$, and (\ref{proba}) reduces to the usual relativistic oscillation probability (\ref{nosc2}) in Lorentz-invariant vacuum, as expected. On the other hand, if $M$ is kept finite, by not taking the Lorentz-symmetric limit, the second term in~(\ref{endiff}), linearly suppressed by $M$, may have phenomenological consequences, for $M$ as large as Planck mass, $M_{\rm Pl}$, as discussed in \cite{brustein}.

\section{Majorana fermions: seesaw-type extension}

Based on the possibility that neutrinos are Majorana particles, we present a way of extending the previous results to the case of Majorana neutrinos. Particularly, we are interested in a seesaw-type model, such that we need to consider, in addition to LH (active) neutrinos, RH (sterile) neutrinos. 

We consider here the model given in eq.(\ref{modelbis}), but assuming that, instead of considering two Dirac fermions, the fermion doublet $\Psi$ is made of a LH and a RH Majorana field, $v_L$ and $N_R$. This configuration, as mentioned in section \ref{sec:Seesaw}, allows us to construct two different kinds of mass terms: 
  \be\label{MD}
		{\mathcal L}^{D+M}= -\frac{1}{2} {\overline \nu}_L \, m^L \, (\nu_L)^c - \overline{\nu}_L \, m^D \, N_R  -  
		\frac{1}{2} {\overline N}_R \, m^R \, (N_R)^c + {\rm h.c.}~,
  \ee
where $m^{L,R}$ are Majorana mass terms, and $m^D$ is the usual Dirac mass term.

The mass terms generated here, via the mechanism studied in this chapter, are the same as the ones we found for the original model with Dirac fermions. In particular, one can choose $g_1 = 0$ and $g_3\ll g_2$, so that the mass matrix below is generated
  \be
		\mathcal{M} = \alpha M g_2   \begin{pmatrix} 0 & g_3 \\ g_3 & g_2 
                            \end{pmatrix}
		=   \begin{pmatrix} m_L & m_D \\ m_D & m_R
		\end{pmatrix}~,
  \ee
with the following eigenvalues
  \bea
		m_{+} &\simeq& \alpha M g_2^2 = m_R\\
		m_{-} &\simeq& \alpha M g_3^2 = \frac{m_D^2}{m_R} \ll m_R~.
  \eea
These results are in agreement with the main idea behind the seesaw mechanism: the heavier the sterile fermion, the lighter the active one.

In the original seesaw mechanism, while the Majorana mass $m^R$ needs to be generated via an unknown (non-standard model) mechanism, the Dirac mass is generated via the Higgs mechanism. In the case studied here, however, both mass terms are generated by the mechanism described above. Thus, the Higgs mechanism is not needed to generate the Dirac mass term, but it can be easily included in the model.

\section*{Appendix A:\\ Derivation of the kinetic term for the auxiliary field}

For simplicity, we neglect here higher-order derivative terms, as they only provide corrections of order $1/M$ in the kinetic term for the auxiliary field, and we eventually take $M\to \infty$.\\
In order to derive the kinetic term for the auxiliary field, we take the following non-homogeneous configuration (plane-wave)
  \be
		\phi=\phi_1+\rho\Big(\exp(ik_\mu x^\mu)+\exp(-ik_\mu x^\mu)\Big)~,
  \ee
where $\rho<<\phi_1$. Integrating over fermions, we find the formal expression
  \be
		i\mbox{Tr}\ln\left(i\slashed\partial-g\phi\right)~,
  \ee
which should then be expanded in $\rho$ and $k$, in order to identify the kinetic term, which is proportional to $k^2\rho^2$. First, we expand to the second order in $\rho$
  \bea
		&&\ln\left(\slashed p-g\phi\right)\\
		&=&\ln\left[(\slashed p-g\phi_1)\delta(p+q)-g\rho\left(\delta(p+q+k)+\delta(p+q-k)\right)\right]\nn\\
		&=&\delta(p+q)\ln(\slashed p-g\phi_1)-g\rho\frac{\slashed p+g\phi_1}{p^2-g^2\phi_1^2}\left(\delta(p+q+k)+\delta(p+q-k)\right)\nn\\
		&&+\frac{g^2\rho^2}{2}\frac{\slashed p+g\phi_1}{p^2-g^2\phi_1^2}\left(\frac{\slashed k-\slashed q-g\phi_1}{(k-q)^2-g^2\phi_1^2}+		\frac{-\slashed k-\slashed q-g\phi_1}{(k+q)^2-g^2\phi_1^2}\right)~		\delta(p+q)\nn\\
		&&+\frac{g^2\rho^2}{2}\frac{\slashed p+g\phi_1}{p^2-g^2\phi_1^2}\frac{-\slashed k-\slashed q-g\phi_1}{(k+q)^2-g^2\phi_1^2}~\delta(p+q+2k)\nn\\
		&&+\frac{g^2\rho^2}{2}\frac{\slashed p+g\phi_1}{p^2-g^2\phi_1^2}\frac{\slashed k-\slashed q-g\phi_1}{(k-q)^2-g^2\phi_1^2}~\delta(p+		 q-2k)+{\cal O}(\rho^3)~,\nonumber
  \eea
such that
  {\footnotesize \bea
		\mbox{Tr}\ln\left(\slashed p-g\phi\right)		&=&V\mbox{tr}\int\frac{d^4p}{(2\pi)^4}\ln(\slashed p-g\phi_1)\\
		&&+V\frac{g^2\rho^2}{2}\mbox{tr}\int\frac{d^4p}{(2\pi)^4}\frac{\slashed p+g\phi_1}{p^2-g^2\phi_1^2}\left(\frac{\slashed k+	\slashed p-g\phi_1}{(k+p)^2-g^2\phi_1^2}+\frac{-\slashed k+\slashed p-g\phi_1}{(k-p)^2-g^2\phi_1^2}\right)+{\cal O}(\rho^3)~,		\nonumber
  \eea}
where $V$ is the space-time volume. In the last equation, we can discard the first term because it corresponds to corrections to the potential $V(\phi_1)$, that we have already calculated. We then expand the second term in $k$ to give (ignoring higher orders in $\rho$) to obtain
  \bea
		\mbox{Tr} \ln\left(\slashed p-g\phi\right) &=& 2Vg^2\rho^2\int\frac{d^4p}{(2\pi)^4}\left(\frac{4(pk)^2}{(p^2-g^2\phi_1^2)^3}-\frac{2k^2}{(p^2-g^2\phi_1^2)^2}\right)\\
		&&+~\mbox{$k$		-independent terms}~+~{\cal O}(k^4)~,\nonumber
  \eea
where the $k$-independent terms correspond to corrections to the potential, arising from $\rho\ne0$, and therefore can be omitted here.
Using the following property
  \be
		\int d^4p ~f(p^2) p^\mu p^\nu=\frac{\eta^{\mu\nu}}{4}\int d^4p~ p^2f(p^2)~,
  \ee
we obtain (considering only the relevant terms proportional to $k^2$)
  \be
		\mbox{Tr} \ln\left(\slashed p-g\phi\right)=2Vk^2g^2\rho^2\int\frac{d^4p}{(2\pi)^4}~\frac{-p^2+2g^2\phi_1^2}{(p^2-g^2\phi_1^2)^3}~.
  \ee
To calculate the integral, we perform a Wick rotation, regulate it by $M$ and replace $g\phi_1$ by the mass $m$ to finally obtain
  \bea
		i\mbox{Tr} \ln\left(\slashed p-g\phi\right)&=&V\frac{k^2g^2\rho^2}{8\pi^2}\int_0^{M^2/m^2}xdx\frac{2+x}{(1+x)^3}\\
		&\simeq&V\frac{k^2g^2\rho^2}{4\pi^2}\ln\left(\frac{M}{m}\right)~.\nonumber
  \eea
We are looking for the following kinetic term 
  \be
		\int d^4x~\frac{Z}{2}\partial_\mu\phi\partial^\mu\phi=VZk^2\rho^2~,
  \ee
such that the identification with $i\mbox{Tr} \ln\left(\slashed p-g\phi\right)$ finally leads to
  \be
		Z=\frac{g^2}{4\pi^2}\ln\left(\frac{M}{m}\right)~.
  \ee
Finally, as explained in subsection 2.3, we have that the Lorentz symmetric limit (\ref{gpropto}) freezes the auxiliary field to its vev, since, for a fixed mass, the kinetic term vanishes in this limit
  \be
		\lim_{M\to\infty}Z\propto\lim_{M\to\infty}\frac{1}{M}\ln\left(\frac{M}{m}\right)=0~.
  \ee

\chapter{Lorentz-violating fermion kinematics from modified quantum gravity}\label{chap6}
\mbox{}

This chapter is based upon the paper \cite{AL}.

\section{Introduction}

In this chapter we consider two modifications of Einstein gravity, both invariant under foliation-preserving diffeomorphisms (see section \ref{subsec:HL}), coupled to matter fields and study how these gravity models induce Lorentz violation in the matter sectors. Particularly, in order to test the validity of these modified gravity models, we calculate LIV one-loop quantum corrections to the matter dispersion relations.

The first model is a modification of Einstein gravity which is not invariant under 4-dimensional diffeomorphisms, but keeps the isotropic scaling between space and time ($[x^i]=[t]$), so that no higher-order derivative operators can be included. In this way, as in GR, one-loop corrections are quadratically divergent, and the model is assumed to be valid up to a certain energy scale which can be estimated by comparing our results to the current experimental bounds on Lorentz violation.

The second model, which can be seen as a UV-improved extension of the first, is the $z=3$ non-projectable version of Horava-Lifshitz gravity, described in section \ref{subsec:HL}. The improved UV behaviour follows from the anisotropic scaling between space and time which allows for the introduction of higher-order space derivatives in the theory, making the one-loop corrections to the matter dispersion relations logarithmically divergent. 
	
In both cases, gravity is better described in terms of the Arnowitt-Deser-Misner (ADM) decomposition of the metric, which naturally exhibits a space-time foliation. Moreover, due to their invariance under a reduced symmetry, these modified gravities contain an extra propagating degree of freedom, when compared to GR \cite{Arkani}, known as the scalar graviton. 

These models are minimally coupled to classical complex scalar and fermion fields, and, in order to calculate the one-loop effective dispersion relations seen by particles, we integrate out the graviton components. Different works investigating the introduction, via quantum corrections, of LIV effects in the matter sector by coupling it to HL gravity have also been performed by others. In~\cite{Pospelov}, considering quantum matter fields, the authors derive the effective speed seen by a scalar field and an Abelian gauge field, and compare the results to measure Lorentz symmetry violation. In \cite{Padilla}, using a different method than \cite{Pospelov}, the non-projectable version of HL gravity is considered for the derivation of the effective LIV matter Lagrangian. Another interesting study is done in \cite{AB2}, in which classical scalar and gauge fields are coupled to the covariant version of HL gravity (see section \ref{subsec:HL}), and one-loop dispersion relations for the matter fields are calculated and then compared to give an estimation of Lorentz violation. We do a similar study here, taking into account classical complex scalar and fermionic backgrounds though, and we calculate the difference $\Delta v^2$ between the effective speeds of light seen by these two species.

Furthermore, it is also worth mentioning that studies of effective dispersion relations for Lifshitz-type models in flat space-time have also been performed. Some examples are \cite{ABH}, where the effective dispersion relation for interacting Lifshitz fermions is derived, in the case where flavour symmetry is broken, and \cite{AB}, where, considering a Lifshitz extension of QED, the authors derive the fermion effective dispersion relation. Our aim here is to study similar features, however, in the context of curved space-time and investigate how global Lorentz-symmetry is affected from a local symmetry breaking.

For the first model, which involves quadratic divergences, we use a cut off to calculate one-loop graviton loops. The second 	model, on the other hand, involves logarithmic divergences only, and we choose to make use of dimensional regularisation to perform the loop integrals; however, as one might expect, we show that the same physical result would be obtained if we had chosen to regularise the integrals with a cut off. We then find that although these models behave differently in the UV, the IR phenomenology for matter fields, regarding the LIV one-loop contributions, is comparable. On the one hand, if the parameters present in each model assume generic values, $10^{10}$ GeV emerges as a characteristic scale above which both models are not consistent with current upper bounds on Lorentz symmetry violation. For the first model the limit $10^{10}$ GeV is for the physical cut off of the theory, whereas for the second model this bound is imposed on the mass scale suppressing the higher-order LIV operators. On the other hand, if one accepts to fine-tune the different parameters, it is therefore always possible to choose them in such a way that the indicator for Lorentz symmetry violation vanishes for both models.

This chapter is organised as follows. Section \ref{sec:models} starts by introducing the models and discussing gauge freedom; then, in subsection \ref{subsec:actionexp}, the actions for gravity and matter sectors are expanded around a flat background, and only terms relevant to the one-loop calculations are kept. In section \ref{sec:corrections}, after deriving and making use of constraints which relate non-propagating, auxiliary, fields to propagating ones, we integrate successively the different components of graviton fluctuations. Finally, in section \ref{sec:analysis}, we put all results together and, based on the current upper bounds for LIV parameters, we discuss the phenomenology of the two models. An appendix is included to provide the details of the one-loop calculations discussed in section \ref{sec:corrections}.

\section{Models}\label{sec:models}

We start by introducing the two modified gravity models studied here, and how they couple to matter fields.\\

The first model we consider is described by a modified Einstein-Hilbert action, with new operators being allowed due to the reduced symmetry of the theory. When using the ADM decomposition and omitting the cosmological constant, its action can be written as
	\be\label{SG}
		S_G = M_P^2 \int dt d^3 x \sqrt{g} N (K_{ij} K^{ij} -\lambda K^2 + R^{(3)}+\alpha a_i a^i)~,
  \ee 
where $M_P^2 = (16 \pi G_N)^{-1}$ and
  \bea 
		g &=& |\det(g_{ij})| \\
		K_{ij} &=& \frac{1}{2N}( \partial_t g_{ij} - D_i N_j - D_j N_i)~~~\mbox{with}~~~D_i N_j = \partial_i N_j - 		\Gamma^{k}_{ij} N_k \nn \\
		K &=& K_{ij} g^{ij}~,~~~ R^{(3)} = R_{ijkl} g^{ik} g^{jl}~,~~~~a_i=\partial_i \ln N\nonumber~.
  \eea 
$ G_N$ and $K_{ij}$ are the Newton gravitational constant and the extrinsic curvature, respectively. The possibility to have $\lambda\ne 1$ and $\alpha\neq 0$ is a result of the fact that instead of being invariant under 4-dimensional diffeomorphisms, the model above is invariant under foliation-preserving diffeomorphisms only. Nonetheless, the Hilbert-Einstein action can easily be recovered with $\lambda=1$ and $\alpha=0$.\\

The second model, we consider here, is the non-projectable version of Horava-Lifshitz gravity (npHL)~\cite{non-proj}. As already discussed in section \ref{subsec:HL}, in the npHL the lapse function $N$ depends on space and time and therefore terms containing $a_i = \partial_i\ln N$ should be included in the action. In this context, space and time scale anisotropically such that
  \be 
		\vec{x}\to b\vec{x}~~~\mbox{while}~~~t\to b^z t~, 
  \ee 
and the choice $z=3$ motivates us to introduce dimension 4 and dimension 6 operators, which become important in the UV. As in~\cite{non-proj} though, we choose to reparametrise the time component in order to write the action in terms of the ``physical'' units. The npHL action is then given by~\cite{non-proj}
  \bea\label{SHL}
		S_{HL} &=& M_P^2 \int dt d^3 x \sqrt{g} N \left\{K_{ij} K^{ij} -\lambda K^2 + R^{(3)}+\alpha a_i a^i\right.\\ 
		&&\left. + F_1 R_{ij}R^{ij} +F_2 (R^{(3)})^2 +F_3 R^{(3)} \nabla_i a^i +F_4 a_i \Delta a^i\right.\nonumber\\
		&&\left.+S_1 (\nabla_i R_{jk})^2 +S_2 (\nabla_i R^{(3)})^2 + S_3 (\Delta R^{(3)} \nabla_i a^i) + S_4 (a_i 	\Delta^2 a^i) \right\}~,\nonumber
  \eea 
where $F_i= (f_i/M_{HL}^2)$ and $S_i=(s_i/M_{HL}^4)$ with $M_{HL}$ being the Horava-Lifshitz scale, {\it i.e.} the mass scale suppressing the higher-derivative operators, while $f_i$ and $s_i$ represent dimensionless coupling constants associated with operators of dimension 4 and 6, respectively.\\

Finally, we assume that the modified gravity theories~(\ref{SG}) and (\ref{SHL}) are coupled to complex scalar and fermion fields through minimal couplings. The action for the complex scalar field is
  \be\label{Ss1}
		S_s = -\int dt d^3 x~\sqrt{g}Ng^{\mu\nu}\partial_\mu \phi \partial_\nu \phi^\star~,
  \ee
whereas for the fermion action, we have
  \be\label{Sf1}
		S_f = -\int dt d^3 x~\frac{ie}{2} \left[\bar{\psi} \gamma^\alpha e^{\mu}_{~\alpha} \nabla_\mu \psi-e^{\mu}_{~\alpha}(\nabla_\mu\bar{\psi}) \gamma^\alpha\psi \right]~,
  \ee
where 
	\bea 
		e &=& \det(e^{~\alpha}_{\mu}) = \sqrt{g} N\\
		\nabla_\mu \psi &=& (\partial_\mu + \Gamma_\mu) \psi~~\mbox{and}~~ \nabla_\mu \bar{\psi} = \partial_\mu				\bar{\psi} - \bar{\psi}\Gamma_\mu\nn \\
		\Gamma_\mu &=& \frac{1}{2} w_{\mu\alpha  \beta}\sigma^{\alpha  \beta}~~\mbox{and}~~ \sigma^{\alpha  \beta}=		\frac{1}{4}[\gamma^\alpha,\gamma^  \beta]~, \nn \\
		w_{\mu\alpha  \beta} &=& e^{\lambda}_{~\alpha}( \partial_\mu e_{\lambda   \beta} - \Gamma^{\sigma}_{				\lambda\mu} e_{\sigma  \beta}) 
		=  e^{\lambda}_{~\alpha}(D_\mu e_{\lambda  \beta})~,\nonumber
	\eea
$w_{\mu\alpha  \beta}$ being the spin connection.

\subsection{Gauge invariance and degrees of freedom}

The new parameters $\lambda$, $\alpha$, $F_i$ and $S_i$ are intrinsically related to the explicit breakdown of 4-dimensional diffeomorphisms, for both gravity models, which takes place when $\lambda \neq 1$, $\alpha \neq 0$, $F_i\neq0 $ and $S_i\neq 0$. Instead, as already mentioned, these models are invariant under foliation-preserving diffeomorphisms 
  \bea \label{diffeo}
		\delta t&=&f(t)\\
		\delta x^{i}&=&\xi^i(t,x)\nn \\
		\delta g_{ij}&=&\partial_{i}\xi_j+\partial_{j}\xi_i+\xi^{k}\partial_{k}g_{ij}+f\dot{g}_{ij} \nn \\
		\delta N_{i}&=&\partial_{i}\xi^kN_k+\xi^k\partial_kN_{i}+\dot{\xi^{j}}g_{ij}+\dot{f}N_i+f \dot{N}_{i} \nn \\
		\delta N&=&\xi^k\partial_kN+\dot{f}N+f \dot{N}~.\nonumber
  \eea
Thus, as discussed in section \ref{subsec:HL}, because of the 4-dimensional diffeomorphism breaking, a third physical degree of freedom is present in both models.

In order to investigate the possible LIV one-loop quantum corrections to the matter sectors, we expand the metric $g_{\mu\nu}$ and, consequently, $e^\mu_{~\alpha}$ around a flat background:
  \bea 
		g_{\mu\nu} &=& \eta_{\mu\nu} + h_{\mu\nu}\\
		g^{\mu\nu} &=& \eta^{\mu\nu} -h^{\mu\nu}+h^{\mu\lambda}h_{\lambda}^\nu+\cdots~,\nonumber\\
		e^{~\alpha}_{\mu} &=& \delta^{\alpha}_{\mu} + \frac{1}{2} h^{\alpha}_{\mu} - \frac{1}{8} h_{\mu \lambda } 		h^{\lambda \alpha} + \cdots\nonumber\\
		e^{\mu}_{~\alpha} &=& \delta^{\mu}_{\alpha} - \frac{1}{2} h^{\mu}_{\alpha} + \frac{3}{8} h^{\mu \lambda } h_{\lambda \alpha} + \cdots ~,\nonumber
  \eea
where dots represent higher orders in fluctuations. We can then use the following relations regarding the 4-dimensional metric $g_{\mu\nu}$, the vierbein $e_\mu^{~\alpha}$ and the ADM components $N, N_i$ and $g_{ij}$
  \bea 
		g_{\mu\nu} &=& e_{\mu}^{~\alpha} e_{\nu}^{~  \beta} \eta_{\alpha  \beta}~,\\
		e_{\mu}^{~\alpha}e^{\mu}_{~  \beta}&=&\eta^{\alpha}_{  \beta}~,~~e_{\nu}^{~\alpha}e^{\mu}_{~\alpha}=					g^{\mu}_{\nu}~,\nonumber\\
		g_{\mu\nu} &=& -N^2 d^2 t + g_{ij}(dx^i+N^idt)(dx^j+N^jdt)~,\nonumber
  \eea 
to write everything in terms of the fluctuations of the ADM components 
  \bea\label{ADMfluc}
		N &=& 1+n\\
		N_i &=& n_i \nonumber\\
		g_{ij} &=& \delta_{ij} + h_{ij}~.\nonumber
  \eea
The fluctuations $n_i$ of the shift vector and the 3-dimensional metric $h_{ij}$ can be further decomposed into their different spin components as:
  \bea
		n_i &=& n_i^T + \partial_i \rho~,\label{nidec}\\
		h_{ij} &=& H_{ij}+ (\partial_i W_j+ \partial_j W_i ) + \left(\partial_{i}\partial_{j}-\frac{\delta_{ij}}{3}		\partial^2\right)B +\frac{\delta_{ij}}{3}h~,\label{hdecomp}
  \eea
where $H_{ij}$ is a transverse-traceless tensor, $n_i^T$ and $W_i$ are transverse vectors and $B$, $h$ and $\rho$ are scalar fields, $h$ being the trace of $h_{ij}$.

We can reduce the number of degrees of freedom in our theory, by making use of the gauge freedom shown in eq.(\ref{diffeo}). A possible gauge choice is to set the field fluctuations $W_i$ and $B$ to zero. Consequently, eq.(\ref{hdecomp}) becomes
  \be\label{hdec} 
		h_{ij}=H_{ij} +\frac{\delta_{ij}}{3} h~,
  \ee 
where $H_{ij}$ and $h$ represent the 3 physical degrees of freedom present in both gravity theories, while $n$, $n_i^T$ and $\rho$ are auxiliary fields only.

\subsection{Expanding the actions}\label{subsec:actionexp}

Being interested in one-loop corrections only, it is enough to expand the actions up to quadratic order in the ADM field fluctuations. The flat space metric, from eq.(\ref{ADMfluc}), is $\delta_{ij}$, such that, for simplicity, all the spatial indices can be lowered, {\it i.e.} $h^{ij} \to h_{ij} $. We have then, for example,
  \bea
		\sqrt{g} &=& 1 + \frac{1}{2} h + \frac{1}{8} (h^2 - 2h_{ij}h_{ij})+\cdots\\
		\Gamma_k &=& -\frac{\sigma_{ij}}{2} \left[  \partial_i h_{kj} - \frac{1}{2}\left(h_{il}\partial_l h_{jk} + 		h_{lj}\partial_i h_{kl} - \frac{1}{2}h_{il}\partial_k h_{jl}\right)\right]+\cdots~,\nonumber
  \eea
where $h=h_{ii}$, and dots represent higher orders in fluctuations which do not contribute to the present calculations and can be therefore omitted.

\subsubsection{Matter sector}

We show here which terms can and cannot be neglected in the matter actions for scalars and fermions, describe the ansatz taken for these external fields, and finally obtain the simplified version of the matter field actions that will be used in the following calculations.

The expansion of the scalar action (\ref{Ss1}) up to quadratic order in the graviton field fluctuations gives
  \bea\label{Ss2}
		S_s^{(2)} &=& -\int dt d^3 x~\left\{\left[1+n+\frac{h}{2}+\frac{hn}{2} +\frac{1}{8}(h^2-2h^2_{ij})\right](-		\dot{\phi} \dot{\phi}^\star +\partial_k \phi \partial_k \phi^\star)\right.\nn\\
		&&+\left. 2 n \dot{\phi}\dot{\phi}^\star+2n_i \dot{\phi}\partial_i \phi^\star-h_{ij}\partial_i\phi\partial_j	\phi^\star+2\left(\frac{h n_i}{2}-n n_i - n_j h_{ij}\right)\dot{\phi} \partial_i\phi^\star\right.\nonumber\\
		&&+\left. (nh-n^2)\dot{\phi} \dot{\phi}^\star +\left(h_{il}h_{lj}-n_i n_j -nh_{ij}-\frac{h h_{ij} }{2}\right)		\partial_i\phi\partial_j\phi^\star\right\}~.
  \eea
The terms in the first line which can only generate Lorentz-symmetric contributions are omitted from now on, since we are only interested in one-loop corrections leading to LIV effects. As pointed out in \cite{Pospelov}, quadratic terms in the metric fluctuations can only generate one-loop results when the graviton fields are contracted among themselves. Thus, for any tensor $T_{ij}$ quadratic in the graviton field, one can use the following simplification:  $T_{ij}\partial_i\phi\partial_j\phi^\star\to (T_{ii}/3)  \partial_k\phi\partial_k\phi^\star$, which takes into account the  rotational invariance in space. In addition, linear terms in the metric perturbations can be omitted because, since matter fields are classical, they cannot generate corrections to matter field kinetic terms. Finally, terms of the form $hn_i, nn_i$ or $n_jh_{ij}$ cannot contribute unless they are contracted with another vector metric fluctuations, but it would lead to cubic terms in fluctuations; therefore, we neglect such terms as well. Thus, the relevant part of the action, containing only terms which can generate Lorentz-violating contributions, is 
  \be\label{Ss3}
		S_s^{(2)} = -\int dt d^3 x~\left\{ (nh-n^2)\dot{\phi}\dot{\phi}^\star + \frac{1}{3}\left(h_{ij}^2-n_i^2 -nh-		\frac{h^2}{2}\right)\partial_k\phi\partial_k\phi^\star\right\}~.
  \ee
This action can be further simplified by writing 
	\be\label{ls}
		(nh-n^2)\dot{\phi}\dot{\phi}^\star =- (nh-n^2)\partial_\mu\phi\partial^\mu\phi^\star+(nh-n^2)\partial_k \phi 		\partial_k \phi^\star~,
  \ee
and then neglecting the Lorentz-symmetric term on the right-hand side, such that (\ref{Ss3}) becomes:
  \be\label{Ss4}
		\tilde S_s^{(2)} = -\frac{1}{3}\int dt d^3 x~\left[ h_{ij}^2-n_i^2- 3n^2+2nh-\frac{h^2}{2}\right]\partial_k		\phi\partial_k\phi^\star~.
  \ee

For the fermion sector, the metric expansion leads to
		\bea\label{Sf2'}
		S_f^{(2)} &=& -\frac{i}{2}\int dt d^3 x \left\{\left[1+n+\frac{h}{2}+\frac{hn}{2} +\frac{1}{8}(h^2-2h^2_{ij}	)\right]\bar{\psi} \gamma^{\mu}\overleftrightarrow{\partial_\mu}\psi\right.\\
		&&+\left. \left(n\delta_{ij} -\frac{1}{2}h_{ij}\right)\left( \bar{\psi}\gamma_i \overleftrightarrow{				\partial_j}\psi+\bar{\psi}\{\gamma_i,\Gamma_j^{(1)}\}\psi\right) +\bar{\psi} \{\gamma^\mu,\Gamma_\mu^{(1)}			\}\psi\right.\nonumber\\    
		&&+\left.\frac{1}{2}n_i\left[\bar{\psi}\left(\gamma_i\overleftrightarrow{\partial_0}-\gamma^0					\overleftrightarrow{\partial_i}+\{\gamma_i,\Gamma_0^{(1)}\}-\{\gamma^0,\Gamma_i^{(1)}\} \right) \psi  \right]
		+\bar{\psi}\{\gamma^\mu,\Gamma_\mu^{(2)}\}\psi \right.\nonumber\\
		&&+\left.\left(\frac{h n_i}{4}-\frac{3}{8} n_j h_{ij} - \frac{n n_i}{4}\right)\bar{\psi} \left(\gamma_i 		\overleftrightarrow{\partial_0}-\gamma^0\overleftrightarrow{\partial_i} \right)\psi 
		+\frac{h}{2} \bar{\psi}\{\gamma^\mu,\Gamma_\mu^{(1)}\}\psi+  \right.\nonumber\\
&&+\left. \frac{1}{8}n_i^2\bar{\psi} \gamma^0\overleftrightarrow{\partial_0}\psi +\left(\frac{3}{8}h_{il}h_{lj}-		\frac{h h_{ij}}{4}+\frac{n(h\delta_{ij}-h_{ij})}{2}-\frac{3}{8}n_in_j\right)\bar{\psi} 
		\gamma_i \overleftrightarrow{\partial_j}\psi  \right\}~.\nonumber
  \eea
As in the scalar field case, the terms on the first line can only contribute to Lorentz-symmetric corrections, hence they will be omitted. For the same reasons explained above, linear terms in the graviton fields as well as terms proportional to $\bar{\psi}\gamma_i\overleftrightarrow{\partial_0}\psi$ or  $\bar{\psi}\gamma_0\overleftrightarrow{\partial_i}\psi$ cannot contribute to the kind of corrections we are looking for and therefore will also be omitted from now on. Terms containing the anti-commutator of a $\gamma$ matrix and the spin connection cannot contribute as well, since they involve the contraction of a symmetric tensor with an antisymmetric one. Finally, in the same way as for the scalar action, terms of the form $T_{ij}\bar{\psi}\gamma_i\overleftrightarrow{\partial_j}\psi$, where $T_{ij}$ is quadratic in the graviton fluctuations, contribute to the one-loop LIV corrections and can be replaced by $(T_{ii}/3)(\bar{\psi}\gamma_k\overleftrightarrow{\partial_k}\psi)$. Thus, the Lorentz-violating fermion action becomes
{\footnotesize
  \bea\label{Sf3}
		S_f^{(2)} &=& -\frac{i}{2}\int dt d^3 x \left\{\frac{1}{8}n_i^2\bar{\psi} \gamma^0\overleftrightarrow{		\partial_0}\psi +\frac{1}{3}\left(\frac{3}{8}h_{ij}^2-\frac{h^2}{4}+n h -\frac{3}{8}n_i^2\right)\bar{\psi} 		\gamma_k \overleftrightarrow{\partial_k}\psi\right\}~,
  \eea}
which, after removing Lorentz-symmetric terms, as in (\ref{ls}), reduces to
  \be\label{Sf4}
		\tilde S_f^{(2)} =-\frac{i}{4}\int dt d^3 x \left[\frac{1}{4}h_{ij}^2-\frac{h^2}{6}-\frac{1}{2}n_i^2+				\frac{2}{3}n h\right]\bar{\psi} \gamma_k \overleftrightarrow{\partial_k}\psi~.
  \ee

After reducing the scalar and fermion actions to their simplest forms (\ref{Ss4}) and (\ref{Sf4}), we then consider a plane-wave as an ansatz for the external fields 
  \bea\label{ansatz}
		\phi(x)&=& \phi_0\exp(-ip^\mu x_\mu)\nn \\
		\psi(x)&=& \psi_0\exp(-iq^\mu x_\mu)~ 
  \eea
Thus,  the quantities $\partial_k \phi\partial_k \phi^\star$ and $\bar{\psi}i\gamma_k \overleftrightarrow{\partial_k}\psi$ become constants, which we replace by $(\vec p^2\phi_0^2)$ and $(-2i)[\bar{\psi}_0(\vec{\gamma}\cdot\vec q)\psi_0]$, respectively. Finally, with these classical background matter field configurations and using the field decompositions (\ref{nidec}) and (\ref{hdec}), the matter actions~(\ref{Ss4}) and~(\ref{Sf4}) can be expressed as 
  \bea\label{expm}
		\tilde S_s^{(2)} &=& -\frac{1}{3}\int dt d^3 x~\left[H_{ij}^2-\frac{h^2}{6}-(n_i^T)^2+\rho\partial^2\rho-			3n^2+2nh \right](\vec p^2\phi_0^2)~,\\
		\tilde S_f^{(2)} &=& -\frac{1}{2}\int dt d^3 x~\left[\frac{1}{4}H_{ij}^2-\frac{h^2}{12}-\frac{(n_i^T)^2}{2}+		\frac{1}{2}\rho \partial^2\rho+
		\frac{2}{3}nh\right][\bar{\psi}_0(\vec{\gamma}\cdot\vec q)\psi_0]~.\nonumber
  \eea

\subsubsection{Gravity actions} 

For the gravity actions, expanding~(\ref{SG}) and~(\ref{SHL}) up to quadratic order in the metric fluctuations, making use of the metric decompositions~(\ref{nidec}) and~(\ref{hdec}), and integrating by parts when necessary, we obtain 
  \bea\label{expg}
		S_G^{(2)} &=& M_P^2 \int dt d^3 x \left[ \frac{1}{4}H_{ij}(\partial^2-\partial_t^2)H_{ij} -\frac{1}{2}n_i^T 		\partial^2 n_i^T-(\lambda-1)\rho (\partial^2)^2\rho\right.\\
		&&+\left.\frac{(3\lambda-1)}{12}h \partial_t^2h -\frac{1}{18}h \partial^2h-\alpha n \partial^2 n+\frac{(3		\lambda-1)}{3}\rho \partial^2 \dot{h} - \frac{2}{3}n \partial^2 h\right]~,\nonumber
  \eea
and
  \bea\label{exphl}
		S_{HL}^{(2)} &=& M_P^2 \int dt d^3x \left\{ \frac{1}{4} H_{ij} \left[- \partial_t^2 +\partial^2 + F_1 (		\partial^2)^2 - S_1 (\partial^2)^3 \right] H_{ij}-\frac{1}{2} n_i^T \partial^2 n_i^T\right.\nn \\
		&& + \left. \frac{1}{18} h \left[\frac{3(3\lambda-1)}{2}\partial_t^2-\partial^2 +(3F_1+8F_2)(\partial^2)^2-		(3S_1+8S_2)(\partial^2)^3  \right] h\right.\nonumber\\
		&& - \left.  n[\alpha\partial^2+F_4 (\partial^2)^2+S_4 (\partial^2)^3] n - (\lambda-1)\rho (\partial^2)^2 		\rho +\frac{(3\lambda-1)}{3} \rho \partial^2 \dot{h}\right. \nonumber\\
		&& - \left. \frac{2}{3} h [\partial^2+F_3 (\partial^2)^2 +S_3 (\partial^2)^3] n\right\}~.
  \eea 
Finally, because ghosts do not couple to the matter sector at tree level, contributions related to these fields will only start to appear at two-loop calculations; hence, in the present work, ghosts can be omitted.

\section{One-loop matter effective actions}\label{sec:corrections}

In this section we obtain the effective matter kinetic terms by integrating over the graviton fluctuations. For both gravity models, we derive the Hamiltonian and momentum constraints and then impose them on the path integral. This approach, also used in \cite{Padilla}, implies that no conformal instability arises in our calculations \cite{Gibbons}. 

It is known in perturbative quantum gravity that, when introducing an irreducible decomposition for the metric, auxiliary components may present a propagator with the ``wrong'' sign, leading to a potential problem in defining the partition function. This unstable mode can be traced down to a conformal factor and can be understood as an artifact arising from perturbative expansion \cite{Mazur}. Nevertheless, this can be avoided by making an analytical continuation of the ``pathological'' metric components to imaginary values, simultaneously with the Wick rotation \cite{'tHooft}. With our gauge choice, this feature would arise from the integration of the auxiliary fields in the metric decomposition, but the use of the Hamiltonian and momentum constraints eliminate this problem, since after imposing such constraints we will be left with integrals over the physical fields only.

\subsection{Model I: modified Einstein-Hilbert gravity}

We study now the first model (\ref{SG}), for which one-loop quantum corrections are quadratically divergent and will be regularised with a cut off. After expansion of the relevant actions in terms of the metric fluctuations, the resulting expressions we are interested in are (\ref{expm}) and (\ref{expg}).

\subsubsection{ \underline{Constraints} }

Varying the actions with respect to the shift vector $(n_i^T,\rho)$ and the lapse function $(n)$, which are auxiliary fields, generate the momentum and Hamiltonian constraints which will be substituted back into the action.

Varying (\ref{expm}) and (\ref{expg}) with respect to $n$ leads to the the following constraint
  \be\label{n1constr}
		[-2\alpha M_P^2\partial^2 +2(\vec{p}^2\phi_0^2) ] n = \frac{2}{3}\left[M_P^2 \partial^2+ (\vec{p}^2\phi_0^2) 		+\frac{1}{2} [\bar{\psi}_0(\vec{\gamma}\cdot\vec q)\psi_0]\right]h~,
  \ee 
whereas variations with respect to the shift vector implies the two constraints below
  \bea  
		\left[2M_P^2(1-\lambda)\partial^2 -\frac{2}{3}\left((\vec{p}^2\phi_0^2) + \frac{3}{4}[\bar{\psi}_0(\vec{			\gamma}\cdot\vec q)\psi_0]\right)\right]\partial^2 \rho 
		&=& -\frac{M_P^2(3\lambda-1)}{3}\partial^2 \dot{h}~,\nn\label{rconstr}\\
		\left[-M_P^2\partial^2+\frac{2}{3}\left((\vec{p}^2\phi_0^2) + \frac{3}{4}[\bar{\psi}_0(\vec{\gamma}\cdot\vec 		q)\psi_0]\right)\right]n_i^T &=& 0~.\label{nTconstr}
  \eea  

When substituting the last constraint back into the actions, all contributions coming from $n_i^T$ disappear, and from now on such actions will only depend on the tensor and scalar components of the metric. On the other hand, since the auxiliary scalar fields $n$ and $\rho$ appear mixed with the scalar graviton $h$, we expand the constraints~(\ref{n1constr}) and (\ref{rconstr}) in terms of the matter contributions before putting them back into the actions
  \bea\label{nrconstr}  
		n&=& -\frac{1}{3\alpha}\left[1+\frac{(\vec{p}^2\phi_0^2)}{M_P^2}\left(\frac{\alpha+1}{\alpha}\right)					(\partial^2)^{-1} + \frac{[\bar{\psi}_0(\vec{\gamma}\cdot\vec q)\psi_0]}{2M_P^2}(\partial^2)^{-1}			\right]h+\cdots~,\label{nconstr1}\\
		\rho &=& \frac{(3\lambda-1)}{6(\lambda-1)}\left[1-\frac{(\vec{p}^2\phi_0^2) + \frac{3}{4}[\bar{\psi}_0(\vec{\gamma}\cdot\vec q)\psi_0]}{3(\lambda-1)M_P^2}(\partial^2)^{-1}\right](\partial^2)^{-1}\dot{h}+\cdots,\label{rconst}
  \eea  
where dots represent higher-order terms in the matter fields. Thus, with all constraints substituted back into the actions, the resulting action is	
  \bea\label{SmI}
		S^{(2)}_I &=&\int dt d^3x \left\{ \frac{1}{2} H_{ij} \left[ \frac{M_P^2}{2} (\partial^2- \partial_t^2) -			\frac{2}{3}(\vec p^2 \phi_0^2)- \frac{1}{4}[\bar{\psi}_0(\vec{\gamma}\cdot\vec q)\psi_0]\right] H_{ij}		\right.\\
		&& + \left. \frac{1}{2} h \left[ \frac{M_P^2}{9}\left(-X\partial_t^2+\left(\frac{2-\alpha}{\alpha}\right)		\partial^2\right)
		+\frac{(\vec{p}^2\phi_0^2)}{9}\left(\frac{\alpha^2+4\alpha+2}{\alpha^2}+\frac{X^2}{6}\frac{\partial_t^2}{		\partial^2}\right)\right.\right.\nonumber\\
		&& +\left.\left. \frac{[\bar{\psi}_0(\vec{\gamma}\cdot\vec q)\psi_0]}{9}\left(\frac{3\alpha+8}{4\alpha}+\frac{X^2}{8}\frac{\partial^2_t}{\partial^2}\right)\right] h\right\} \nonumber~,
  \eea 
where 
  \be\label{X} 
		X=\frac{3\lambda-1}{\lambda-1}~.
  \ee
For a consistent propagation of the scalar graviton $h$, one needs $X>0$, implying that the allowed values for $\lambda$ are
  \be\label{allowedlambda}
		\lambda<1/3~~\mbox{or}~~\lambda>1~.
  \ee
Finally, from the action (\ref{SmI}), the scalar graviton ($h$) dispersion relation is
  \be\label{hdrIR}
		\omega^2 = \left(\frac{\lambda -1}{3\lambda-1}\right) \left(\frac{2-\alpha}{\alpha}\right) \vec{k}^2~,
  \ee 
such that, when taking into account that the allowed values for $\lambda$ are the ones in eq.(\ref{allowedlambda}), it presents real energies for $0<\alpha<2$. This is consistent with the low-energy behaviour of the scalar graviton in the non-projectable version of HL gravity, as seen in section \ref{subsec:HL}.

\subsubsection{ \underline{Loop integration} }

We reserve the Appendix to provide the relevant details about the integration over the graviton components leading to the results discussed in this section.\\

\nin{\bf Spin-2 component}\\

The integration over the spin-2 component $H_{ij}$ gives 
  \be
		\mathcal{C}_{H_{ij}}^I = \exp\left\{-\frac{1}{M_P^2}\left[\frac{4}{3}(\vec p^2 \phi_0^2) + \frac{1}{2}[\bar{\psi}_0(\vec{\gamma}		\cdot\vec q)\psi_0]\right]
		\frac{\delta(0)}{2(2\pi)^2} \Lambda^2+\cdots\right\}~,
  \ee 
where $\delta(0)$ is the space-time volume, and dots represent either field-independent contributions or higher-order terms in $(\vec p^2 \phi_0^2)$ and $[\bar{\psi}_0(\vec{\gamma}\cdot\vec q)\psi_0]$.

\vspace{0.5cm}

\nin{\bf Spin-0 component}\\

Integrating over the spin-0 component $h$, we obtain 
  \bea 
	\mathcal{C}_{h}^I	&=&\exp\left\{\frac{1}{M_P^2}\left[\frac{\alpha^2+4\alpha+2}{2\alpha^2}(\vec p^2\phi_0^2)+\frac{3\alpha+8}{8		\alpha}[\bar{\psi_0}(\vec{\gamma}\cdot \vec q)\psi_0]\right.\right. \\
		&&\left.\left.-\frac{X^2}{4}\left(\frac{(\vec p^2\phi_0^2)}{3}+\frac{[\bar{\psi_0}(\vec{		\gamma}\cdot \vec q)\psi_0]}{4}\right)\right]Y\frac{\delta(0)\Lambda^2}{2(2\pi)^2} +\cdots \right\}~,\nn
  \eea
where
  \be\label{Y}
		Y = \sqrt{ \frac{\alpha(\lambda-1)}{(2-\alpha)(3\lambda-1)} }~.
  \ee

\subsubsection{ \underline{Total Lorentz-violating contributions}}

Considering the results obtained above, we put together here all contributions and write the total Lorentz-violating corrections for both scalar and fermion fields. We also note that
  \be\label{dpf}
		(\vec p^2 \phi_0^2) = \frac{1}{\delta(0)}\int dt d^3x~\partial_k \phi \partial_k \phi^\star~~~\mbox{and}~~~		[\bar{\psi}_0(\vec{\gamma}\cdot\vec q)\psi_0] = \frac{1}{\delta(0)}\int dt d^3x~\bar{\psi}i\gamma_k 		\partial_k\psi~.
  \ee 

According to the results (\ref{resHI}) and (\ref{reshI}), the total contributions to the matter fields can be written as 
  \be\label{sresI}  
		S^I_{LIV}(\phi)=\frac{1}{2(2\pi)^2}\frac{\Lambda^2}{M_P^2}\left[\frac{4}{3}+ \frac{Y}{2}\left(\frac{X^2}{6}-\frac{\alpha^2		+4\alpha+2}{\alpha^2}\right)\right]
		\partial_k \phi\partial_k \phi^\star~
  \ee
in the scalar case, and 
  \be\label{fresI}  
		S^I_{LIV}(\psi) =\frac{1}{2(2\pi)^2}\frac{\Lambda^2}{M_P^2}\left[\frac{1}{2}-\frac{Y}{8}\left(\frac{3\alpha+8}{\alpha}-				\frac{X^2}{2}\right)\right]\bar{\psi}i\gamma_k\partial_k\psi~,
  \ee
in the fermion case, with the definitions (\ref{X}) and (\ref{Y}) for $X$ and $Y$, respectively.

\subsection{Model II: non-projectable Horava-Lifshitz gravity}\label{sec:mii}

We turn here to the non-projectable version of Horava-Lifshitz gravity~(\ref{SHL}) coupled to matter fields, for which the relevant actions are given by eqs.(\ref{expm}) and (\ref{exphl}). 

In the absence of matter, due to the inclusion of higher derivative operators, one-loop quantum corrections for this model diverge logarithmically, but in the presence of dynamical matter, it has been shown in \cite{Pospelov,Padilla} that quadratic divergences arise from the coupling gravity-bosonic matter. In the present case, since matter is classical and thus does not involve any new loop momentum, it cannot induce further divergences compared to the single gravity case. However, the use of Hamiltonian and momentum constraints generate an artificial quartic divergence, as a result of the introduction of additional space derivatives in the decompositions (\ref{nidec}) and (\ref{hdecomp}) of the graviton: as can be seen from the action (\ref{SmII}) below, the presence of matter is then accompanied by the derivative operator $\partial_t^2/\partial^2$, arising from the coupling between $\rho$ and $h$ in the gravity sector. This divergence is thus a gauge artifact, on which we will give more details in section \ref{sec:reg}.

As a consequence, it is natural to use dimensional regularisation, where the naively power-law divergent integrals actually vanish \cite{Leibbrandt}. We note that in some cases the vanishing or finiteness of a regularised integral which otherwise would naively be divergent can be explained in the following way: in the regularised integral, divergences associated to different regions of the domain of integration cancel each other, such that the integral is finite when the regulator is removed \cite{Weinzierl}.

\subsubsection{ \underline{Constraints} }

The momentum constraints in (\ref{rconstr}), obtained from the variation of the action with respect to the shift vector, are the same as in the previous modified gravity model because no higher derivative operators depending on $\rho$ and $n_i^T$ are added to the action. On the other hand, the additional contributions to the lapse function fluctuations $n$ lead to a new Hamiltonian constraint
  \be 
		[-2M_P^2 \mathcal{D}_2 + 2 (\vec{p}^2\phi_0^2)]n =\frac{2}{3}\left[M_P^2 \mathcal{D}_1 + (\vec{p}^2\phi_0^2) + \frac{1}{2}[\bar{		\psi}_0(\vec{\gamma}\cdot\vec q)\psi_0] \right] h ~,
  \ee 
which can be written as
  \be\label{nconstr}
		n = -\frac{1}{3}\left[\frac{\mathcal{D}_1}{\mathcal{D}_2} + \frac{(\vec{p}^2 \phi_0^2)}{M_P^2}\frac{1}{\mathcal{D}_2}\left(1+\frac{		 \mathcal{D}_1}{\mathcal{D}_2}\right) + \frac{1}{2M_P^2}[\bar{\psi}_0(\vec{\gamma}\cdot\vec q)\psi_0]\frac{1}{\mathcal{D}_2} \right]    h+\cdots~,
  \ee 
where dots represent higher-order terms in $[\bar{\psi}_0(\vec{\gamma}\cdot\vec q)\psi_0]$ and $(\vec{p}^2\phi_0^2)$, and
  \bea 
		\mathcal{D}_1 &=& [\partial^2+ F_3 (\partial^2)^2+S_3 (\partial^2)^3]~,\\
		\mathcal{D}_2 &=& [\alpha\partial^2+ F_4 (\partial^2)^2+S_4 (\partial^2)^3]~.\nonumber
  \eea 
We can now use the constraints (\ref{rconstr}) and (\ref{nconstr}) to rewrite the original actions~(\ref{expm}) and~(\ref{exphl}) and arrive at the expression below, which only depends on the physical metric fluctuations $H_{ij}$ and $h$,
  {\footnotesize
	\bea\label{SmII} 
		S^{(2)}_{II} &=&\int dt d^3x \left\{ \frac{1}{2} H_{ij} \left[ \frac{M_P^2}{2} (- \partial_t^2 +\partial^2 + F_1 (\partial^2)^2 - 	 S_1 (\partial^2)^3 ) -\frac{2}{3}(\vec p^2 \phi_0^2)- \frac{1}{4}[\bar{\psi}_0(\vec{\gamma}\cdot\vec q)\psi_0]\right] H_{ij}					\right.\nonumber \\
		&& + \left. \frac{1}{2} h \left[ \frac{M_P^2}{9}\left(-X\partial_t^2-\partial^2+(3F_1+8F_2)(\partial^2)^2-(3S_1+8S_2)(\partial^2)^3
		+2\left(\frac{\mathcal{D}_1^2}{\mathcal{D}_2}\right)\right)\right.\right.\nonumber\\
		&&+\left.\left. \frac{(\vec{p}^2\phi_0^2)}{9}\left(1+4\left(\frac{\mathcal{D}_1}{\mathcal{D}_2}\right)+2\left(\frac{\mathcal{D}_1}{		\mathcal{D}_2}\right)^2
		+\frac{X^2}{6}\frac{\partial_t^2}{\partial^2}\right)\right.\right.\nonumber\\
		&& +\left.\left. \frac{[\bar{\psi}_0(\vec{\gamma}\cdot\vec q)\psi_0]}{9}\left(\frac{3}{4}+2\left(\frac{\mathcal{D}_1}{				\mathcal{D}_2}\right)+\frac{X^2}{8}\frac{\partial^2_t}{\partial^2}\right)\right] h\right\}~.
  \eea}

\subsubsection{ \underline{Loop integration} }

As for the first model, we give in the Appendix the details of the integration over graviton, which, for the present model, is done using dimensional regularisation, with $d=3-\epsilon$.\\

\nin{\bf Spin-2 component}\\
The integration over the spin-2 component $H_{ij}$ gives
  \be 
\mathcal{C}_{H_{ij}}^{II}=\exp\left\{-\frac{1}{M_P^2}\left[\frac{4}{3}(\vec p^2 \phi_0^2) + \frac{1}{2}[\bar{\psi}_0(\vec{\gamma}\cdot\vec q)\psi_0]\right]
\frac{\delta(0)}{(2\pi)^2\sqrt{|S_1|}}\frac{\mu^\epsilon}{\epsilon}+\cdots\right\}~.
  \ee 

\vspace{0.5cm}

\nin{\bf Spin-0 component}\\

Finally, integrating over the spin-0 component $h$, we find
  \bea 
	\mathcal{C}_{h}^{II}	&=& \exp\left\{\frac{1}{M_P^2}\left[\frac{1}{2\sqrt{C_6^{(0)}}}\left((\vec p^2 \phi_0^2)+\frac{3}{4}[\bar{\psi}_0(\vec{\gamma}		\cdot\vec q)\psi_0]\right)+\frac{(\vec p^2 \phi_0^2)}{\sqrt{C_6^{(2)}}}\right.\right.\\
		&&\left.\left.+\frac{1}{\sqrt{C_6^{(1)}}}\left(2(\vec p^2 \phi_0^2)+[\bar{\psi}_0(\vec{\gamma}\cdot				\vec q)\psi_0]\right)\right]\frac{\delta(0)}{(2\pi)^2\sqrt{X}}\frac{\mu^\epsilon}{\epsilon}+\cdots\right\}\nonumber~,
  \eea
where the constants $C_6^{(n)}$, with $n=0,1,2$, are also given in the Appendix.

\subsubsection{ \underline{Total Lorentz-violating contributions} }\label{sec:miires}

As can be seen in the Appendix, the following integral appears repeatedly when calculating the LIV corrections 
  \be\label{intd}
		\mathcal{I}\left(\Delta\right)=\int \frac{d^d k}{(2\pi)^d} \frac{1}{\vec{k}^2\sqrt{\vec{k}^2+ \Delta }}~.
  \ee
Solving the integral above with dimensional regularisation ($d=3-\epsilon$), we find
  \be\label{Isol}  
		\mathcal{I}(\Delta) = \frac{1}{2 \pi^2} \frac{\mu^\epsilon}{\epsilon}+\mathcal{O}(\epsilon)~.
  \ee 
Thus, in the limit $\epsilon\to0$, we obtain
  \be 
		\mu \frac{\partial}{\partial \mu} \mathcal{I}(\Delta)= \frac{1}{2 \pi^2}~,
  \ee 
such that 
  \be\label{Isol2}
		\mathcal{I}(\Delta) = \frac{1}{2 \pi^2} \ln\left(\frac{\mu}{\mu_0}\right)~,
  \ee 
where $\mu_0$ is a mass scale. 

In order to choose $\mu_0$ accordingly, we calculate the same integral $\mathcal{I}(\Delta)$ using a cut off $\Lambda$ in dimension $d=3$ and find:
  \be\label{Ico} 
		\mathcal{I}(\Delta) = \frac{1}{2\pi^2}\ln\left(\frac{\Lambda+\sqrt{\Lambda^2 +\Delta}}{\sqrt{\Delta}}\right)~.
  \ee 
From the form of $\Delta$ in~(\ref{trH}) and (\ref{3dint}), we note that it has the following generic form $a M_{HL}^2$, where $a$ represents a dimensionless constant of order 1 for each of the different cases. Then, we expand~(\ref{Ico}) for $\Lambda\gg M_{HL}$ to find
  \be 
		\mathcal{I}(\Delta) = \frac{1}{2\pi^2} \ln\left(\frac{\Lambda}{M_{HL}}\right) ~,
  \ee 
where finite terms were omitted in the expression above. Comparing eq.(\ref{Isol2}) with the result above, it is natural to choose $\mu_0 = M_{HL}$ and $\mu=\Lambda$.\\

In this way, with the results obtained above, we can finally write the total Lorentz-violating contributions for both scalar and fermion fields. Using the relations (\ref{dpf}) and assuming~(\ref{Isol2}) with $\mu_0=M_{HL}$, the total LIV one-loop corrections for scalar and fermion fields are, respectively,

{\scriptsize
  \bea \label{sfresII}
		S^{II}_{LIV}(\phi) &=& \left[-\frac{1}{2(2\pi)^2}\left(\frac{4}{3}\frac{1}{\sqrt{|s_1|}} - \frac{1}{2}\frac{1}{\sqrt{X c_6^{(0)}}} -\frac{2}{\sqrt{X c_6^{(1)}}}-\frac{1}{\sqrt{X c_6^{(2)}}}  \right)\frac{M_{HL}^2}{M_P^2} \ln\left(\frac{M_{HL}^2}{\Lambda^2 }\right)\right] \partial_k \phi 		\partial_k \phi^\star~,\nonumber\\
		S^{II}_{LIV}(\psi) &=& \left[-\frac{1}{2(2\pi)^2}\left(\frac{1}{2}\frac{1}{\sqrt{|s_1|}} - \frac{3}{8}\frac{1}{\sqrt{X c_6^{(0)}}} -\frac{1}{\sqrt{X c_6^{(1)}}} \right) \frac{M_{HL}^2}{M_P^2}\ln\left(\frac{M_{HL}^2}{\Lambda^2 }\right)\right]\bar{\psi} i \gamma_k \partial_k \psi~.
		\eea}

\subsection{Non-minimal coupling}

As the models studied here involve minimal couplings between matter and gravity, one could ask what effects could non-minimal couplings have on the effective dispersion relations. In this subsection we show why we do not expect non-minimal couplings to change our results. For sake of simplicity, we focus on the second model only, which allows for more possibilities. Nonetheless, similar arguments can be easily applied to the first model as a particular case.

We start with the scalar field. Given that the terms in the action cannot have mass dimension greater than 6, and that the scalar field is dimensionless for $z=3$ in $d=3$ space dimensions, its non-minimal coupling to gravity would contain, for example, the following terms
  \be
		\left[\xi_1 R^{(3)}+\xi_2(R^{(3)})^2+\xi_3(a_i a^i)\right](\phi\phi^\star)~,
  \ee
where the coefficients $\xi_i$ have the correct mass dimension for the terms inside the square bracket to be of dimension 6. Similarly, for the fermion of mass dimension 3/2, we could have terms like
  \be
		\left[\zeta_1 R^{(3)}+\zeta_2(a_i a^i)\right](\ol\psi\psi)~.
  \ee
In both cases, after integration by parts, the space derivatives appearing in $R^{(3)}$ and $a_i$ could become space derivatives with respect to the matter fields, and thus naively contribute to the effective dispersion relation. However, the latter is obtained with matter plane wave configurations, such that $\phi\phi^\star$ and $\ol\psi\psi$ are actually constants and therefore do not give additional contributions to the dispersion relation. If a more general field configuration was chosen, it would also contribute to all the other terms calculated here, in such a way that the final dispersion relation would not change: the functional for matter fields obtained after integrating gravitons is unique, and the corresponding dispersion relations are obtained by plugging a plane wave solution.

The conclusion above is valid at one-loop though, as non-minimal coupling can radiatively generate terms which modify the matter kinetic terms, with an impact on the dispersion relation at higher-order loops.

\subsection{Comments on regularisation}\label{sec:reg}

All integrals in this chapter have been calculated by first integrating over frequency and then over momentum. For most of the integrals, the integration over frequency is finite, and has been performed without any regularisation. We are then left with a 3-dimensional integration over momentum, which is regularised either with a cut off or with dimensional regularisation. But there is also the situation where the integration over frequency is divergent, and we discuss here few details for both models.\\

\nin {\bf Model I }

The first model, as previously discussed, involves quadratic divergences and therefore should not be treated with dimensional regularisation, which sees only logarithmic divergences. A typical example which appears in our calculations where the integration over frequencies is finite is
  \be 
		\int \frac{d^4k}{(2\pi)^4} \frac{1}{\omega^2+z^2\vec{k}^2} =\frac{1}{4\pi^3} \int_0^\Lambda dk\int_{-\infty}^{+\infty} d\omega 		 	\frac{\vec{k}^2}{\omega^2+z^2\vec{k}^2}=\frac{\Lambda^2}{8\pi^2z}~.\\
  \ee
where $z$ is a positive dimensionless constant.

However, we also find integrals for which the integration over frequencies is divergent, and these are calculated with the same cut off as the one used for momentum integration in the other integrals. Such integrals take the form
  \bea
		\int \frac{d^4k}{(2\pi)^4} \frac{\omega^2}{\vec{k}^2(\omega^2+z^2\vec{k}^2)}
		&=&\frac{1}{4\pi^3} \int_0^{\infty} dk\int_{-\Lambda}^{+\Lambda} d\omega \frac{\omega^2}{\omega^2+z^2\vec{k}^2}\\
		&=&\frac{1}{4\pi^3}\int_0^{\infty} dk\int_{-\Lambda}^{+\Lambda} d\omega \left[1-\frac{z^2\vec{k}^2}{\omega^2+z^2\vec{k}^2}\right]		\nn \\
		&=&\frac{1}{2\pi^3}\int_0^{\infty} dk\left[\Lambda- zk \arctan\left(\frac{\Lambda}{z k}\right)\right]\nn \\
		&=&\frac{\Lambda^2}{8\pi^2z}\nonumber~.
  \eea
It is worth noting that, although the commutativity of the order of integration is not obvious when the integrals are divergent, the same result can be obtained if one performs first the finite integration over momentum, and then uses the cut off for frequency.\\

\nin {\bf Model II}

For the second model, which is logarithmic divergent, artificial quartic divergences also appear as a result of the graviton decomposition (\ref{hdecomp}) in terms of auxiliary fields. The artificial divergences are associated with integrals of the form:
  \bea
		\int \frac{d^4k}{(2\pi)^4} \frac{\omega^2}{\vec{k}^2(\omega^2+z^6\vec{k}^6)} &=& \frac{1}{4\pi^3}
		\int_0^{\infty} dk\int_{-\Lambda^3}^{+\Lambda^3} d\omega \left[1-\frac{z^6\vec{k}^6}{\omega^2+z^6\vec{k}^6}\right]\\
		&=&\frac{1}{2\pi^3}\int_0^{\infty} dk \left[\Lambda^3-z^3k^3 \arctan\left(\frac{\Lambda^3}{z^3k^3}\right)  \right]\nn \\
		&=&\frac{\Lambda^{4}}{8\pi^2z}\nonumber~,
  \eea
where we regularised the integral over frequency with the cut off $\Lambda^3$, since $[\omega]=3$ in $z=3$ HL gravity. As explained at the beginning of section \ref{sec:mii}, this divergence is artificial and is thus omitted. The other integrals for this model behave as expected, they involve a finite integration over frequency and a logarithmically divergent momentum integral. As shown in the Appendix, the latter is regularised with dimensional regularisation, but, as explained in section \ref{sec:miires}, if we use a cut off to calculate such integrals instead, the same logarithmic divergent results are found. 

\section{Analysis}\label{sec:analysis}

The measurable deviation from the Lorentz-symmetric case which cannot be scaled away by field or coordinate redefinitions is the difference between the propagation speed of the massless scalar and fermion fields: $|v_s^2-v_f^2|$, as pointed out in \cite{Pospelov}. This quantity, according to the current upper bounds on Lorentz-symmetry violation \cite{bounds}, should be typically smaller than $10^{-20}$. We note here that \cite{Pospelov} consider dynamical matter fields, instead of classical ones, which implies in the cancellation of the above mentioned quartic divergence, as expected for a gauge artifact. Indeed, the graviton loop giving rise to this divergence is cancelled by the equivalent matter loop contribution. In our case though, this cancellation does not take place because the matter loop is not present. This suggests the non-trivial fact that a classical matter background can be consistently considered only if one removes this specific gauge artifact by hand.

When taking into account the one-loop results obtained in previous sections, we find the effective kinetic terms
  \bea
		&&-i\bar{\psi} \gamma^0 \partial_t\psi+v_f^{m} i\bar{\psi}\gamma_k \partial_k\psi\\
		&&-\partial_t\phi\partial_t\phi^\star+(v_s^{m})^2\partial_k\phi\partial_k\phi^\star~,\nonumber
  \eea
where $v_f^{m} = 1+\delta v_f^{m}$ and $(v_s^{m})^2=1+(\delta v_s^{m})^2$ with $m =I$ for the first model (\ref{SG}) and $m = II$ for the second one (\ref{SHL}).

\subsection{Model I}

For the first model, we identify $(\delta v_s^{I})^2$ and $\delta v_f^{I}$ with the results in eqs.(\ref{sresI}) and (\ref{fresI}), respectively, {\it i.e.}
  \bea
		(\delta v_s^I)^2 &\simeq& \frac{1}{2(2\pi)^2}\left[\frac{4}{3}+\frac{Y}{2}\left(\frac{X^2}{6}-\frac{\alpha^2+4\alpha+2}{\alpha^2}		\right)\right]\frac{\Lambda^2}{M_P^2}~,\\
		(\delta v_f^I)^2&\simeq& 2 \delta v_f^I \simeq  \frac{1}{2(2\pi)^2}\left[1-\frac{Y}{4}\left(\frac{3\alpha+8}{\alpha}-\frac{X^2}{2}		\right)\right]\frac{\Lambda^2}{M_P^2}~.
  \eea
Subtracting the fermion contribution from the scalar one, we finally obtain the measurable departure from Lorentz symmetry
  \bea\label{finalresI}
		&&|(\delta v_s^I)^2-(\delta v_f^I)^2| = \frac{1}{2(2\pi)^{2}}\left|F(\lambda,\alpha)\right|\frac{\Lambda^2}{ M_P^2}~, 						 ~~~\mbox{where}\\
		&&F(\lambda,\alpha) = -\frac{1}{3}+\frac{1}{4}\sqrt{\frac{\alpha(\lambda-1)}{(2-\alpha)(3\lambda-1)}}\left[\frac{(3\lambda-1)^2}{6(		\lambda-1)^2}-\frac{\alpha^2-4}{\alpha^2}\right].\nonumber
  \eea
	
If we do not impose any specific values for $\lambda$ and $\alpha$, $F(\lambda,\alpha)$ is generically of order $1$, and the current experimental bounds on Lorentz violation are satisfied with the difference (\ref{finalresI}) if the graviton loop momentum is cut off by $\Lambda\lesssim10^{10}$ GeV. 

Nevertheless, it is easy to see that there are specific values for $\lambda$ and $\alpha$ such that the difference (\ref{finalresI}) vanishes. An example is 
  \be
F(\lambda_0,\alpha_0)=0 ~~~\mbox{for}~~~\lambda_0 \simeq 0.332~~~ \mbox{and}~~~ \alpha_0 \simeq 1.995~,
  \ee
which is consistent with the allowed values for these parameters. These specific values for $\lambda$ and $\alpha$ have been chosen here because they are close to the boundary values $\lambda_b=1/3$ and $\alpha_b=2$, which may suggest that quantum corrections point towards this specific point in parameter space. This idea seems to be supported in \cite{Dodorico} where $\lambda_b=1/3$ is found to be an IR fixed point for the Wilsonian renormalisation flow.

\subsection{Model II}

For the second model, the results found in~(\ref{sfresII}) can be related to $(\delta v_s^{II})^2$ and $\delta v_f^{II}$ by
 {\footnotesize 
	\bea   
		(\delta v_s^{II})^2 &\simeq& -\frac{1}{2(2\pi)^2}\left(\frac{4}{3}\frac{1}{\sqrt{|s_1|}} - \frac{1}{2}\frac{1}{\sqrt{X c_6^{(0)}}} 
		-\frac{2}{\sqrt{X c_6^{(1)}}}-\frac{1}{\sqrt{X c_6^{(2)}}}  \right) \frac{M_{HL}^2}{M_P^2} \ln\left(\frac{M_{HL}^2}{\Lambda^2 }		\right)~,\nonumber\\
		(\delta v_f^{II})^2 &\simeq& 2\delta v_f^{II} \simeq -\frac{1}{2(2\pi)^2}\left(\frac{1}{\sqrt{|s_1|}} - \frac{3}{4}\frac{1}{\sqrt{		X c_6^{(0)}}} 
		-\frac{2}{\sqrt{X c_6^{(1)}}} \right) \frac{M_{HL}^2}{M_P^2} \ln\left(\frac{M_{HL}^2}{\Lambda^2 }\right)~.
  \eea}
Consequently, the difference between these two contributions is
  {\footnotesize
	\bea\label{finalresII} 
		|(\delta v_s^{II})^2 - (\delta v_f^{II})^2| \simeq\left|\frac{1}{2(2\pi)^2}\left(\frac{1}{3}\frac{1}{\sqrt{|s_1|}} 
		+ \frac{1}{4}\frac{1}{\sqrt{X c_6^{(0)}}} -\frac{1}{\sqrt{X c_6^{(2)}}}  \right) \frac{M_{HL}^2}{M_P^2} \ln\left(\frac{M_{HL}^2}{		\Lambda^2 }\right)\right|~.
  \eea}
	
If we assume now that $\Lambda \sim M_P$ and $s_1$, $c_6^{(n)}$ are of order 1, the bounds on Lorentz violation~\cite{bounds} are then satisfied for $M_{HL}\lesssim10^{10}$ GeV. Note that, by construction of HL gravity, $M_{HL}$ should also be large enough to suppress higher space derivatives in the IR. The fact that quantum corrections lead to an upper bound on $M_{HL}$ follows from imposing that the UV regime, above $M_{HL}$, dominate the loop integrals ``early'' enough, for quantum corrections not to be too large.

Finally, similarly to what we have found for the first model, in the present case it is possible that the effective LIV correction~(\ref{finalresII}) vanishes if the coupling constants are chosen accordingly. For instance, this happens when setting $s_1$, $c_6^{(n)}$ to 1, and taking $\lambda\approx 1.969$, which is in the allowed regime for this parameter.

\section*{Appendix: loop calculations}

\subsection*{ \underline{Model I} }

\nin{\bf Spin-2 component}

From the action (\ref{SmI}) we consider only the terms depending on $H_{ij}$ to obtain 
  \bea
		\tilde S_I^{(2)}(H_{ij}) = \int dt d^3 x~ \frac{1}{2}H_{ij}\left[ \frac{M_P^2}{2} (\partial^2-\partial_t^2) -\frac{2}{3}(\vec p^2 		\phi_0^2) - \frac{1}{4}[\bar{\psi}_0(\vec{\gamma}\cdot\vec q)\psi_0]\right]H_{ij}~.
  \eea
The action above in terms of the Fourier transform $\tilde H_{ij}(k)$ of $H_{ij}(x)$, after a Wick rotation ($t\to it$, $\omega\to-i\omega$), can be expressed as
  {\footnotesize 
	\bea
		 &&\int \mathcal{D}H_{ij} \exp\left\{i \tilde S_{I}^{(2)}(H_{ij})\right\}\to\int \mathcal{D} \tilde{H}_{ij}\exp\left\{-\int \frac{		d^4 k_1d^4 k_2}{(2\pi)^8}
		~\frac{1}{2}\tilde{H}_{ij}(k_2) \left[\mathcal{A}^{(I)}_{\tilde H_{ij}} \right]  \tilde{H}_{ij}(k_1)\right\}, \nonumber\\
		&& \mbox{where}~~ \mathcal{A}^{(I)}_{\tilde H_{ij}} = \left[ \frac{M_P^2}{2} k_1^2 +\frac{2}{3}(\vec p^2 \phi_0^2)
		+ \frac{1}{4}[\bar{\psi}_0(\vec{\gamma}\cdot\vec q)\psi_0]\right]\delta(k_1+k_2)
  \eea}
with $k_1^2=\omega_1^2+\vec{k}_1^2$ in Euclidean space. Then, we perform the functional integration, taking into account that $\tilde{H}_{ij}$ has two components, to find
  \bea\label{resHI}
		&&\left\{\det \left[\frac{1}{(2\pi)^8} \left(\frac{M_P^2}{2} k_1^2 +\frac{2}{3}(\vec p^2 \phi_0^2)+ 
		\frac{1}{4}[\bar{\psi}_0(\vec{\gamma}\cdot\vec q)\psi_0] \right)\delta(k_1+k_2) \right]\right\}^{-1} \\
		&=& \exp\left\{-\mbox{Tr} \ln\left[\frac{1}{(2\pi)^8}\left( \frac{M_P^2 k_1^2}{2} +\frac{2}{3}(\vec p^2 \phi_0^2) 
		+ \frac{1}{4}[\bar{\psi}_0(\vec{\gamma}\cdot\vec q)\psi_0]\right) \right]\delta(k_1+k_2)\right\}\nonumber\\
		&=& \exp\left\{-\frac{1}{M_P^2}\left[\frac{4}{3}(\vec p^2 \phi_0^2) + \frac{1}{2}[\bar{\psi}_0(\vec{\gamma}\cdot\vec q)\psi_0]\right]
		\mbox{Tr}\left(\frac{\delta(k_1+k_2)}{k_1^2}\right)+\cdots\right\}~,\nonumber
  \eea 
where dots represent field-independent terms or higher orders in $(\vec p^2 \phi_0^2)$ and $[\bar{\psi}_0(\vec{\gamma}\cdot\vec q)\psi_0]$. The trace is finally calculated using a momentum cut off $\Lambda$ which leads to
  \be\label{trace}
		\mbox{Tr}\left(\frac{\delta(k_1+k_2)}{k_1^2}\right) = \int \frac{d^4 k_1d^4 k_2}{(2\pi)^8}\left(\frac{\delta(k_1+k_2)}{k_1^2}				\right)\delta(k_1+k_2)= \frac{\delta(0)}{2(2\pi)^2} \Lambda^2~,
  \ee 
where $\delta(0)$ is the space-time volume.

\vspace{0.5cm}

\nin{\bf Spin-0 component}

For the scalar graviton we follow similar steps. We start by considering the terms which depend on $h$ in the action (\ref{SmI}), write them in terms of the Fourier transform $\tilde h$ of $h$ and perform a Wick rotation, to obtain
  \bea
		\int \mathcal{D}h \exp\left\{i S_{I}^{(2)}(h)\right\}\to\int \mathcal{D} \tilde{h}\exp\left\{-\int \frac{d^4 k_1}{(2\pi)^4}					\frac{d^4 k_2}{(2\pi)^4}~\frac{1}{2}\tilde{h}(k_2) \left[\mathcal{A}^{(I)}_{\tilde h}\right]\tilde{h}(k_1)\right\}~,
  \eea 
where
  \bea\label{AhI} 
		\mathcal{A}^{(I)}_{\tilde h} &=& \frac{1}{9} \left\{M_P^2\left(X\omega_1^2+\frac{2-\alpha}{\alpha}\vec{k}_1^2\right)+(\vec p^2 \phi_0^2)\left[\frac{X^2}{6}\frac{\omega_1^2}{\vec{k}_1^2}-\frac{\alpha^2+4\alpha+2}{\alpha^2}\right]\right.\nonumber\\
		&&+\left.[\bar{\psi}_0(\vec{\gamma}\cdot\vec q)\psi_0]\left[\frac{X^2}{8}\frac{\omega_1^2}{\vec{k}_1^2}-\frac{3\alpha+8}{4\alpha}	\right]\right\}\delta(k_1+k_2)~.
  \eea 
Evaluating the functional integral, we have
{\footnotesize
  \bea \label{reshI}
		&&\exp\left\{\frac{1}{M_P^2}\left[\left(\frac{\alpha^2+4\alpha+2}{2\alpha^2}(\vec p^2\phi_0^2)+\frac{3\alpha+8}{8\alpha}
		[\bar{\psi_0}(\vec{\gamma}\cdot \vec q)\psi_0]\right)\mbox{Tr}\left(\frac{\delta(k_1+k_2)}{X\omega_1^2+\alpha^{-1}(2-\alpha)				\vec{k}_1^2}\right)\right.\right.\nonumber\\
		&&-\left.\left.\frac{X^2}{4}\left(\frac{(\vec p^2\phi_0^2)}{3}+\frac{[\bar{\psi_0}(\vec{\gamma}\cdot \vec q)\psi_0]}{4}\right)
		\mbox{Tr}\left(\frac{\delta(k_1+k_2)\omega_1^2}{\vec{k}_1^2 \left(X\omega_1^2+\alpha^{-1}(2-\alpha)\vec{k}_1^2 \right)}\right) 			\right] \right\}~.
  \eea}
Finally, we evaluate both traces above with the common cut off $\Lambda$ for frequencies and wave vectors, leading to the same result  
  \bea 
		&&Tr\left[\frac{\delta(k_1+k_2)\omega_1^2}{\vec{k}_1^2 \left(X\omega_1^2+\alpha^{-1}(2-\alpha)\vec{k}_1^2 \right)}\right]
		=Tr\left[\frac{\delta(k_1+k_2)}{ \left(X\omega_1^2+\alpha^{-1}(2-\alpha)\vec{k}_1^2 \right)}\right]\\
		&& = Y\frac{\delta(0)\Lambda^2}{2(2\pi)^2}~~~~~\mbox{with}~~~~~Y = \sqrt{ \frac{\alpha(\lambda-1)}{(2-\alpha)(3\lambda-1)} }~.
  \eea

\subsection*{ \underline{Model II} }

\nin{\bf Spin-2 component}

When comparing the terms which depend on $H_{ij}$ in the expressions~(\ref{SmI}) and (\ref{SmII}), we see that the only difference is the presence of the higher-order space derivatives in the propagator of the tensor field in the non-projectable HL case. Thus, the integration over the spin-2 component of the metric is analogous to~(\ref{resHI}) when $k_1^2=\omega^2+\vec{k}^2_1$ (in Euclidean space) is replaced by $\omega_1^2+\vec{k}_1^2-F_1(\vec{k}_1^2)^2-S_1(\vec{k}_1^2)^3$. Assuming that $F_1,S_1<0$, we obtain 
{\footnotesize
  \be\label{resHII} 
\exp\left\{-\frac{1}{M_P^2}\left[\frac{4}{3}(\vec p^2 \phi_0^2) + \frac{1}{2}[\bar{\psi}_0(\vec{\gamma}\cdot\vec q)\psi_0]\right]
\mbox{Tr}\left(\frac{\delta(k_1+k_2)}{\omega_1^2+\vec{k}_1^2+|F_1|(\vec{k}_1^2)^2+|S_1|(\vec{k}_1^2)^3}\right)+\cdots\right\}~.
  \ee}
Let us calculate the trace above, starting by integrating over the frequencies
  \bea \label{trH}
\mbox{Tr}\left(\frac{\delta(k_1+k_2)}{\omega_1^2+\vec{k}_1^2+|F_1|(\vec{k}_1^2)^2+|S_1|(\vec{k}_1^2)^3}\right) &=& \frac{\delta(0)}{2} \int \frac{d^3k}{(2\pi)^3} 
\frac{1}{\sqrt{\vec{k}^2+|F_1| \vec{k}^4 +|S_1|\vec{k}^6}}\nn \\
&\approx&\frac{\delta(0)}{2\sqrt{|S_1|}} ~\mathcal{I}\left(\frac{|F_1|}{|S_1|}\right)~,
  \eea 
where
  \be 
 \mathcal{I}\left(\Delta\right)=\int \frac{d^3 k}{(2\pi)^3} \frac{1}{\vec{k}^2\sqrt{\vec{k}^2+ \Delta }}~.
  \ee 
Using dimensional regularisation, this integral becomes
  \be
\mathcal{I}\left(\Delta\right) = \frac{2 \pi^{d/2}}{\Gamma(d/2)} \frac{\mu^{3-d}}{(2\pi)^d}\int_0^\infty d k \frac{k^{d-3}}{(k^2+\Delta)^{1/2}}~,
  \ee 
which, when calculated according to \cite{Ryder}, yields
  \be\label{intGamma} 
\int_0^\infty d k \frac{k^  \beta}{(k^2+\Delta)^\alpha} = \frac{\Gamma(\frac{1+  \beta}{2})\Gamma(\alpha-\frac{1+  \beta}{2})}{2(\Delta)^{\alpha-\frac{1+  \beta}{2}}\Gamma(\alpha)}~.
  \ee 
Writing $d=3-\epsilon$, we then find 
  \be
\mathcal{I}(\Delta) = \frac{1}{2 \pi^2} \frac{\mu^\epsilon}{\epsilon}+\mathcal{O}(\epsilon)~,
  \ee 
which finally leads to
  \be
\mbox{Tr}\left(\frac{\delta(k_1+k_2)}{\omega_1^2+\vec{k}_1^2+|F_1|(\vec{k}_1^2)^2+|S_1|(\vec{k}_1^2)^3}\right) \approx \frac{\delta(0)}{(2\pi)^2\sqrt{|S_1|}}\frac{\mu^\epsilon}{\epsilon}+\cdots
  \ee
where dots represent finite terms. We note here that the result above does not present any dependence on $|F_1|$, since this coupling is associated with $\vec{k}^4$ in~(\ref{resHII}), which plays a sub-dominant role in the UV.

\vspace{0.5cm}

\nin{\bf Spin-0 component}

After gathering the terms which depend on $h$ in the action (\ref{SmII}), we can write the equivalent action in terms of the Fourier transform $\tilde{h}$ of $h$ and, after performing a Wick rotation, obtain
  \bea\label{hwr}
		\int \mathcal{D}h \exp\left\{i S_{II}^{(2)}(h)\right\} = \int \mathcal{D} \tilde{h}\exp\left\{-\int \frac{d^4 k_1}{(2\pi)^4}\frac{		d^4 k_2}{(2\pi)^4}~\frac{1}{2}\tilde{h}(k_2) \left[\mathcal{A}^{(II)}_{\tilde h}\right]\tilde{h}(k_1)\right\}~,
  \eea 
where
  \bea\label{AhII}
		\mathcal{A}^{(II)}_{\tilde h} &=& \frac{1}{9} \left\{M_P^2\left[ \mathcal{P}^{-1}_h(k_1)\right]-(\vec{p}^2\phi_0^2)\left(1+4\left(		\frac{\mathcal{D}_1(\vec{k}_1^2)}{\mathcal{D}_2(\vec{k}_1^2)}\right)+2\left(\frac{\mathcal{D}_1(\vec{k}_1^2)}{\mathcal{D}_2(				\vec{k}_1^2)}\right)^2-\frac{X^2}{6}\frac{\omega_1^2}{\vec{k}_1^2}\right)\right.\nonumber\\
		&&-\left.[\bar{\psi}_0(\vec{\gamma}\cdot\vec q)\psi_0]\left(\frac{3}{4}+2\left(\frac{\mathcal{D}_1(\vec{k}_1)}{\mathcal{D}_2(				\vec{k}_1)}\right)-\frac{X^2}{8}\frac{\omega^2_1}{\vec{k}_1^2}\right)\right\}\delta(k_1+k_2)~,
  \eea 
and
  \bea
		\mathcal{P}^{-1}_h(k_1) &=& X \omega_1^2 +\frac{ \vec{k}_1^2\left[1+(3F_1+8F_2)\vec{k}_1^2+(3S_1+8S_2)(\vec{k}_1^2)^2\right]			\mathcal{D}_2(\vec{k}_1)+2\left[\mathcal{D}_1(\vec{k}_1)\right]^2}{-\mathcal{D}_2(\vec{k}_1)}\nn \\
		\mathcal{D}_1(\vec{k}_1) &=& -[\vec{k}_1^2- F_3 (\vec{k}_1^2)^2+S_3 (\vec{k}_1^2)^3]\nn \\
		\mathcal{D}_2(\vec{k}_1) &=& -[\alpha\vec{k}_1^2- F_4 (\vec{k}_1^2)^2+S_4 (\vec{k}_1^2)^3]~.
  \eea 

In the IR limit, for which higher-order operators are neglected, the expression (\ref{AhII}) becomes (\ref{AhI}), and the dispersion relation for the scalar graviton is given by (\ref{hdrnp}) or (\ref{hdrIR}), as expected. 

The functional integration over $h$ gives
 {\footnotesize
 \bea 
		&&\exp\left\{\frac{1}{2M_P^2}\left[(\vec p^2 \phi_0^2)+\frac{3}{4}[\bar{\psi}_0(\vec{\gamma}\cdot\vec q)\psi_0]\right]
		\mbox{Tr}\left(\frac{\delta(k_1+k_2)}{\mathcal{P}^{-1}_h(k_1)}\right)\right.\\
		&&\left.+\frac{2}{M_P^2}\left[(\vec p^2 \phi_0^2)+\frac{1}{2}[\bar{\psi}_0(\vec{\gamma}\cdot\vec q)\psi_0]\right]
		\mbox{Tr}\left(\frac{\delta(k_1+k_2)\mathcal{D}_1(\vec{k}_1)}{\mathcal{P}^{-1}_h(k_1) \mathcal{D}_2(\vec{k}_1)}\right)+\frac{				(\vec p^2 \phi_0^2)}{M_P^2}\mbox{Tr}\left(\frac{\delta(k_1+k_2)(\mathcal{D}_1(\vec{k}_1))^2}{\mathcal{P}^{-1}_h(k_1) (\mathcal{D}_2		(\vec{k}_1))^2} \right)\right.\nonumber\\
		&&\left. -\frac{X^2}{12 M_P^2}\left[(\vec p^2 \phi_0^2) +\frac{ 3}{4}[\bar{\psi}_0(\vec{\gamma}\cdot\vec q)\psi_0]\right]\mbox{Tr}		\left(\frac{\delta(k_1+k_2)\omega_1^2}{\mathcal{P}^{-1}_h(k_1) \vec{k_1}^2}\right)+\cdots\right\}~.\nonumber
  \eea}
We can write the first three traces in the following generic form
  \be\label{logsc}
		\mbox{Tr}\left[\frac{\delta(k_1+k_2)}{\mathcal{P}^{-1}_h(k_1)}\left(\frac{\mathcal{D}_1(\vec{k}_1)}{ \mathcal{D}_2(\vec{k}_1)}			\right)^n\right],~~~\mbox{for}~n=0,1,2~,
  \ee  
which, after integrating over the frequencies, leads to
  \bea\label{trn}
		&&\mbox{Tr}\left[\frac{\delta(k_1+k_2)}{\mathcal{P}^{-1}_h(k_1)}\left(\frac{\mathcal{D}_1(\vec{k}_1)}{ \mathcal{D}_2(\vec{k}_1)}		\right)^n\right]=\frac{\delta(0)}{2\sqrt{X}}\int \frac{d^3 k}{(2\pi)^3} \left[\mathcal{G}(\vec{k})\left(\frac{\mathcal{D}_2(\vec{k})}{\mathcal{D}_1(\vec{k})}\right)^{2n}\right]^{-\frac{1}{2}}~,
  \eea 
with
  \bea \label{Gk}
		\mathcal{G}(\vec{k})&=& (\mathcal{P}_h^{-1} -X \omega^2)\\
		&=&\frac{ \vec{k}^2\left[1+(3F_1+8F_2)\vec{k}^2+(3S_1+8S_2)(\vec{k}^2)^2\right]	\mathcal{D}_2(\vec{k})
		+2\left[\mathcal{D}_1(\vec{k})\right]^2}{-\mathcal{D}_2(\vec{k})}~.\nn
  \eea 
As we are mainly interested in the UV dominant contributions, we expand the terms inside the square brackets in~(\ref{trn}), keeping only contributions of at least quartic order in the momentum, such that the right-hand side of eq.(\ref{trn}) reduces to  
	\be\label{3dint} 
		\frac{\delta(0)}{2\sqrt{XC_6^{(n)}}}~ \mathcal{I}\left(\frac{C_4^{(n)}}{C^{(n)}_6}\right),
  \ee 
with the integral $\mathcal{I}$ given by eq.(\ref{intd}) and, for $n=0,1,2$,
  \bea
		C_4^{(n)}&=&(c_4^{(n)}/M_{HL}^{2})\\ 
		C_6^{(n)}&=&(c_6^{(n)}/M_{HL}^{4})\nn \\ 
		c_4^{(0)} &=& -(2 f_1+8f_2)-  \frac{2s_3( f_4 s_3-2 f_3 s_4)}{s_4^2}\nn \\
		c_4^{(1)} &=& 2f_4+\frac{2s_4(2s_1+8s_2)(f_4s_3-f_3 s_4)-s_3s_4^2(2f_1+8f_2)}{s_3^3} \nn \\
		c_4^{(2)} &=& \frac{s_4^2\left[-s_3 s_4^2(2f_1+8f_2) + 4s_4(f_4s_3 -f_3 s_4)(2s_1+8s_2)+2s_3^2(3f_4s_3-2f_3s_4)\right]}{s_3^5}\nn \\
		c_6^{(0)}&=& -(2 s_1+ 8s_2)-\frac{2 s_3^2}{s_4} \nn \\
		c_6^{(1)} &=& -2s_4 -\frac{ (2 s_1+8s_2) s_4^2}{s_3^2}\nn \\
		c_6^{(2)} &=& -2\frac{s_4^3}{s_3^2}-\frac{s_4^4(2s_1+8s_2)}{s_3^4}~.\nonumber 
  \eea
This integral is again solved with dimensional regularisation using the result~(\ref{Isol}): 
  \be\label{convint} 
		\mbox{Tr}\left[\frac{\delta(k_1+k_2)}{\mathcal{P}^{-1}_h(k_1)}\left(\frac{\mathcal{D}_1(\vec{k}_1)}{ \mathcal{D}_2(\vec{k}_1)}		\right)^n\right] = \frac{\delta(0)}{(2\pi)^2\sqrt{X C_6^{(n)}}}\frac{\mu^\epsilon}{\epsilon}+\cdots,
  \ee 
where dots represent finite terms. 

Finally, we are left with the last trace
  \bea\label{trwk}
		\mbox{Tr}\left(\frac{\delta(k_1+k_2)\omega_1^2}{\mathcal{P}^{-1}_h(k_1) \vec{k_1}^2}\right)&=&\delta(0)\int \frac{d^4k}{(2\pi)^4}		 \frac{\omega^2}{P_h^{-1}(k)\vec{k}^2}\\
		&=&\frac{\delta(0)}{ X}\left\{\int \frac{d^4k}{(2\pi)^4}\frac{1}{\vec{k}^2}-\int \frac{d^4k}{(2\pi)^4}\frac{\mathcal{G}(\vec{k})		/X}{[\omega^2+\mathcal{G}(\vec{k})/X]\vec{k}^2}\right\}~,\nonumber
  \eea 
with $\mathcal{G}$ defined in~(\ref{Gk}). While the first integral in the last line of~(\ref{trwk}) vanishes with dimensional regularisation, the second one, after integration over the frequencies, leads to
  \be 
		\mbox{Tr}\left(\frac{\delta(k_1+k_2)\omega_1^2}{\mathcal{P}^{-1}_h(k_1) \vec{k_1}^2}\right)=\frac{\delta(0)}{2 X^{3/2}}\int \frac{		d^3k}{(2\pi)^3}\frac{\sqrt{\mathcal{G}(\vec{k})}}{\vec{k}^2}~.
  \ee 
We then expand $\mathcal{G}(\vec{k})$, keeping only the UV dominant contribution
  \be 
		\mbox{Tr}\left(\frac{\delta(k_1+k_2)\omega_1^2}{\mathcal{P}^{-1}_h(k_1) \vec{k_1}^2}\right)=\frac{\delta(0)\sqrt{C_6^{(0)}}}{2 X^{		3/2}}\int \frac{d^3k}{(2\pi)^3} |\vec{k}| ~ + \cdots
  \ee 
with dots representing finite terms. This integral also vanishes with dimensional regularisation, hence the result is finite.

\chapter{Conclusions}\label{Conclusions}
\mbox{}

In this thesis we have investigated the influence that Lorentz violation can have on fermion fields, especially neutrinos. We have proposed different models with Lorentz symmetry violation to tackle important questions in particle physics, such as the origin of neutrino masses and oscillations. We have made use of non-perturbative methods and studied the generation of masses and mixing for two neutrino flavours in the case of Dirac as well as Majorana fermions. In addition, inspired by another exciting research area: the search for a perturbatively renormalisable theory of quantum gravity, we have considered the coupling of matter fields (fermions and scalars) to modified gravity models, such as Horava-Lifshitz gravity, in order to study how quantum gravity can induce LIV effects in the matter sector. 

In the following paragraphs we summarise our main findings and briefly discuss possible future work.
\\

In chapter \ref{chap4} we have considered the coupling of flavoured fermion fields to LIV vector gauge bosons, with Lorentz invariance being violated in the gauge sector at a mass scale $M$. We have focused on the limiting case where the gauge couplings go to zero, while the LIV mass scale $M \to \infty $, in such a way that the Schwinger-Dyson dynamically generated fermion mass matrix remains finite. We have shown that the arrangement of the couplings is such that no vector boson mass is generated, and therefore, as the LIV vector bosons completely decouple from the fermions in the limit of interest, the former can be viewed as regulator fields. We have then solved the Schwinger-Dyson equations for different cases and shown that, although fermion masses are generated dynamically in various cases, oscillation among fermion flavours only takes place in one specific situation. In this situation, one of the fermion mass eigenstates remains massless, and the mixing angle is necessarily maximal, $\theta = \pm \pi/4$. Additionally, we have extended our analysis to explore the possibility of neutrinos being Majorana fermions by discussing two scenarios. In the first case, using the properties of Majorana fields, we have studied the coupling of two left-handed neutrinos with a LIV regulator gauge field and demonstrated that different masses can be obtained in the Lorentz symmetric limit, which then leads to standard oscillations among the neutrino flavours. In the second scenario, where the fermion doublet is formed by a left-handed (active) neutrino and a right-handed (sterile) neutrino field, we have shown that if a Dirac mass term is generated via the usual Higgs mechanism, our approach can then be used to dynamically generate a heavy Majorana mass for the sterile neutrino. Thus, the seesaw mechanism takes over: the large Majorana mass associated with the right-handed field suppresses the Dirac mass term, resulting in a light active neutrino. 
\\

In chapter \ref{chap5} we have proposed another alternative to study the generation of neutrino masses and oscillations within the framework of Lorentz violation. In this case, however, instead of coupling fermions to other fields, we have discussed the coupling of fermions, with LIV kinematics, among themselves through four-fermion interactions. By using the effective potential approach, we have shown that fermion masses and oscillations are generated, and that the presence of these, contrary to what happens in the Lorentz symmetric case, are not dependent on the size of the coupling constants $g_i^2$ governing the four-fermion interactions. Thus, since fermion masses are generated for any coupling strength, we have demonstrated that the limits $g_i^2\to 0$ and $M\to \infty$, where $M$ is the mass scale suppressing the LIV operators, can be simultaneously taken after quantisation, in such a way that the masses and mixing generated by quantum corrections remain finite, and their values can be chosen according to phenomenology. Finally, assuming Majorana neutrinos, we have shown that our approach can be used to generate both Majorana and Dirac mass terms, thus reproducing the seesaw mechanism features when a appropriate choice of parameters is made.
\\

In both chapter \ref{chap4} and chapter \ref{chap5} we have considered two fermion flavours at most. Nevertheless, we know of three different neutrino flavours in nature and that oscillations take place among all of them. It would therefore be interesting to generalise these, as well as other future proposals, to the case of three flavours. In addition to being more realistic, another advantage of working with three neutrino generations is that, by doing so, we could obtain some information about the origin and size of the CP-violating phases in the neutrino sector. Consequently, since CP violation in the lepton sector leads to lepton asymmetry which can then be converted into baryon asymmetry by non-perturbative effects, known as sphalerons \cite{BilenkyBook}, this generalisation could help us with yet another big open question in particle physics: the origin of the baryon asymmetry in the universe.
\\ 

In addition, in chapter \ref{chap6}, considering the interaction of matter fields (scalars and fermions) with gravitons described by two modified gravity models, we have studied how quantum gravity corrections can induce LIV contributions in the matter sector. On the one hand, both gravity models are similar in that they do not share the same space-time symmetries with GR (4-dimensional diffeomorphism invariance), leading to local Lorentz violation. On the other hand, these models present a different UV behaviour: one gives quadratic divergences when calculating one-loop corrections to the matter fields, while the other, the non-projectable HL gravity, gives logarithmic divergences only. Using both models, we have calculated the LIV one-loop corrections to the matter field dispersion relations and then compared our results with the current bounds on Lorentz symmetry violation. We found that, if one wishes to conclude with generic values for the different parameters, both models lead to the same order of magnitude $10^{10}$ GeV for the typical scale above which the predicted Lorentz symmetry violation is too large. For the first model, this limiting energy scale is associated with the cut off of the theory, whereas for the second model $10^{10}$ GeV is identified as the maximum value for the HL scale, {\it i.e.} the energy scale suppressing the higher-order operators. Moreover, we also found that, in both cases, the LIV corrections to the matter field dispersion relations vanish if their respective parameters are fine-tuned accordingly. 

It is worth pointing out that the typical scale $10^{10}$ GeV is consistent with other results, such as in \cite{Pospelov,AB2}. We note that this scale also corresponds to the Higgs potential instability \cite{Higgsinst}, which could be avoided when curvature effects are taken into account \cite{rajantie}. Therefore, it would be interesting to look for a stabilising mechanism based on non-relativistic gravity models.\\

Moreover, another possibility for future work would be the derivation of LIV operators for the matter sector from modified gravity models, such as Horava-Lifshitz gravity, in order to study the generation of neutrino masses and oscillations. Particularly, we could consider the coupling of relativistic Abelian gauge fields and neutrinos to a modified gravity model and derive, through quantum corrections, the LIV higher-order spatial derivatives present in the effective models studied in chapters \ref{chap4} and \ref{chap5}. We could also extend the work done in chapter \ref{chap6} by taking into account higher-order contributions to the matter sector so that, in addition to keeping lowest order corrections which contribute to the matter field dispersion relations, it would be possible to investigate the generation of four-fermion interaction terms, such as the one considered in the original model in chapter \ref{chap5}. 

\singlespace



\begin{thebibliography}{99}



\bibitem{ALM1}
  J.~Alexandre, J.~Leite and N.~E.~Mavromatos,
  ``Lorentz-violating regulator gauge fields as the origin of dynamical flavor oscillations,''
  Phys.\ Rev.\ D {\bf 87} (2013) 12,  125029
  [arXiv:1304.7706 [hep-ph]];
	
	\bibitem{ALM2}
  J.~Alexandre, J.~Leite and N.~E.~Mavromatos,
  ``Quasirelativistic fermions and dynamical flavor oscillations,''
  Phys.\ Rev.\ D {\bf 90} (2014) 4,  045026
  [arXiv:1404.7429 [hep-th]].
	
	
\bibitem{AL} 
  J.~Alexandre and J.~Leite,
  ``Effective fermion kinematics from modified quantum gravity,''
  Class.\ Quant.\ Grav.\  {\bf 33}, no. 19, 195005 (2016)
  doi:10.1088/0264-9381/33/19/195005
  [arXiv:1506.03755 [hep-ph]].
	
	\bibitem{seesaw1}
  P.~Minkowski,
  ``$\mu \to e\gamma$ at a Rate of One Out of $10^{9}$ Muon Decays?,''
  Phys.\ Lett.\ B {\bf 67} (1977) 421.
  doi:10.1016/0370-2693(77)90435-X
	\bibitem{seesaw2}	
J.~Schechter and J.~W.~F.~Valle,
  ``Neutrino Masses in SU(2) x U(1) Theories,''
  Phys.\ Rev.\ D {\bf 22} (1980) 2227;
\bibitem{seesaw3}	
For reviews see also: R.~N.~Mohapatra, S.~Antusch, K.~S.~Babu, G.~Barenboim, M.~-C.~Chen, A.~de Gouvea, P.~de Holanda and B.~Dutta {\it et al.},
  ``Theory of neutrinos: A White paper,''
  Rept.\ Prog.\ Phys.\  {\bf 70} (2007) 1757
  [hep-ph/0510213] and references therein. 
	
	\bibitem{Higgs1}
  G.~Aad {\it et al.}  [ATLAS Collaboration],
  ``Observation of a new particle in the search for the standard model Higgs boson with the ATLAS detector at the LHC,''
  Phys.\ Lett.\ B {\bf 716} (2012) 1
  [arXiv:1207.7214 [hep-ex]];
	\bibitem{Higgs2}
  S.~Chatrchyan {\it et al.}  [CMS Collaboration],
  ``Observation of a new boson at a mass of 125 GeV with the CMS experiment at the LHC,''
  Phys.\ Lett.\ B {\bf 716} (2012) 30
  [arXiv:1207.7235 [hep-ex]].

	
\bibitem{BilenkyBook} 
  S.~Bilenky,
  ``Introduction to the physics of massive and mixed neutrinos,''
  Lect.\ Notes Phys.\  {\bf 817}, 1 (2010).
  doi:10.1007/978-3-642-14043-3
	
\bibitem{GiuntiBook} 
  C.~Giunti and C.~W.~Kim,
  ``Fundamentals of Neutrino Physics and Astrophysics,''
  Oxford, UK: Univ. Pr. (2007) 710 p
	
\bibitem{ZuberBook} 
  K.~Zuber,
  ``Neutrino physics,''
  Boca Raton: USA: CRC Pr. (2012) 448p
	
\bibitem{Fermibeta} 
  E.~Fermi,
  ``An attempt of a theory of beta radiation. 1.,''
  Z.\ Phys.\  {\bf 88}, 161 (1934).
  doi:10.1007/BF01351864;

\bibitem{1stneutdet1} 
  C.~L.~Cowan, F.~Reines, F.~B.~Harrison, H.~W.~Kruse and A.~D.~McGuire,
  ``Detection of the free neutrino: A Confirmation,''
  Science {\bf 124}, 103 (1956).
  doi:10.1126/science.124.3212.103
\bibitem{1stneutdet2}	
  F.~Reines and C.~L.~Cowan,
  ``The neutrino,''
  Nature {\bf 178}, 446 (1956).
  doi:10.1038/178446a0
	
\bibitem{muonneut} 
  G.~Danby, J.~M.~Gaillard, K.~A.~Goulianos, L.~M.~Lederman, N.~B.~Mistry, M.~Schwartz and J.~Steinberger,
  ``Observation of High-Energy Neutrino Reactions and the Existence of Two Kinds of Neutrinos,''
  Phys.\ Rev.\ Lett.\  {\bf 9}, 36 (1962).
  doi:10.1103/PhysRevLett.9.36
	
	
\bibitem{tauneut} 
  K.~Kodama {\it et al.},
  ``Detection and analysis of tau neutrino interactions in DONUT emulsion target,''
  Nucl.\ Instrum.\ Meth.\ A {\bf 493}, 45 (2002).
  doi:10.1016/S0168-9002(02)01555-3
	
\bibitem{solarneutpro} 
  R.~Davis,
  ``A review of measurements of the solar neutrino flux and their variation,''
  Nucl.\ Phys.\ Proc.\ Suppl.\  {\bf 48}, 284 (1996).
  doi:10.1016/0920-5632(96)00263-0
	
	
\bibitem{SNO} 
  Q.~R.~Ahmad {\it et al.} [SNO Collaboration],
  ``Direct evidence for neutrino flavor transformation from neutral current interactions in the Sudbury Neutrino Observatory,''
  Phys.\ Rev.\ Lett.\  {\bf 89}, 011301 (2002)
  doi:10.1103/PhysRevLett.89.011301
  [nucl-ex/0204008].
	
\bibitem{KamLAND} 
  K.~Eguchi {\it et al.} [KamLAND Collaboration],
  ``First results from KamLAND: Evidence for reactor anti-neutrino disappearance,''
  Phys.\ Rev.\ Lett.\  {\bf 90}, 021802 (2003)
  doi:10.1103/PhysRevLett.90.021802
  [hep-ex/0212021].
	
	 \bibitem{white2} K.~N.~Abazajian, M.~A.~Acero, S.~K.~Agarwalla, A.~A.~Aguilar-Arevalo, C.~H.~Albright, S.~Antusch, 
C.~A.~Arguelles and A.~B.~Balantekin {\it et al.},
 ``Light Sterile Neutrinos: A White Paper,''
  arXiv:1204.5379 [hep-ph].
  
	
\bibitem{Weinberg}
  S.~Weinberg,
  ``Baryon and Lepton Nonconserving Processes,''
  Phys.\ Rev.\ Lett.\  {\bf 43} (1979) 1566.
  doi:10.1103/PhysRevLett.43.1566
	
	
\bibitem{PDG} 
  K.~A.~Olive {\it et al.} [Particle Data Group Collaboration],
  ``Review of Particle Physics,''
  Chin.\ Phys.\ C {\bf 40}, no. 10, 100001 (2016).
  doi:10.1088/1674-1137/40/10/100001
  


\bibitem{IZ}
  C.~Itzykson, J.~B.~Zuber
  ``Quantum Field Theory,''
  New York, Usa: Mcgraw-hill (1980) 705 P.(International Series In Pure and Applied Physics)
	
	\bibitem{SDE2+1QED1} 
  R.~D.~Pisarski,
  ``Chiral Symmetry Breaking in Three-Dimensional Electrodynamics,''
  Phys.\ Rev.\ D {\bf 29}, 2423 (1984).
  doi:10.1103/PhysRevD.29.2423
	\bibitem{SDE2+1QED2} 
  T.~W.~Appelquist, M.~J.~Bowick, D.~Karabali and L.~C.~R.~Wijewardhana,
  ``Spontaneous Chiral Symmetry Breaking in Three-Dimensional QED,''
  Phys.\ Rev.\ D {\bf 33}, 3704 (1986).
  doi:10.1103/PhysRevD.33.3704
	
	
\bibitem{ExtMagLad1}
	V.~P.~Gusynin, V.~A.~Miransky and I.~A.~Shovkovy,
  ``Dimensional reduction and catalysis of dynamical symmetry breaking by a
   magnetic field,''
  Nucl.\ Phys.\  B {\bf 462} (1996) 249
  [arXiv:hep-ph/9509320];
\bibitem{ExtMagLad2}
  V.~P.~Gusynin, V.~A.~Miransky and I.~A.~Shovkovy,
  ``Dynamical chiral symmetry breaking by a magnetic field in QED,''
  Phys.\ Rev.\ D {\bf 52}, 4747 (1995)
  doi:10.1103/PhysRevD.52.4747
  [hep-ph/9501304].
	
	\bibitem{ExtMagImpLad1}
	V.~P.~Gusynin, V.~A.~Miransky and I.~A.~Shovkovy,
  ``Dynamical chiral symmetry breaking in QED in a magnetic field: Toward exact results,''
  Phys.\ Rev.\ Lett.\  {\bf 83}, 1291 (1999)
  doi:10.1103/PhysRevLett.83.1291
  [hep-th/9811079].
	\bibitem{ExtMagImpLad2}
	V.~P.~Gusynin, V.~A.~Miransky and I.~A.~Shovkovy,
  ``Theory of the magnetic catalysis of chiral symmetry breaking in QED,''
  Nucl.\ Phys.\ B {\bf 563}, 361 (1999)
  doi:10.1016/S0550-3213(99)00573-8
  [hep-ph/9908320].
  

	\bibitem{NJL}
  Y.~Nambu and G.~Jona-Lasinio,
  ``Dynamical Model of Elementary Particles Based on an Analogy with Superconductivity. 1.,''
  Phys.\ Rev.\  {\bf 122} (1961) 345.

\bibitem{BCS} 
  J.~Bardeen, L.~N.~Cooper and J.~R.~Schrieffer,
  ``Microscopic theory of superconductivity,''
  Phys.\ Rev.\  {\bf 106}, 162 (1957).
  doi:10.1103/PhysRev.106.162
	
\bibitem{GN}
  D.~J.~Gross and A.~Neveu,
  ``Dynamical Symmetry Breaking in Asymptotically Free Field Theories,''
  Phys.\ Rev.\ D {\bf 10} (1974) 3235.
  doi:10.1103/PhysRevD.10.3235
	
\bibitem{Miranskybook} 
  V.~A.~Miransky,
  ``Dynamical symmetry breaking in quantum field theories,''
  Singapore, Singapore: World Scientific (1993) 533 p

\bibitem{ABH}
  J.~Alexandre, J.~Brister and N.~Houston,
  ``On higher-order corrections in a four-fermion Lifshitz model,''
  Phys.\ Rev.\ D {\bf 86} (2012) 025030
  [arXiv:1204.2246 [hep-ph]].
	

\bibitem{Anselmi4ferm} 
  D.~Anselmi and E.~Ciuffoli,
  ``Renormalization Of High-Energy Lorentz Violating Four Fermion Models,''
  Phys.\ Rev.\ D {\bf 81}, 085043 (2010)
  doi:10.1103/PhysRevD.81.085043
  [arXiv:1002.2704 [hep-ph]].
		
\bibitem{string} 
  V.~A.~Kostelecky and S.~Samuel,
  ``Spontaneous Breaking of Lorentz Symmetry in String Theory,''
  Phys.\ Rev.\ D {\bf 39}, 683 (1989).
  doi:10.1103/PhysRevD.39.683
	
\bibitem{foam} 
  G.~Amelino-Camelia, J.~R.~Ellis, N.~E.~Mavromatos and D.~V.~Nanopoulos,
  ``Distance measurement and wave dispersion in a Liouville string approach to quantum gravity,''
  Int.\ J.\ Mod.\ Phys.\ A {\bf 12}, 607 (1997)
  doi:10.1142/S0217751X97000566
  [hep-th/9605211].

\bibitem{brane} 
  C.~P.~Burgess, J.~M.~Cline, E.~Filotas, J.~Matias and G.~D.~Moore,
  ``Loop generated bounds on changes to the graviton dispersion relation,''
  JHEP {\bf 0203}, 043 (2002)
  doi:10.1088/1126-6708/2002/03/043
  [hep-ph/0201082].
	
\bibitem{non-comm} 
  S.~M.~Carroll, J.~A.~Harvey, V.~A.~Kostelecky, C.~D.~Lane and T.~Okamoto,
  ``Noncommutative field theory and Lorentz violation,''
  Phys.\ Rev.\ Lett.\  {\bf 87}, 141601 (2001)
  doi:10.1103/PhysRevLett.87.141601
  [hep-th/0105082].

\bibitem{VSL} 
  J.~Magueijo,
  ``New varying speed of light theories,''
  Rept.\ Prog.\ Phys.\  {\bf 66}, 2025 (2003)
  doi:10.1088/0034-4885/66/11/R04
  [astro-ph/0305457].

\bibitem{Liberatireview} 
  T.~Jacobson, S.~Liberati and D.~Mattingly,
  ``Lorentz violation at high energy: Concepts, phenomena and astrophysical constraints,''
  Annals Phys.\  {\bf 321}, 150 (2006)
  doi:10.1016/j.aop.2005.06.004
  [astro-ph/0505267].

\bibitem{SME1} 
  D.~Colladay and V.~A.~Kostelecky,
  ``CPT violation and the standard model,''
  Phys.\ Rev.\ D {\bf 55}, 6760 (1997)
  doi:10.1103/PhysRevD.55.6760
  [hep-ph/9703464];
\bibitem{SME2}
	D.~Colladay and V.~A.~Kostelecky,
  ``Lorentz violating extension of the standard model,''
  Phys.\ Rev.\ D {\bf 58}, 116002 (1998)
  doi:10.1103/PhysRevD.58.116002
  [hep-ph/9809521].
	
\bibitem{nmSME1} 
  V.~A.~Kostelecky and M.~Mewes,
  ``Electrodynamics with Lorentz-violating operators of arbitrary dimension,''
  Phys.\ Rev.\ D {\bf 80}, 015020 (2009)
  doi:10.1103/PhysRevD.80.015020
  [arXiv:0905.0031 [hep-ph]];
\bibitem{nmSME2} 
  A.~Kostelecky and M.~Mewes,
  ``Neutrinos with Lorentz-violating operators of arbitrary dimension,''
  Phys.\ Rev.\ D {\bf 85}, 096005 (2012)
  doi:10.1103/PhysRevD.85.096005
  [arXiv:1112.6395 [hep-ph]];
\bibitem{nmSME3} 
  A.~Kostelecky and M.~Mewes,
  ``Fermions with Lorentz-violating operators of arbitrary dimension,''
  Phys.\ Rev.\ D {\bf 88}, no. 9, 096006 (2013)
  doi:10.1103/PhysRevD.88.096006
  [arXiv:1308.4973 [hep-ph]].
	
\bibitem{SUSYLV} 
  S.~Groot Nibbelink and M.~Pospelov,
  ``Lorentz violation in supersymmetric field theories,''
  Phys.\ Rev.\ Lett.\  {\bf 94}, 081601 (2005)
  doi:10.1103/PhysRevLett.94.081601
  [hep-ph/0404271].

\bibitem{bounds}
  V.~A.~Kostelecky and N.~Russell,
  ``Data Tables for Lorentz and CPT Violation,''
  Rev.\ Mod.\ Phys.\  {\bf 83} (2011) 11
  [arXiv:0801.0287 [hep-ph]].
	
\bibitem{mSMEneutrino1}
	V.~A.~Kostelecky and M.~Mewes,
  ``Lorentz and CPT violation in neutrinos,''
  Phys.\ Rev.\ D {\bf 69}, 016005 (2004)
  doi:10.1103/PhysRevD.69.016005
  [hep-ph/0309025];
\bibitem{mSMEneutrino2}
	V.~A.~Kostelecky and M.~Mewes,
  ``Lorentz and CPT violation in the neutrino sector,''
  Phys.\ Rev.\ D {\bf 70}, 031902 (2004)
  doi:10.1103/PhysRevD.70.031902
  [hep-ph/0308300].


\bibitem{SMEneutosc2} 
  T.~Katori, V.~A.~Kostelecky and R.~Tayloe,
  ``Global three-parameter model for neutrino oscillations using Lorentz violation,''
  Phys.\ Rev.\ D {\bf 74}, 105009 (2006)
  doi:10.1103/PhysRevD.74.105009
  [hep-ph/0606154];
\bibitem{SMEneutosc3} 
  V.~Barger, D.~Marfatia and K.~Whisnant,
  ``Challenging Lorentz noninvariant neutrino oscillations without neutrino masses,''
  Phys.\ Lett.\ B {\bf 653}, 267 (2007)
  doi:10.1016/j.physletb.2007.07.047
  [arXiv:0706.1085 [hep-ph]];
\bibitem{SMEneutosc4} 
  V.~Barger, J.~Liao, D.~Marfatia and K.~Whisnant,
  ``Lorentz noninvariant oscillations of massless neutrinos are excluded,''
  Phys.\ Rev.\ D {\bf 84}, 056014 (2011)
  doi:10.1103/PhysRevD.84.056014
  [arXiv:1106.6023 [hep-ph]].

\bibitem{puma1}
  J.~S.~Diaz and V.~A.~Kostelecky,
  ``Three-parameter Lorentz-violating texture for neutrino mixing,''
  Phys.\ Lett.\ B {\bf 700} (2011) 25
  [arXiv:1012.5985 [hep-ph]].
\bibitem{puma2}
  J.~S.~Diaz and A.~Kostelecky,
  ``Lorentz- and CPT-violating models for neutrino oscillations,''
  Phys.\ Rev.\ D {\bf 85} (2012) 016013
  [arXiv:1108.1799 [hep-ph]].

\bibitem{tribi}
  P.~F.~Harrison, D.~H.~Perkins and W.~G.~Scott,
  ``Tri-bimaximal mixing and the neutrino oscillation data,''
  Phys.\ Lett.\ B {\bf 530} (2002) 167
  doi:10.1016/S0370-2693(02)01336-9
  [hep-ph/0202074].
	

\bibitem{T131} 
  F.~P.~An {\it et al.} [Daya Bay Collaboration],
  ``Observation of electron-antineutrino disappearance at Daya Bay,''
  Phys.\ Rev.\ Lett.\  {\bf 108}, 171803 (2012)
  doi:10.1103/PhysRevLett.108.171803
  [arXiv:1203.1669 [hep-ex]];
\bibitem{T132} 	
	J.~K.~Ahn {\it et al.} [RENO Collaboration],
  ``Observation of Reactor Electron Antineutrino Disappearance in the RENO Experiment,''
  Phys.\ Rev.\ Lett.\  {\bf 108}, 191802 (2012)
  doi:10.1103/PhysRevLett.108.191802
  [arXiv:1204.0626 [hep-ex]].
	
\bibitem{Stelle} 
  K.~S.~Stelle,
  ``Renormalization of Higher Derivative Quantum Gravity,''
  Phys.\ Rev.\ D {\bf 16}, 953 (1977).
  doi:10.1103/PhysRevD.16.953
	
\bibitem{Ostrog}
  M.~Ostrogradsky,
  ``Memoires sur les equations differentielles, relatives au probleme des isoperimetres,''
  Mem.\ Acad.\ St.\ Petersbourg {\bf 6} (1850) no.4,  385.
	
\bibitem{Ostrog2} 
  R.~P.~Woodard,
  ``Avoiding dark energy with 1/r modifications of gravity,''
  Lect.\ Notes Phys.\  {\bf 720}, 403 (2007)
  [astro-ph/0601672].
	


\bibitem{Lifreview}
  J.~Alexandre,
  ``Lifshitz-type Quantum Field Theories in Particle Physics,''
  Int.\ J.\ Mod.\ Phys.\ A {\bf 26} (2011) 4523
  [arXiv:1109.5629 [hep-ph]].

\bibitem{Anselmi1}
  D.~Anselmi and M.~Halat,
  ``Renormalization of Lorentz violating theories,''
  Phys.\ Rev.\ D {\bf 76}, 125011 (2007)
  doi:10.1103/PhysRevD.76.125011
  [arXiv:0707.2480 [hep-th]];
\bibitem{Anselmi2}
  D.~Anselmi,
  ``Weighted scale invariant quantum field theories,''
  JHEP {\bf 0802}, 051 (2008)
  doi:10.1088/1126-6708/2008/02/051
  [arXiv:0801.1216 [hep-th]].
\bibitem{Anselmi3} 
  D.~Anselmi,
  ``Weighted power counting and Lorentz violating gauge theories. I. General properties,''
  Annals Phys.\  {\bf 324}, 874 (2009)
  doi:10.1016/j.aop.2008.12.005
  [arXiv:0808.3470 [hep-th]];
\bibitem{Anselmi4}
  D.~Anselmi,
  ``Weighted power counting and Lorentz violating gauge theories. II. Classification,''
  Annals Phys.\  {\bf 324}, 1058 (2009)
  doi:10.1016/j.aop.2008.12.007
  [arXiv:0808.3474 [hep-th]].

\bibitem{Visser} 
  M.~Visser,
  ``Lorentz symmetry breaking as a quantum field theory regulator,''
  Phys.\ Rev.\ D {\bf 80}, 025011 (2009)
  doi:10.1103/PhysRevD.80.025011
  [arXiv:0902.0590 [hep-th]].

\bibitem{deviation}
  R.~Iengo, J.~G.~Russo and M.~Serone,
  ``Renormalization group in Lifshitz-type theories,''
  JHEP {\bf 0911} (2009) 020
  [arXiv:0906.3477 [hep-th]];
		
\bibitem{Horava}
  P.~Horava,
  ``Quantum Gravity at a Lifshitz Point,''
  Phys.\ Rev.\ D {\bf 79} (2009) 084008
  [arXiv:0901.3775 [hep-th]].

\bibitem{MagueijoHL} 
  S.~Alexander, J.~Magueijo and A.~Marciano,
  ``Horava-Lifshitz theory as a Fermionic Aether in Ashtekar gravity,''
  Phys.\ Rev.\ D {\bf 86}, 064025 (2012)
  doi:10.1103/PhysRevD.86.064025
  [arXiv:1206.6296 [hep-th]].
	
	
\bibitem{StrongCpHL1} 
  C.~Charmousis, G.~Niz, A.~Padilla and P.~M.~Saffin,
  ``Strong coupling in Horava gravity,''
  JHEP {\bf 0908}, 070 (2009)
  doi:10.1088/1126-6708/2009/08/070
  [arXiv:0905.2579 [hep-th]];
\bibitem{StrongCpHL2} 	
	D.~Blas, O.~Pujolas and S.~Sibiryakov,
  ``On the Extra Mode and Inconsistency of Horava Gravity,''
  JHEP {\bf 0910}, 029 (2009)
  doi:10.1088/1126-6708/2009/10/029
  [arXiv:0906.3046 [hep-th]].

\bibitem{non-proj} 
  D.~Blas, O.~Pujolas and S.~Sibiryakov,
  ``Consistent Extension of Horava Gravity,''
  Phys.\ Rev.\ Lett.\  {\bf 104}, 181302 (2010)
  [arXiv:0909.3525 [hep-th]].

	
\bibitem{StrongCnpHL} 
  D.~Blas, O.~Pujolas and S.~Sibiryakov,
  ``Comment on `Strong coupling in extended Horava-Lifshitz gravity',''
  Phys.\ Lett.\ B {\bf 688}, 350 (2010)
  doi:10.1016/j.physletb.2010.03.073
  [arXiv:0912.0550 [hep-th]].
	
\bibitem{Einstein-aether} 
  T.~Jacobson,
  ``Extended Horava gravity and Einstein-aether theory,''
  Phys.\ Rev.\ D {\bf 81}, 101502 (2010)
  Erratum: [Phys.\ Rev.\ D {\bf 82}, 129901 (2010)]
  doi:10.1103/PhysRevD.82.129901, 10.1103/PhysRevD.81.101502
  [arXiv:1001.4823 [hep-th]].
	
	
\bibitem{covariant}
  P.~Horava and C.~M.~Melby-Thompson,
  ``General Covariance in Quantum Gravity at a Lifshitz Point,''
  Phys.\ Rev.\ D {\bf 82} (2010) 064027
  [arXiv:1007.2410 [hep-th]].

\bibitem{membranes} 
  P.~Horava,
  ``Membranes at Quantum Criticality,''
  JHEP {\bf 0903}, 020 (2009)
  [arXiv:0812.4287 [hep-th]].
	
	
\bibitem{JA1}
  J.~Alexandre,
  ``Dynamical mass generation in Lorentz-violating QED,''
  arXiv:1009.5834 [hep-ph];
\bibitem{JA2}
  J.~Alexandre and A.~Vergou,
  ``Properties of a consistent Lorentz-violating Abelian gauge theory,''
  Phys.\ Rev.\ D {\bf 83} (2011) 125008
  [arXiv:1103.2701 [hep-th]].

\bibitem{AM1}
  J.~Alexandre and N.~E.~Mavromatos,
  ``Mass Hierarchies in Lorentz-Violation-induced Dynamical Mass Models,''
  Phys.\ Rev.\ D {\bf 83} (2011) 127703
  [arXiv:1104.1583 [hep-th]].
  

\bibitem{mag2}
V.~P.~Gusynin, V.~A.~Miransky and I.~A.~Shovkovy,
``Catalysis of dynamical flavor symmetry breaking by a magnetic field in
  (2+1)-dimensions,''
  Phys.\ Rev.\ Lett.\  {\bf 73} (1994) 3499
  [Erratum-ibid.\  {\bf 76} (1996) 1005]
  [arXiv:hep-ph/9405262];
\bibitem{mag4}
J.~Alexandre, K.~Farakos and G.~Koutsoumbas,
  ``Remark on the momentum dependence of the magnetic catalysis in QED,''
  Phys.\ Rev.\  D {\bf 64} (2001) 067702;
  
\bibitem{mag5}
J.~Alexandre, K.~Farakos, S.~J.~Hands, G.~Koutsoumbas and S.~E.~Morrison,
  ``QED(3) with dynamical fermions in an external magnetic field,''
  Phys.\ Rev.\  D {\bf 64} (2001) 034502
  [arXiv:hep-lat/0101011];
\bibitem{mag6}
J.~Alexandre,
  ``Vacuum polarization in thermal QED with an external magnetic field,''
  Phys.\ Rev.\ D {\bf 63} (2001) 073010
  [hep-th/0009204].

 \bibitem{PT} For a review see: D.~Binosi and J.~Papavassiliou,
  ``Pinch Technique: Theory and Applications,''
  Phys.\ Rept.\  {\bf 479} (2009) 1
  [arXiv:0909.2536 [hep-ph]] and references therein. 
	
\bibitem{NM}
  N.~E.~Mavromatos,
  ``Quantum-Gravity Induced Lorentz Violation and Dynamical Mass Generation,''
  Phys.\ Rev.\ D {\bf 83} (2011) 025018
  [arXiv:1011.3528 [hep-ph]].

	
\bibitem{Anselmi5}
  A.~Dhar, G.~Mandal, S.~R.~Wadia and ,
  ``Asymptotically free four-fermi theory in 4 dimensions at the z=3 Lifshitz-like fixed point,''
  Phys.\ Rev.\ D {\bf 80} (2009) 105018
  [arXiv:0905.2928 [hep-th]];
\bibitem{Anselmi6}
  D.~Anselmi and E.~Ciuffoli,
  ``Low-energy Phenomenology Of Scalarless Standard-Model Extensions With High-Energy Lorentz Violation,''
  Phys.\ Rev.\ D {\bf 83} (2011) 056005
  [arXiv:1101.2014 [hep-ph]].

  
\bibitem{AFMP1}
  J.~Alexandre, K.~Farakos, N.~E.~Mavromatos and P.~Pasipoularides,
  ``Neutrino oscillations in a stochastic model for space-time foam,''
  Phys.\ Rev.\ D {\bf 77} (2008) 105001
  [arXiv:0712.1779 [hep-ph]];
\bibitem{AFMP2}
	J.~Alexandre, K.~Farakos, N.~E.~Mavromatos and P.~Pasipoularides,
  ``Neutrino oscillations in a Robertson-Walker Universe with space time foam,''
  Phys.\ Rev.\ D {\bf 79} (2009) 107701
  [arXiv:0902.3386 [hep-ph]].
	
	
\bibitem{dynmassA1}
  R.~Jackiw, K.~Johnson and ,
  ``Dynamical Model of Spontaneously Broken Gauge Symmetries,''
  Phys.\ Rev.\ D {\bf 8} (1973) 2386;
\bibitem{dynmassA2}
  J.~M.~Cornwall, R.~E.~Norton and ,
  ``Spontaneous Symmetry Breaking Without Scalar Mesons,''
  Phys.\ Rev.\ D {\bf 8} (1973) 3338.

  
\bibitem{AM2}
  J.~Alexandre and N.~E.~Mavromatos,
  ``A Lorentz-Violating Alternative to Higgs Mechanism?,''
  Phys.\ Rev.\ D {\bf 84} (2011) 105013
  [arXiv:1108.3983 [hep-ph]].
  
  
\bibitem{feynman} 
	H.~Hellmann, 
	``Einfuhrung in die Quantenchemie''
  (Leipzig: Franz Deuticke 1937), p. 285;
\bibitem{feynman2} 
  R.~P.~Feynman,
  ``Forces in Molecules,''
  Phys.\ Rev.\  {\bf 56}, 340 (1939).
  doi:10.1103/PhysRev.56.340
  	
\bibitem{numsm1} 
	M.~Shaposhnikov,
  ``The nuMSM, leptonic asymmetries, and properties of singlet fermions,''
  JHEP {\bf 0808} (2008) 008
  [arXiv:0804.4542 [hep-ph]];
\bibitem{numsm2} 
	M.~Shaposhnikov,
	``The nuMSM, dark matter and neutrino masses,''
  J.\ Phys.\ Conf.\ Ser.\  {\bf 39} (2006) 176 and references therein.
	
  
\bibitem{mavroLV}  N.~E.~Mavromatos,
  ``Quantum-Gravity Induced Lorentz Violation and Dynamical Mass Generation,''
  Phys.\ Rev.\ D {\bf 83}, 025018 (2011)
  [arXiv:1011.3528 [hep-ph]].
	    
	\bibitem{Lif4fermion}
  D.~Anselmi,
  ``Standard Model Without Elementary Scalars And High Energy Lorentz Violation,''
  Eur.\ Phys.\ J.\ C {\bf 65} (2010) 523
  [arXiv:0904.1849 [hep-ph]];
  
\bibitem{AB}  
 J.~Alexandre and J.~Brister,
  ``Fermion effective dispersion relation for z=2 Lifshitz QED,''
  Phys.\ Rev.\ D {\bf 88} (2013) 065020
  [arXiv:1307.7613 [hep-th]].

\bibitem{brustein} 
  R.~Brustein, D.~Eichler and S.~Foffa,
  ``Probing the Planck scale with neutrino oscillations,''
  Phys.\ Rev.\ D {\bf 65}, 105006 (2002)
  [hep-ph/0106309].
  
\bibitem{Arkani}
  N.~Arkani-Hamed, H.~C.~Cheng, M.~Luty and J.~Thaler,
  ``Universal dynamics of spontaneous Lorentz violation and a new spin-dependent inverse-square law force,''
  JHEP {\bf 0507} (2005) 029
  [hep-ph/0407034].
  
  
\bibitem{Pospelov}
  M.~Pospelov and Y.~Shang,
  ``On Lorentz violation in Horava-Lifshitz type theories,''
  Phys.\ Rev.\ D {\bf 85} (2012) 105001
  [arXiv:1010.5249 [hep-th]].
  
\bibitem{Padilla} 
  I.~Kimpton and A.~Padilla,
  ``Matter in Horava-Lifshitz gravity,''
  JHEP {\bf 1304}, 133 (2013)
  [arXiv:1301.6950 [hep-th]].
	
\bibitem{AB2} 
  J.~Alexandre and J.~Brister,
  ``Effective matter dispersion relation in quantum covariant Horava-Lifshitz gravity,''
  Phys.\ Rev.\ D {\bf 92} (2015) 2,  024025
  [arXiv:1505.01392 [hep-th]].
 
  
\bibitem{Gibbons}  
  G.~W.~Gibbons, S.~W.~Hawking and M.~J.~Perry,
  ``Path Integrals and the Indefiniteness of the Gravitational Action,''
  Nucl.\ Phys.\ B {\bf 138} (1978) 141;

  
\bibitem{Mazur}  
  P.~O.~Mazur and E.~Mottola,
  ``The Gravitational Measure, Solution of the Conformal Factor Problem and Stability of the Ground State of Quantum Gravity,''
  Nucl.\ Phys.\ B {\bf 341} (1990) 187.
  
  
\bibitem{'tHooft}
  G.~'t Hooft,
 ``Perturbative quantum gravity'', International School of Subnuclear Physics, C02-08-29.1
  

  
\bibitem{Leibbrandt}
  G.~Leibbrandt,
  ``Introduction to the Technique of Dimensional Regularization,''
  Rev.\ Mod.\ Phys.\  {\bf 47} (1975) 849.


\bibitem{Weinzierl}
  S.~Weinzierl,
  ``Review on loop integrals which need regularization but yield finite results,''
  Mod.\ Phys.\ Lett.\ A {\bf 29} (2014) 15,  1430015
  [arXiv:1402.4407 [hep-ph]].


\bibitem{Dodorico}
  G.~D'~Odorico, J.~W.~Goossens and F.~Saueressig,
  ``Covariant computation of effective actions in Horava-Lifshitz gravity,''
  JHEP {\bf 1510} (2015) 126
  [arXiv:1508.00590 [hep-th]].


\bibitem{Ryder} 
  L.~H.~Ryder,
  ``Quantum Field Theory,''
  ISBN-9780521237642.


\bibitem{Higgsinst} 
  G.~Degrassi, S.~Di Vita, J.~Elias-Miro, J.~R.~Espinosa, G.~F.~Giudice, G.~Isidori and A.~Strumia,
  ``Higgs mass and vacuum stability in the Standard Model at NNLO,''
  JHEP {\bf 1208}, 098 (2012)
  [arXiv:1205.6497 [hep-ph]].
  
\bibitem{rajantie}
  M.~Herranen, T.~Markkanen, S.~Nurmi and A.~Rajantie,
  ``Spacetime curvature and the Higgs stability during inflation,''
  Phys.\ Rev.\ Lett.\  {\bf 113} (2014) 21,  211102
  [arXiv:1407.3141 [hep-ph]].
 
\end{thebibliography}
\end{document}